\documentstyle[amsfonts,11pt]{article}
\evensidemargin\oddsidemargin
\def\text#1{\mbox{ #1}\,}
\def\stackunder#1#2{\mathrel{\mathop{#2}\limits_{#1}}}

\newcommand{\cont}{\,\raisebox{0.3ex}{\tiny \underline{\quad}\hskip%
 -0.25ex \tiny {$\mid$}}\,}

\newtheorem{theorem}{Theorem}
\newtheorem{lemma}{Lemma}
\newtheorem{proposition}{Proposition}
\newtheorem{corollary}[theorem]{Corollary}

\begin{document}

\hfill
KL--TH--97/7 \bigskip

\begin{center}
{\LARGE Functional Representations for Fock Superalgebras}

\bigskip

{\large Joachim Kupsch\footnote{%
e-mail: kupsch@physik.uni-kl.de} and Oleg G. Smolyanov\footnote{On
leave of absense from Faculty of Mechanics and Mathematics, Moscow State
University, 119899 Moscow, Russia. e-mail: smolian@nw.math.msu.su}} \\ 

{\large Fachbereich Physik der Universit\"at Kaiserslautern\\D-67663
Kaiserslautern, Germany}

\medskip

{\large July 1997}
\end{center}

\begin{abstract}
 The Fock space of bosons and fermions and its underlying superalgebra are
represented by algebras of functions on a superspace. We define Gaussian
integration on infinite dimensional superspaces, and construct superanalogs
of the classical function spaces with a reproducing kernel -- including the
Bargmann-Fock representation -- and of the Wiener-Segal representation. The
latter representation requires the investigation of Wick ordering on ${Z }_2$%
-graded algebras. As application we derive a Mehler formula for the
Ornstein-Uhlenbeck semigroup on the Fock space.
\end{abstract}

\section{Introduction}

The representations which we consider in this paper are superanalogs of the
Bargmann-Fock \cite{Bargmann:1961}\cite{Segal:1962} and of the Wiener-Segal 
\cite{Wiener:1930}\cite{Ito:1951}\cite{Segal:1956} representations. In the
literature the former representation is also called Bargmann-Segal, or
holomorphic, or complex wave representation, and the latter is denoted as It{%
\^o}-Segal-Wiener or real wave representation. Moreover we include
reproducing kernel spaces \cite{Aronszajn:1950} which can be considered as
an abstract version of the Bargmann-Fock representation. These
representations realize the Fock space of a theory that includes bosons and
fermions as some spaces of functions defined on an infinite dimensional
superspace, and taking values in an auxiliary infinite dimensional Grassmann
algebra $\Lambda $ with a Hilbert norm \cite{Smolyanov/Shavgulidze:1988}\cite
{Smolyanov/Shavgulidze:1990}. The physical Fock space is the completion of a
superalgebra with the ${\Bbb Z}_2$-graded tensor product which combines the
symmetric tensor product of its bosonic subspace and the antisymmetric
tensor product of its fermionic subspace. The superspaces are subspaces of
Grassmann or $\Lambda $-modules generated by the one-particle subspaces of
the Fock space. The inner product of the spaces corresponding to the
Bargmann-Fock and to the Wiener-Segal representations can be defined by an
integration with respect to a so called Gaussian supermeasure, which is
defined on the whole complex Hilbert space (Bargmann-Fock representation) or
on its real subspace (Wiener-Segal representation). Actually in both cases
one has to use some extensions of these spaces to get $\sigma $-additive
measures. The version of Gaussian supermeasure presented in this paper
slightly deviates from the supermeasure which was first introduced in \cite
{Smolyanov/Shavgulidze:1988} and also used in \cite{Kupsch/Smolyanov:1996}.

A great part of the constructions of our paper is devoted to the methods of
reproducing kernel spaces. This more abstract approach has been used for the
study of stochastic processes and of the bosonic Fock space for a long time,
see e.g. \cite{Kallianpur:1970}\cite{Hida:1980}; but it is also an effective
tool for the investigation of the fermionic Fock space \cite{Kupsch:1991}.
The superanalog of the Bargmann-Fock representation will therefore be
developed in the more general context of reproducing kernel spaces. As in
the classical case the inner product is only defined in an abstract way
(without reference to an integral representation). But there is a great
flexibility in choosing the domain of the functions, including the full
analyticity domain used for the Bargmann-Fock representation and also the
real subdomain used for the Wiener-Segal representation.

Already within superalgebra of the the Fock space Wick (or normal) ordering
and a Wiener-Segal picture of the Fock space can be defined \cite{Kree:1978}%
\cite{Kree:1980}\cite{Kupsch:1990}. These algebraic constructions are the
starting point for the Wiener-Segal representation by functions on
superspaces. It is exactly the Wiener-Segal representation which allows to
transfer formulas of usual analysis on Gaussian measure spaces to
superanalysis. Our paper gives a systematic treatment of this
representation, which has scarcely been used in superanalysis.

For all these constructions one needs to consider not only the original Fock
space but the $\Lambda $-supermodule generated by it; just this supermodule
contains coherent states defined on the superspace of the one-particle
states. The representations of the Fock space by superfunctions (whose
values are elements of $\Lambda $) allows to represent operators in Fock
space by operators acting on the functional spaces and having a form similar
to operators acting on the classical function spaces, but now the fermionic
part is included.

The paper is organized as follows. In the short Sect.\ref{Term} we introduce
some terminology and notations of infinite dimensional superanalysis. The
adopted version of superanalysis that is often called functional
superanalysis, takes its origin (essentially for finite dimensional spaces)
in the papers of Rogers \cite{Rogers:1980}\cite{Rogers:1986}\cite
{Rogers:1995}, of Jadczyk and Pilch \cite{Jadczyk/Pilch:1981}, of Vladimirov
and Volovich \cite{Vladimirov/Volovich:1984}\cite{Vladimirov/Volovich:1985a}
and in the book of DeWitt \cite{DeWitt:1984}; our approach to the infinite
dimensional case is most close to the papers \cite
{Smolyanov/Shavgulidze:1988}\cite{Smolyanov/Shavgulidze:1990}, see also \cite
{Khrennikov:1988}. We should mention as well the books \cite{Berezin:1966}%
\cite{Berezin:1987} of Berezin whose work had a great influence on
superanalysis.

The next two Sects. \ref{Fock} and \ref{Gauss} are devoted to a description
of the Fock space ${\cal S}({\cal H}),$ generated by a ${\Bbb Z}_2$-graded
Hilbert space ${\cal H}$, of its underlying superalgebra, and of Wick
(normal) ordering within this superalgebra. These sections contain some
results of the papers \cite{Kree:1978}\cite{Kree:1980} and \cite{Kupsch:1990}
in a form suited to an extension to superanalysis. In the next Sect.5 we
define a Grassmann module extension of the Fock space, simply called $%
\Lambda $-extension, and denoted by ${\cal S}^\Lambda ({\cal H})$. This
extension plays an essential role in all our further investigations. Within $%
{\cal S}^\Lambda ({\cal H})$ we can define a generalization of the usual
(bosonic) coherent states. These states are used to represent the space $%
{\cal S}^\Lambda ({\cal H})$ -- and hence its subspace ${\cal S}({\cal H})$
-- by $\Lambda $-valued functions on the superspace ${\cal H}_\Lambda $,
which, in the sense of \cite{Smolyanov/Shavgulidze:1988}, is generated by
the Hilbert space ${\cal H}$. In Sect.\ref{Smeasure} we introduce the notion
of Gaussian supermeasures, which take over the role of Gaussian measures for
superanalysis. Actually we describe a connection between an algebraic
picture of integration based on Gaussian functionals (as done without
superanalysis in Sect.\ref{Gauss} and in \cite{Kupsch:1990}), and a more
analytic version of integration presented in the language of infinite
dimensional distributions. The latter type of integration can be considered
also as an infinite dimensional version of the Berezin integration \cite
{Berezin:1966}\cite{DeWitt:1984}\cite{Rogers:1984}\cite
{Vladimirov/Volovich:1985a}\cite{Vladimirov/Volovich:1985b}.

The main results of the paper are contained in Sects.\ref{RKS-BF} and \ref
{WS}. Applying techniques, used for the fermionic Fock space in \cite
{Kupsch:1991}, we construct a reproducing kernel space for the space ${\cal S%
}^\Lambda ({\cal H}),$ develop a superanalog of the Bargmann-Fock
representation, and discuss kernels for some linear operators in the Fock
space ${\cal S}({\cal H})$ (or also in ${\cal S}^\Lambda ({\cal H})$).
Moreover, with help of a generalized Wick ordering we derive a superanalog
of the Wiener-Segal representation, and we prove a superanalog of the Mehler
formula for the Ornstein-Uhlenbeck semigroup operating on the extended Fock
space ${\cal S}^\Lambda ({\cal H})$.

The paper has several appendices. In App.\ref{NormEst} we investigate some
norm estimates of tensor algebras. We prove in particular that there does
not exist any Hilbert norm on a tensor algebra (including the Grassmann
algebra) that satisfies the estimate $\left\| xy\right\| \leq c\left\|
x\right\| \left\| y\right\| $ for all elements $x,y$ with a constant $c=1,$
but we construct such a norm for $c=\sqrt{3}$. The further appendices
include some additional properties of Grassmann algebras and of coherent
states. In App.\ref{Integration} we indicate a method of fermionic
integration that uses an ordering prescription on the basic Hilbert space
without any $\Lambda $-extension.

A remark about supersymmetry should be added. Superanalysis as presented in
this paper deals with a theory which combines bosons and fermions, but it is
not necessarily related to a supersymmetric theory. Superanalysis just
provides an appropriate kinematical basis for such a theory. Supersymmetry
is an additional constraint for functionals or operators. We give an example
of such a constraint at the end of Sect.\ref{Gauss}.

We would like to mention that the paper is self-contained, and the reader
need not to have any prior knowledge of superanalysis.

\section{Terminology and notations\label{Term}}

A vector space ${\cal E}$ over the complex or the real field is called $%
{\Bbb Z}_2$-graded if it is a direct sum of two its vector subspaces, which
are called -- and their elements also -- even and odd, ${\cal E}={\cal E}_{%
\overline{0}}\oplus {\cal E}_{\overline{1}}$. In this case one denotes by $%
\pi $ the function $\pi :x\in ({\cal E}_{\overline{0}}\cup {\cal E}_{%
\overline{1}})\backslash \{0\}\rightarrow \left\{ 0,1\right\} ,$ which is
the indicator of $\Lambda _1\backslash \{0\},$ i.e. $\pi (x)=k$ if $x\in 
{\cal E}_{\overline{k}}\backslash \{0\},k=0,1,$ and the value $\pi (x)$ is
called the parity of $x$. A superalgebra is a ${\Bbb Z}_2$-graded vector
space $\Lambda =\Lambda _{\overline{0}}\oplus \Lambda _{\overline{1}}$
equipped with an associative multiplication having the following properties:
if $a,b\in {\cal E}_{\overline{0}}\cup {\cal E}_{\overline{1}}$ with $ab\not
=0,$ then $\pi (ab)=|\pi (a)-\pi (b)|$. A superalgebra is called
(super)commutative if $ab=(-1)^{\pi (a)\pi (b)}ba$ for any $a,b\in (\Lambda
_{\overline{0}}\cup \Lambda _{\overline{1}})\backslash \{0\}$. We always
assume that any superalgebra has a unit that we denote by $e_0$ (or also by $%
\kappa _0$ if Greek letters are used for the elements of the superalgebra).
A ${\Bbb Z}_2$-graded locally convex space (LCS), respectively normed,
Banach, Hilbert etc. space is a ${\Bbb Z}_2$-graded vector space ${\cal E}_{%
\overline{0}}\oplus {\cal E}_{\overline{1}}$ equipped with a structure of a
LCS, respectively, of normed, Banach, Hilbert etc. space, such that ${\cal E}
$ is a topological sum (Hilbert sum in the Hilbert space case) of its
topological vector subspaces (Banach, Hilbert etc. subspaces) ${\cal E}_{%
\overline{0}}$ and ${\cal E}_{\overline{1}}$. In this case one also says
that the structure of a LCS (respectively, normed, Hilbert space) is
compatible with the structure of the ${\Bbb Z}_2$-graded vector space.

A locally convex superalgebra is a superalgebra $\Lambda =\Lambda _{%
\overline{0}}\oplus \Lambda _{\overline{1}}$, equipped with a structure of a
LCS that is compatible with the structure of the ${\Bbb Z}_2$ graded vector
space $\Lambda $ and is such that there exists a family ${\cal P}$ of
seminorms on $\Lambda $ defining the topology of $\Lambda ,$ the
multiplication $\Lambda \times \Lambda \to \Lambda $, $(ab)\to ab$ being
continuous with respect to each $p\in {\cal P}$. A superalgebra with a
Hilbert structure is a locally convex superalgebra $\Lambda $ whose topology
can be defined by a Hilbert norm $p(a)=\left\| a\right\| $ (such that $%
p(e_0)=1$). If this superalgebra is complete it is called a Hilbert
superalgebra. As a general rule such a Hilbert superalgebra is not a Banach
algebra in the usual sense, because a norm estimate $\parallel ab\parallel
\le \parallel a\parallel \parallel b\parallel $ for all $a,b\in \Lambda $ is
not admitted, see App.\ref{NormEst}. If not specified otherwise, we assume
that the annihilator $\Lambda _{\overline{1}}^{\perp }=\{x\in \Lambda
;\;\forall z\in \Lambda _{\overline{1}},xz=0\}$ of the odd part of any
superalgebra $\Lambda $ is equal to zero, and hence this superalgebra is
infinite dimensional with $\dim \Lambda _{\overline{1}}=\infty $.\footnote{%
If $\dim \Lambda <\infty $ let $\left\{ e_j\right\} $ be a basis of the
vector space $\Lambda _{\overline{1}}$ and let $n$ be the maximal natural
number for which there exist $n$ elements, $e_{j_1},\ldots ,e_{j_n}$ of the
basis, such that $e_{j_1}\cdot \cdot \cdot e_{j_n}\not =0$. Then for any
element $e_k$ of the basis $e_ke_{j_1}\cdot \cdot \cdot e_{j_n}=0$ and hence 
$e_{j_1}\cdot \cdot \cdot e_{j_n}\in \Lambda _{\overline{1}}^{\perp }$.}

A Grassmann algebra is an algebra $\Lambda $ with a unit $e_0$ that contains
a non-empty finite or countable set of elements $\{e_j\mid j\in {\bf B}%
\subset {\Bbb N}\}$, which are called generators, having the following
properties: the set $\{e_0\}\cup \{e_j\}$ is linearly independent; $\forall
i,j\in {\Bbb N},e_i,e_j=-e_je_i$ and $e_ie_j\not =0$ if $i\not =j$, the
minimal subalgebra of $\Lambda $ containing the set $\{e_0\}\cup \{e_j\}$
coincides with $\Lambda $. Each Grassmann algebra can be equipped with the
following structure of superalgebra. Let $\Lambda _{\overline{1}}$ be a
vector subspace of $\Lambda $ generated by all products $e_{j_1}\ldots
e_{j_n},j_n\in {\bf B}$ where the number $n$ of multipliers is odd and let $%
\Lambda _{\overline{0}}$ be a vector subspace of $\Lambda $ generated by $%
e_0 $ and by all products $e_{j_1}\ldots e_{j_n}$, where the number of
multipliers is even. Then the decomposition $\Lambda =\Lambda _{\overline{0}%
}\oplus \Lambda _{\overline{1}}$ together with the function $\pi :(\Lambda _{%
\overline{0}}\cup \Lambda _{\overline{1}})\backslash \{0\}\to \{0,1\}$ which
is the indicator of $\Lambda _{\overline{1}}\backslash \{0\}$, defines the
structure of a superalgebra. In App.\ref{NormEst} we prove that any
Grassmann algebra can be equipped with a Hilbert structure, and in the
following we always assume that a Grassmann algebra has such a structure.
(This structure does not depend on the choice of the set of generators.) The
completion of an infinite dimensional Grassmann algebra with a Hilbert
structure with respect to the corresponding Hilbert norm is called a
Hilbert-Grassmann algebra; hence a Hilbert-Grassmann algebra is a Hilbert
superalgebra with vanishing annihilator of the odd part. Everywhere below we
take as superalgebra $\Lambda =\Lambda _{\overline{0}}\oplus \Lambda _{%
\overline{1}}$ a Hilbert-Grassmann algebra over the complex numbers. But
many definitions and statements are valid for any infinite dimensional
superalgebra.

If ${\cal E}$ and ${\cal G}$ are vector spaces then the symbol ${\cal E}%
\otimes {\cal G}$ denotes their algebraic tensor product. If ${\cal E}$ and $%
{\cal G}$ are equipped with Hilbert norms then ${\cal E}\otimes {\cal G}$ is
assumed to be equipped with the (uniquely defined) corresponding Hilbert
cross-norm, and by ${\cal E}\widehat{\otimes }{\cal G}$ one denotes the
completion of ${\cal E}\otimes {\cal G}$. For any vector space ${\cal E}$
the vector space $\Lambda \otimes {\cal E}$ can be provided with different
structures of a $\Lambda $-module; we will specify them later. If ${\cal E}$
is a ${\Bbb Z}_2$-graded vector space, then $\Lambda \otimes {\cal E}$ is
also equipped with the natural structures of ${\Bbb Z}_2$-graded $\Lambda $
module; in particular, the ${\Bbb Z}_2$-graded vector space $\Lambda \otimes 
{\cal E}$ with the structure of a $\Lambda $-module will be denoted by $%
{\cal E}^\Lambda $ and it will be called $\Lambda $-supermodule over ${\cal E%
}$. The even $\Lambda _{\overline{0}}$--submodule $\Lambda _{\overline{0}%
}\otimes {\cal E}_{\overline{0}}\oplus \Lambda _{\overline{1}}\otimes {\cal E%
}_{\overline{1}}$ of ${\cal E}^\Lambda $ is called the $\Lambda $-{\it %
superspace} over ${\cal E}$ and is denoted by ${\cal E}_\Lambda $. In the
usual superanalysis of finite dimensional spaces, $\dim {\cal E}_{\overline{0%
}}=m,\dim {\cal E}_{\overline{1}}=n,$ the space ${\cal E}_\Lambda $ can be
represented in another way. Choosing a basis $\left\{ e_a\right\}
_{a=1,...,m}$ of ${\cal E}_{\overline{0}}$ and a basis $\left\{ f_b\right\}
_{b=1,...,n}$ of ${\cal E}_{\overline{1}}$ the variable $\zeta =\zeta _{%
\overline{0}}+\zeta _{\overline{1}}\in \Lambda _{\overline{0}}\otimes {\cal E%
}_{\overline{0}}\oplus \Lambda _{\overline{1}}\otimes {\cal E}_{\overline{1}%
} $ can be decomposed as $\zeta =\sum_{a=1}^m\lambda _a\otimes
e_a+\sum_{b=1}^n\mu _b\otimes f_b$ with $\lambda _a\in \Lambda _{\overline{0}%
}$ and $\mu _b\in \Lambda _{\overline{1}},$ and $\zeta $ is mapped onto an
element $\left( \lambda _1,...,\lambda _m;\mu _1,...,\mu _n\right) \in
\Lambda ^{m,n}=\left( \Lambda _{\overline{0}}\right) ^m\times \left( \Lambda
_{\overline{1}}\right) ^n$ The space $\Lambda ^{m,n}\cong \Lambda _{%
\overline{0}}\otimes {\Bbb C}^m\oplus \Lambda _{\overline{1}}\otimes {\Bbb C}%
^n$ is then denoted as superspace.\cite{Rogers:1980}\cite{Jadczyk/Pilch:1981}%
\cite{DeWitt:1984}\cite{Vladimirov/Volovich:1984}

If ${\cal E}$ is equipped with a Hilbert norm then ${\cal E}_\Lambda $ and $%
{\cal E}^\Lambda $ are assumed to be equipped with corresponding Hilbert
norms. Moreover, if ${\cal E}$ is a Hilbert space, we always assume that the
spaces ${\cal E}_\Lambda $ and ${\cal E}^\Lambda $ are completed.

If ${\cal E}$ and ${\cal G}$ are normed spaces then by ${\cal L}({\cal E},%
{\cal G})$ one denotes the space of all continuous mappings of ${\cal E}$
into ${\cal G}$. If ${\cal E}$ and ${\cal G}$ are Hilbert spaces, by ${\cal L%
}({\cal E}_\Lambda ,{\cal G}^\Lambda )$ one denotes the topological $\Lambda 
$-module space of $\Lambda _{\overline{0}}$-linear mappings of ${\cal E}%
_\Lambda $ into ${\cal G}^\Lambda $ equipped with the topology of bounded
convergence. A mapping $f:{\cal E}_\Lambda \to {\cal G}^\Lambda $ is called
(Fr\'echet) differentiable at $x_0\in {\cal E}_\Lambda $ if there exists an
element of ${\cal L}({\cal E}_\Lambda ,{\cal G}^\Lambda ),$ called the
(Fr\'echet) derivative of $f$ at $x_0$ and denoted by $f^{\prime }(x_0)$
such that for any bounded sequence $\eta _n\in {\cal E}_\Lambda $ the limit $%
t^{-1}\left( f(x_0+t_n\eta _n)-f(x_0)\right) -f^{\prime }(x_0)\eta _n\to 0$
if $t_n\to 0$ with $t_n\in R^1,t_n\not =0$ exists. The derivatives of higher
order $f^{(n)}(x),n\in {\Bbb N,}$ are defined by induction, such that $%
f^{(n)}(x)\in {\cal L}_n({\cal E}_\Lambda ,{\cal G}^\Lambda )$ where ${\cal L%
}_n({\cal E}_\Lambda ,{\cal G}^\Lambda )$ is the vector space of all $n$%
-fold $\Lambda _{\overline{0}}$--linear continuous mappings of ${\cal E}%
_\Lambda \times \ldots \times {\cal E}_\Lambda $ into $G^\Lambda $ equipped
with the topology of bounded convergence.

\section{Fock Spaces and Superalgebras\label{Fock}}

\subsection{${\Bbb Z}_2$-graded Hilbert spaces\label{Hilbert}}

Let ${\cal E}$ be a complex separable Hilbert space and ${\cal E}^{*}$ its
dual space. Then the direct sum ${\cal F}={\cal E}\oplus {\cal E}^{*}$ is
again a Hilbert space, the inner product denoted by $(f,g)$, with the
structure given by

\begin{enumerate}
\item  An antiunitary involution $f\in {\cal F}\rightarrow f^{*}\in {\cal F}$%
, $f^{**}=f$, which maps ${\cal E}$ into ${\cal E}^{*}$ and vice versa.

\item  The bilinear symmetric form $\omega _{+}(f,g)=\left\langle
f|g\right\rangle $ on ${\cal F}\times {\cal F}$ that is defined by $\omega
_{+}(f,g)\equiv \left\langle f|g\right\rangle :=(f^{*},g).$ The spaces $%
{\cal E}$ and ${\cal E}^{*}$ are isotropic subspaces of this form.

\item  The canonical antisymmetric bilinear form on ${\cal F}\times {\cal F}%
:\omega _{-}(f,g)\equiv \left\langle f|j_{-}g\right\rangle ,$where $j_{-}$
is the mapping of ${\cal F}$ into ${\cal F}$ defined by $j_{-}(u+v):=u-v$ if 
$u\in {\cal E}$ and $v\in {\cal E}^{*},$ i.e. $j_{-}$ interchanges the
relative phases between ${\cal E}$ and ${\cal E}^{*}.$
\end{enumerate}

\noindent To have shorter notations we define by $j_{+}$ the identical
transformation of ${\cal F}$( $j_{+}f=f,f\in {\cal F}$) then $\omega _{\pm
}(f,g):=\left\langle f|j_{\pm }g\right\rangle .$ The diagonal subspace 
\begin{equation}
{\cal F}^D:=\{f|f=f^{*},f\in {\cal F}\}.  \label{h.4}
\end{equation}
is a real Hilbert space, which is isomorphic to ${\cal E}_{{\Bbb R}},$ the
underlying real space of the complex space ${\cal E},\quad u\in {\cal E}_{%
{\Bbb R}}\leftrightarrow f=\frac 1{\sqrt{2}}(u+u^{*})\in {\cal F}^D.$ The
space ${\cal F}$ is the complexification of both the spaces ${\cal F}^D$ and 
${\cal E}_{{\Bbb R}}.$ If $u,v\in {\cal E}_{{\Bbb R}}$ and $f=\frac 1{\sqrt{2%
}}(u+u^{*}),g=\frac 1{\sqrt{2}}(v+v^{*})\in {\cal F}^D$, then $\omega
_{+}(f,g)=\left\langle f\mid g\right\rangle ={\rm Re}\left( u,v\right) $ and 
$\omega _{-}(f,g)=i{\rm Im}\left( u,v\right) $.

In the following we denote a Hilbert space ${\cal F}$ provided with the
symmetric bilinear form $\omega _{+}(f,g)$ as ${\cal H}_{\overline{0}}$, and
a Hilbert space ${\cal F}$ provided with the antisymmetric form $\omega
_{-}(f,g)$ as ${\cal H}_{\overline{1}}.$ The spaces ${\cal H}_{\overline{0}}$
and ${\cal H}_{\overline{1}}$ can be interpreted as spaces of bosons and of
fermions. We do not assume an isomorphism between the spaces ${\cal H}_{%
\overline{0}}$ and ${\cal H}_{\overline{1}}$; but see the end of Sect.\ref
{Gauss}. The direct sum ${\cal H}={\cal H}_{\overline{0}}\oplus {\cal H}_{%
\overline{1}}$ of these spaces is of crucial importance in the following.
With the parity $\pi (f)=k$ if $f\in {\cal H}_{\overline{k}}\backslash
\left\{ 0\right\} ,k=0,1,$ the space ${\cal H}={\cal H}_{\overline{0}}\oplus 
{\cal H}_{\overline{1}}$ is a ${\Bbb Z}_2$-graded Hilbert space. The
corresponding isotropic subspaces will be denoted as ${\cal E}_{\overline{k}%
} $ or ${\cal E}_{\overline{k}}^{*}$ , $k=0,1,$ and the involution $%
f\rightarrow f^{*}$ maps ${\cal E}_{\overline{k}}$ into ${\cal E}_{\overline{%
k}}^{*}.$ One can interpret the spaces ${\cal E}_{\overline{k}}$ and ${\cal E%
}_{\overline{k}}^{*},k=0,1,$ as the spaces of particles and of
antiparticles. The space ${\cal H}$ is then provided with a bilinear form $%
\omega ,$ defined by 
\begin{equation}
\omega (f,g)=\omega _{+}(f_0,g_0)+\omega _{-}(f_1,g_1)  \label{h.7}
\end{equation}
for $f=f_0+f_1$ and $g=g_0+g_1,$ with $f_k,g_k\in {\cal H}_{\overline{k}%
},k=0,1.$ This form has the symmetry property $\omega (f,g)=(-1)^{\pi (f)\pi
(g)}\omega (g,f)$ for all $f,g\in {\cal H}_{\overline{0}}\cup {\cal H}_{%
\overline{1}}\setminus \left\{ 0\right\} .$ On the other hand we define the
bilinear symmetric form 
\begin{equation}
\left\langle f\mid g\right\rangle =\left( f^{*}\mid g\right)  \label{h.9}
\end{equation}
and the forms (\ref{h.7}) and (\ref{h.9}) are related by $\omega
(f,g)=\left\langle f\mid jg\right\rangle $ where $j$ is defined by 
\begin{equation}
j(u+b)=u-v\text{ if }u\in {\cal H}_{\overline{0}}\oplus {\cal E}_{\overline{1%
}}\text{ and }v\in {\cal E}_{\overline{1}}^{*}.  \label{h.11}
\end{equation}

\subsection{Tensor algebras and superalgebra\label{alg}}

The incomplete (algebraic) tensor product of $n$ copies of a Hilbert space $%
{\cal H}$ is denoted as ${\cal H}^{\otimes n}.$ The algebraic sum of these
spaces is the tensor algebra ${\cal T}_{fin}{\cal (H)=}\stackunder{n\geq 0}{%
\oplus }{\cal H}^{\otimes n}.$ We provide the spaces ${\cal H}^{\otimes n}$
with the standard Hilbert norm $\parallel f_1\otimes ...\otimes f_n\parallel
_n^2:=\prod_{a=1}^n\parallel f_a\parallel ^2.$ The completed tensor spaces $%
{\cal H}^{\otimes n}$are denoted by ${\cal T}_n({\cal H}).$ The space ${\cal %
T}_{fin}{\cal (H)}$ is equipped with the Hilbert space norm 
\begin{equation}
\parallel F\parallel ^2=\sum_{n=0}^\infty \parallel F_n\parallel _n^2\text{
if }F=\sum_{n=0}^\infty F_n\text{ with }F_n\in {\cal H}^{\otimes n}.
\label{alg1a}
\end{equation}

Let ${\cal H}={\cal H}_{\overline{0}}\oplus {\cal H}_{\overline{1}}$ be the
graded Hilbert space of the preceding section$.$ The subspace ${\cal H}_{%
\overline{0}}$ (${\cal H}_{\overline{1}}$) carries the bosonic (fermionic)
degrees of freedom. We define a bounded linear operator ${\bf P}^{(n)}$ on
the set of decomposable tensors $f_1\otimes ...\otimes f_n$ of vectors $f_a$
with a defined parity, $f_a\in {\cal H}_{\overline{0}}\cup {\cal H}_{%
\overline{1}},$ by 
\begin{equation}
{\bf P}^{(n)}f_1\otimes ...\otimes f_n=\frac 1{n!}\stackunder{\sigma }{\sum }%
\chi _\sigma (f_1,...,f_n)f_{\sigma (1)}\otimes ...\otimes f_{\sigma (n)}
\label{alg2}
\end{equation}
with a sign function $\chi _\sigma (f_1,...,f_n)=(-1)^{N_\sigma }$ where

\noindent $N_\sigma =\#\left\{ (a_i,a_j)\mid \pi (f_{a_i})=\pi (f_{a_j})=1,%
\text{ with }i<j\text{ and }a_i>a_j\right\} $ counts the inversions of the
fermionic arguments. The sum extends over all permutations $\sigma $ of the
numbers $\left\{ 1,...,n\right\} $. This mapping has a unique extension to a
projection operator on the space ${\cal T}_n({\cal H})$. The image of the
space ${\cal H}^{\otimes n}$ is denoted as ${\cal H}^{\odot n}={\bf P}^{(n)}%
{\cal H}^{\otimes n}$. The algebraic (finite) sum of the spaces ${\cal H}%
^{\odot n}$%
\begin{equation}
{\cal S}_{fin}({\cal H}):=\stackunder{n}{\oplus }{\cal H}^{\odot n}.
\label{alg3}
\end{equation}
is a commutative superalgebra with the graded product $F\in {\cal H}^{\odot
p},G\in {\cal H}^{\odot q}\rightarrow F\odot G\in {\cal H}^{\odot (p+q)}$%
\begin{equation}
F\odot G:{\cal =}\sqrt{\frac{(p+q)!}{p!q!}}{\bf P}^{(p+q)}F\otimes G
\label{alg4}
\end{equation}
This product reduces to the symmetric tensor product if only tensors of the
space ${\cal H}_{\overline{0}}$ are used, and it coincides with the
antisymmetric tensor product, if all factors are tensors of the space ${\cal %
H}_{\overline{1}}.$\footnote{%
We also use the notation $F\vee G$ for the symmetric tensor product, and $%
F\wedge G$ for the antisymmetric tensor product.} The definitions (\ref
{alg1a}) and (\ref{alg4}) yield the usual norm for the
symmetric/antisymmetric tensor product $\parallel f_1\odot ...\odot
f_n\parallel _n^2=\parallel f_1\vee ...\vee f_n\parallel _n^2={\rm per}%
\left\langle f_a\mid f_b\right\rangle $ if $f_a\in {\cal H}_{\overline{0}},$ 
$a=1,...,n,$ and $\parallel f_1\odot ...\odot f_n\parallel _n^2=\parallel
f_1\wedge ...\wedge f_n\parallel _n^2=\det \left\langle f_a\mid
f_b\right\rangle $ if $f_a\in {\cal H}_{\overline{1}}$ , $a=1,...,n.$ If $%
{\cal G}\subset {\cal H}$ is a subset of ${\cal H}$ we use the notation $%
{\cal S}_{fin}{\cal (G)}$ for the subalgebra generated by vectors from $%
{\cal G}$ with the product (\ref{alg4}). Hence ${\cal S}_{fin}{\cal (H}_{%
\overline{0}}{\cal )}$ is the symmetric tensor algebra of the Hilbert space $%
{\cal H}_{\overline{0}},$ and ${\cal S}_{fin}{\cal (H}_{\overline{1}}{\cal )}
$ is the antisymmetric tensor algebra of the Hilbert space ${\cal H}_{%
\overline{1}}.$ These symmetric/antisymmetric tensor algebras will also be
denoted as ${\cal T}_{fin}^{+}{\cal (H}_{\overline{0}}{\cal )}$ or ${\cal T}%
_{fin}^{-}{\cal (H}_{\overline{1}}{\cal )},$ respectively.

With the product (\ref{alg4}) the space ${\cal S}_{fin}({\cal H})={\cal S}%
_{fin}({\cal H}_{\overline{0}})\odot {\cal S}_{fin}({\cal H}_{\overline{1}})$
is a ${\Bbb Z}_2$-graded algebra with even subspace ${\cal S}_{fin}({\cal H}%
_{\overline{0}})\odot \left( \oplus _{q\geq 0,even}{\cal H}_{\overline{1}%
}^{\odot q}\right) $ and with the odd subspace ${\cal S}_{fin}({\cal H}_{%
\overline{0}})\odot \left( \oplus _{q\geq 0,odd}{\cal H}_{\overline{1}%
}^{\odot q}\right) .$ The completion of ${\cal H}^{\odot n}$ with the norm $%
\left\| .\right\| _n$ is denoted as ${\cal S}_n({\cal H})$. The completion
of ${\cal S}_{fin}{\cal (H)}$ with the norm (\ref{alg1a}) $\left\| \cdot
\right\| $ is the Fock space 
\begin{equation}
{\cal S(H)}=\stackunder{n=0}{\stackrel{\infty }{\oplus }}{\cal S}_n({\cal H}%
).  \label{alg7}
\end{equation}
The product of two homogeneous elements $F\in {\cal H}^{\odot p}$ and $G\in 
{\cal H}^{\odot q}$ is continuous and can be extended to the completed
spaces $F\in {\cal S}_p{\cal (H)},G\in {\cal S}_q{\cal (H)}\rightarrow
F\odot G\in {\cal S}_{p+q}{\cal (H)}$ with the norm estimate, see (\ref
{alg1a}) and (\ref{alg4}), 
\begin{equation}
\left\| F\odot G\right\| ^2\leq \frac{(p+q)!}{p!q!}\left\| F\right\|
^2\left\| G\right\| ^2.  \label{alg8}
\end{equation}
In agreement with the notations given above we shall also write ${\cal S(H}_{%
\overline{0}}{\cal )}$ for the closed Fock space of symmetric tensors of the
bosonic space ${\cal H}_{\overline{0}}$ and ${\cal S}({\cal H}_{\overline{1}%
})$ for the closed Fock space of antisymmetric tensors of the fermionic
space ${\cal H}_{\overline{1}}.$

The superalgebra ${\cal S}_{fin}{\cal (H)}$ is isomorphic to the tensor
product 
\begin{equation}
{\cal S}_{fin}{\cal (H)}\cong {\cal S}_{fin}({\cal H}_{\overline{0}})\otimes 
{\cal S}_{fin}({\cal H}_{\overline{1}})={\cal T}_{fin}^{+}({\cal H}_{%
\overline{0}})\otimes {\cal T}_{fin}^{-}({\cal H}_{\overline{1}})
\label{alg6}
\end{equation}
of the symmetric tensor algebra ${\cal T}_{fin}^{+}({\cal H}_{\overline{0}})$
and of the antisymmetric tensor algebra ${\cal T}_{fin}^{-}({\cal H}_{%
\overline{1}}).$ In this representation the product (\ref{alg4}) of
factorizing tensors $F=F_0\otimes F_1,G=G_0\otimes G_1$ with $F_0,G_0\in 
{\cal S}_{fin}{\cal (H}_{\overline{0}}{\cal )}$ and $F_1,G_1\in {\cal S}%
_{fin}{\cal (H}_{\overline{1}}{\cal )}$ is given by $F\odot G=(F_0\vee
G_0)\otimes (F_1\wedge G_1).$

The involution $f\rightarrow f^{*}$ on the space ${\cal H}$ can be uniquely
extended to an isometric involution $F\rightarrow F^{*}$ on ${\cal S(H)}$,
which satisfies $(F\odot G)^{*}=G^{*}\odot F^{*}$ for all $F,G\in {\cal S(H)}
$ for which the graded tensor product exists. This involution induces a
bilinear pairing on ${\cal S(H)}$ 
\begin{equation}
\left\langle F\mid G\right\rangle =\left( F^{*}\mid G\right) .  \label{alg11}
\end{equation}
This duality has been called {\it twisted duality} in \cite{Kree:1980}
because the order of the left argument is inverted 
\begin{equation}
\left\langle f_m\odot ...\odot f_1\mid x_1\odot ...\odot x_n\right\rangle
=\delta _{mn}\sum \chi _\sigma (x_1,...,x_n)\left\langle f_1\mid x_{\sigma
(1)}\right\rangle \cdot \cdot \cdot \left\langle f_n\mid x_{\sigma
(n)}\right\rangle  \label{alg11a}
\end{equation}
The normalizations given in (\ref{alg1a}) and (\ref{alg4}) lead to the
following important formula. Let ${\cal G}_1,{\cal G}_2\subset {\cal H}$ be
orthogonal subspaces, ${\cal G}_1\perp {\cal G}_2$, then 
\begin{equation}
\left\langle G\odot F\mid X\odot Y\right\rangle =\left\langle F\mid
X\right\rangle \left\langle G\mid Y\right\rangle  \label{alg12}
\end{equation}
holds if $F\in {\cal S}({\cal G}_1),$ $X\in {\cal S(G}_1^{*}{\cal )}$ and $%
G\in {\cal S}({\cal G}_2),Y\in {\cal S}({\cal G}_2^{*}).$

The {\it interior product} or {\it contraction} of a tensor $F\in {\cal S}%
_{fin}({\cal H})$ with a tensor $Y\in {\cal S}({\cal H})$ is the unique
element $Y{\cont }F$ of ${\cal S}_{fin}({\cal H}{\cal )}$ for which the
identity 
\begin{equation}
\left\langle X\mid Y\cont F\right\rangle =\left\langle X\odot Y\mid
F\right\rangle  \label{alg13}
\end{equation}
is valid for all $X\in {\cal S}_{fin}({\cal H})$. The identity (\ref{alg13})
is equivalent to $\left( X\mid Y\cont F\right) =\left( Y^{*}\odot X^{*}\mid
F\right) .$

The norm (\ref{alg1a}) can be generalized to 
\begin{equation}
\parallel F\parallel _{(\gamma )}^2=\sum_{n=0}^\infty (n!)^\gamma \parallel
F_n\parallel _n^2\text{ if }F=\sum_{n=0}^\infty F_n\text{ with }F_n\in {\cal %
H}^{\otimes n}  \label{alg15}
\end{equation}
with a parameter $\gamma \in {\Bbb R}.$ The completion of ${\cal S}_{fin}(%
{\cal H})$ with the norm $\parallel .\parallel _{(\gamma )}$ is denoted by $%
{\cal S}_{(\gamma )}({\cal H})$. These spaces satisfy the obvious inclusions 
${\cal S}_{(\alpha )}({\cal H})\subset {\cal S}_{(\beta )}({\cal H})$ if $%
\alpha \geq \beta ,$ and the Fock space ${\cal S}({\cal H})$ is ${\cal S}%
_{(0)}({\cal H}).$ The tensor product (\ref{alg4}) and the contraction (\ref
{alg13}) have no continuous extension to the Fock space ${\cal S(H)}$, but
in App.\ref{NormEst} we derive the following statements.

\begin{lemma}
\label{TensorNorm}If $F,G\in {\cal S}_{(\gamma )}({\cal H}),\gamma >0,$ then 
$F\odot G\in {\cal S(H)}$ with $\left\| F\odot G\right\| \leq c_\gamma \cdot
\left\| F\right\| _{(\gamma )}\left\| G\right\| _{(\gamma )}.$\\ If $Y\in 
{\cal S(H)}$ and $F\in {\cal S}_{(\gamma )}({\cal H}),\gamma >0,$ then $Y%
\cont F\in {\cal S(H)}$ with $\left\| Y\cont F\right\| \leq c_\gamma \cdot
\left\| Y\right\| \left\| F\right\| _{(\gamma )}.$
\end{lemma}

Finally we remind a standard notation for operators on Fock spaces. If $T$
is a (bounded) linear operator on ${\cal H},$ then its so called second
quantization $\Gamma (T)$ is a linear (bounded) operator on ${\cal S}%
_{(\gamma )}({\cal H}),$ defined by $\Gamma (T)e_0=e_0$ (unit = vacuum
state), and $\Gamma (T)\left( f_1\odot ...\odot f_n\right) =\left(
Tf_1\right) \odot ...\odot \left( Tf_n\right) $ for $f_a\in {\cal H}%
,a=1,...,n.$ The linear operator $d\Gamma (T)$ is defined as the derivation
of the algebra ${\cal S}_{fin}({\cal H})$ generated by the operator $T,$
i.e. $d\Gamma (T)e_0=0,d\Gamma (T)f=Tf$ if $f\in {\cal H},$ and $d\Gamma
(T)\left( F\odot G\right) =\left( d\Gamma (T)F\right) \odot G+F\odot \left(
d\Gamma (T)G\right) $ for $F,G\in {\cal S}_{fin}({\cal H}).$

\section{Gaussian functionals and Wick ordering\label{Gauss}}

In this section we present an algebraic formulation of the integration with
respect to a Gaussian measure. These algebraic methods, developed in \cite
{Kree:1980} and \cite{Kupsch:1990}, have the advantage that bosonic and
fermionic integration can be treated with the same methods, such that the
generalization to the superalgebra ${\cal S}_{fin}({\cal H})$ is
straightforward. Wick ordering and the Wiener-Segal representation, well
known for the bosonic Fock space, have within this framework a transparent
generalization for the Fock space ${\cal S}({\cal H}).$

The starting point of a Gaussian integration is the bilinear form (\ref{h.7}%
). This form is continuous on ${\cal H}\times {\cal H}$, and it can be
written as 
\begin{equation}
\omega (f,g)=\left\langle \Omega \mid f\odot g\right\rangle  \label{Gauss1}
\end{equation}
with a tensor $\Omega =\Omega ^{*},$which is an element of $({\cal H}\odot 
{\cal H})^{^{\prime }}$, the algebraic dual of ${\cal H}\odot {\cal H}$. We
shall give another characterization of $\Omega $ bellow. The tensor $\Omega $
separates into a bosonic and a fermionic part $\Omega =\Omega _0+\Omega _1$
with $\Omega _k\in ({\cal E}_{\overline{k}}\odot {\cal E}_{\overline{k}%
}^{*})^{^{\prime }}\subset ({\cal H}\odot {\cal H})^{^{\prime }},k=0,1.$ The
exponential $\exp \Omega =I+\Omega +\frac 1{2!}\Omega \odot \Omega +...$
yields the Gaussian combinatorics on tensors $f_1\odot f_2\odot ...\odot
f_{2n}$ with $f_a\in {\cal H}_{\overline{0}}\cup {\cal H}_{\overline{1}%
},a=1,...,2n,$%
\begin{equation}
\left\langle \exp \Omega |f_1\odot f_2\odot ...\odot f_{2n}\right\rangle
=\frac 1{2^nn!}\sum_\sigma \chi _\sigma (f_1,...,f_{2n})\omega (f_{\sigma
(1)},f_{\sigma (2)})...\omega (f_{\sigma (2n-1)},f_{\sigma (2n)})
\label{Gauss2a}
\end{equation}
where the sum extends over all permutations $\sigma $ of the numbers $%
\left\{ 1,...,2n\right\} $ and the sign function $\chi _\sigma $ has been
defined for (\ref{alg2}). If all $f_a\in {\cal H}_{\overline{0}}$ , $%
a=1,...,2n,$ the sign function is $\chi _\sigma =1,$ and the sum (\ref
{Gauss2a}) agrees with the hafnian haf$(\omega (f_a,f_b))$, and if $f_a\in 
{\cal H}_{\overline{1}}$ , $a=1,...,2n,$ the sum (\ref{Gauss2a}) agrees with
the pfaffian pf$(\omega (f_a,f_b))$, see e.g.\cite{Caianiello:1973}. The
functional 
\begin{equation}
L(F):=\left\langle \exp \Omega |F\right\rangle  \label{Gauss2}
\end{equation}
is then linearly extended to ${\cal S}_{fin}({\cal H})$. Moreover, if $%
F,G\in {\cal S}_{fin}({\cal E})$ a simple calculation leads to the identity 
\begin{equation}
L(F_1^{*}\odot F_2)=(F_1|F_2).  \label{Gauss3}
\end{equation}
The functional (\ref{Gauss2}) is therefore also defined on products $%
F_1^{*}\odot F_2$ with $F_{1,2}\in {\cal S}({\cal E})$, but it cannot be
extended to the full Fock space ${\cal S}({\cal H})$.

The interior product (\ref{alg13}) can be used to define the linear operator 
$W:{\cal S}_{fin}({\cal H})\rightarrow {\cal S}_{fin}({\cal H})$%
\begin{equation}
WF:=\exp (-\Omega )\cont F  \label{Gauss4}
\end{equation}
This mapping is denoted as {\it Wick ordering} or {\it normal ordering} of
the algebra ${\cal S}_{fin}({\cal H})$. It satisfies the reality condition $%
WF^{*}=(WF)^{*}$. Since $\left\langle \exp \Omega \odot F|\exp (-\Omega )%
\cont G\right\rangle =\left\langle F|G\right\rangle $ for all $F,G\in {\cal S%
}_{fin}({\cal H})$, the inverse operator of $W$ is $W^{-1}F=\exp \Omega %
\cont F$. The operators $W$ and $W^{-1}$ cannot be extended to continuous
mapping on the Fock space ${\cal S}({\cal H})$. If ${\cal S}_{fin}({\cal H})$
is restricted to ${\cal S}_{fin}({\cal H}_{\overline{0}})={\cal T}_{fin}^{+}(%
{\cal H}_{\overline{0}}{\cal )}$ we obtain the bosonic Wick ordering $%
W_0F:=\exp (-\Omega _0)\cont F$, and if ${\cal S}_{fin}({\cal H})$ is
restricted to ${\cal S}_{fin}({\cal H}_{\overline{1}})={\cal T}_{fin}^{-}(%
{\cal H}_{\overline{1}}{\cal )}$ we obtain the fermionic Wick ordering $%
W_1F:=\exp (-\Omega _1)\cont F$ in agreement with the use of Wick ordering
in quantum field theory, see \cite{Kupsch:1990}.

The fundamental algebraic identity to derive an ${\cal L}^2$-picture
(Wiener-Segal picture) for superalgebras is \cite{Kree:1978}\cite{Kree:1980}%
\cite{Kupsch:1990} 
\begin{equation}
\left\langle \exp \Omega |F^{*}\odot G\right\rangle =(W^{-1}F|JW^{-1}G)
\label{Gauss5}
\end{equation}
with $F,G\in {\cal S}_{fin}({\cal H})$, where the linear operator $J$ is
defined on ${\cal S}_{fin}({\cal H})$ by 
\begin{equation}
JF=(-1)^qF\text{ if }F\in {\cal S}_{fin}({\cal H}_{\overline{0}})\odot {\cal %
E}_{\overline{1}}^{\odot p}\odot ({\cal E}_{\overline{1}}^{*})^{\odot q},
\label{Gauss6}
\end{equation}
i.e. $J=\Gamma (j)$ with $j$ given in (\ref{h.11}). Since (\ref{Gauss5}) is
the basis of the Wiener-Segal representation in Sect.\ref{WS}, we give a
proof of (\ref{Gauss5}) in App. \ref{CohStates}. If we restrict the tensors $%
F$ and $G$ to the symmetric tensor algebra ${\cal S}_{fin}({\cal H}_{%
\overline{0}})$ then $J_{+}=id,$ and the sesquilinear form 
\begin{equation}
\left\langle \exp \Omega _0|F^{*}\vee G\right\rangle =(W_0^{-1}F|W_0^{-1}G)
\label{Gauss6a}
\end{equation}
is positive definite and can be diagonalized by Wick ordering $\left\langle
\exp \Omega _0|W_0F^{*}\vee W_0G\right\rangle =(F|G)$ as well known for
Gaussian integrals. If we choose tensors $F$ and $G$ of the exterior algebra 
${\cal S}_{fin}({\cal H}_{\overline{1}})$ the sesquilinear form $%
\left\langle \exp \Omega _1|F^{*}\wedge G\right\rangle =(W^{-1}F|JW^{-1}G)$
is not positive. A positive sesquilinear form on ${\cal S}_{fin}({\cal H})$
can be derived with the help of the antilinear invertible mapping \cite
{Kree:1980}\cite{Kupsch:1990} 
\begin{equation}
F\rightarrow F^{\dagger }:=(WJW^{-1}F)^{*}=WJ_{*}W^{-1}F^{*},  \label{Gauss7}
\end{equation}
where $J_{*}$ is defined by $J_{*}F:=(-1)^pF=(-1)^{p+q}JF$ for $F\in {\cal S}%
_{fin}({\cal H}_{\overline{0}})\odot {\cal E}_{\overline{0}}^{\odot p}\odot (%
{\cal E}_{\overline{1}}^{*})^{\odot q}.$ The mapping (\ref{Gauss7}) reduces
to $F\rightarrow F^{*}$ if $F\in {\cal S}_{fin}({\cal H}_{\overline{0}%
}\oplus {\cal E}_{\overline{1}}),$ but it is not an involution\footnote{%
The more complicated structure of $F^{\dagger }$ is related to the fact that 
${\cal T}_{fin}^{-}({\cal H}_{\overline{1}})$ is not a C$^{\text{*}}$%
-algebra.} on ${\cal S}_{fin}({\cal H}_{\overline{1}})$. For $F\in {\cal S}%
_{fin}({\cal H}_{\overline{0}})\odot {\cal E}_{\overline{0}}^{\odot p}\odot (%
{\cal E}_{\overline{1}}^{*})^{\odot q}$ we calculate $F^{\dagger \dagger
}:=(F^{\dagger })^{\dagger }=WJ_{*}JW^{-1}F=(-1)^{p+q}F,$ hence $F^{\dagger
\dagger }=(-1)^nF$ if $F\in {\cal S}_{fin}({\cal H}_{\overline{0}})\odot 
{\cal H}_{\overline{1}}{}^{\odot n}$. Inserted in (\ref{Gauss5}) the
conjugation (\ref{Gauss7}) yields the positive definite form $\left\langle
\exp \Omega |F^{\dagger }\odot G\right\rangle =(W^{-1}F|W^{-1}G)$ for $%
F,G\in {\cal S}_{fin}({\cal H})$. When this form is diagonalized by Wick
ordering, we exactly recover the inner product of the Fock space ${\cal S}(%
{\cal H})$%
\begin{equation}
\left\langle \exp \Omega |(WF)^{\dagger }\odot (WG)\right\rangle =(F|G).
\label{Gauss9}
\end{equation}
An identity equivalent to (\ref{Gauss5}) is $\left\langle \exp \Omega
|F\odot G\right\rangle =\left\langle W^{-1}F|JW^{-1}G\right\rangle ,$ which
yields that the bilinear form 
\begin{equation}
F,G\in {\cal S}_{fin}({\cal H})\rightarrow \left\langle \exp \Omega
|(WF)\odot (WG)\right\rangle =\left\langle F\mid JG\right\rangle
\label{Gauss10}
\end{equation}
is continuous in the norm (\ref{alg1a}).

So far all arguments started from the algebraic tensor space ${\cal S}_{fin}(%
{\cal H}).$ There is another approach using an extension of the Hilbert
space ${\cal H}$ to a triplet of Hilbert spaces ${\cal H}^{+}\subset {\cal H}%
\subset {\cal H}^{-}$ with Hilbert-Schmidt embeddings. The continuity of the
bilinear form (\ref{Gauss1}) implies that the tensor $\Omega $ is an element
of ${\cal S}_2({\cal H}^{-}),$ and therefore $\exp \Omega $ converges within 
${\cal S}_{(-\gamma )}({\cal H}^{-})$ if $\gamma >0,$ see App.\ref{NormEst}.
The subscript $(\gamma )$ indicates the modified norm (\ref{alg15}). The
functional (\ref{Gauss2}) is therefore a continuous mapping $F\in {\cal S}%
_{(\gamma )}({\cal H}^{+})\rightarrow L(F)\in {\Bbb C}$.$.$ Moreover, normal
ordering is a continuous mapping $F\in {\cal S}_{(\gamma )}({\cal H}%
^{+})\rightarrow WF\in {\cal S}({\cal H})$, see Corollary \ref{Wick} in App.%
\ref{NormEst}.

We would like to add a remark about positive functionals on antisymmetric
tensor algebras. Let ${\cal T}_{fin}^{-}({\cal K})$ be the algebra of
antisymmetric tensors of a complex Hilbert space ${\cal K}$ with involution
as introduced in Sect \ref{Hilbert}. If $L(F)$ is a linear functional on $%
{\cal T}_{fin}^{-}({\cal K})$, which satisfies $L(F^{*}\wedge F)\geq 0$ for
all $F\in {\cal T}_{fin}^{-}({\cal K}),$ then $L(F)$ has the trivial form $%
L(F)=\alpha \left\langle e_0\mid F\right\rangle ,$ where $e_0$ is the unit
of the algebra and $\alpha $ is a non-negative real number. More general
functionals are possible, if we demand the positivity condition $%
L(F^{*}\wedge F)\geq 0$ only for a restricted class of tensors $F\in {\cal T}%
_{fin}^{-}({\cal V})$ where ${\cal V}\subset {\cal H}$ is a closed linear
subspace such that ${\cal V}^{*}\cap {\cal V}=\left\{ 0\right\} .$ A
non--trivial example of such a functional is the Gaussian functional (\ref
{Gauss2}) with ${\cal V}={\cal E}_{\overline{1}}\subset {\cal H}_{\overline{1%
}}={\cal K},$ see (\ref{Gauss3}) restricted to ${\cal S}_{fin}({\cal E}_{%
\overline{1}})={\cal T}_{fin}^{-}({\cal E}_{\overline{1}}).$ Another
important example of this type of positivity condition is the
Osterwalder-Schrader positivity of the Euclidean quantum field theory of
fermions, see e.g. \cite{Haba/Kupsch:1995}, Sect.3.2.

A final remark about {\it supersymmetry} should be added. The Fock space $%
{\cal S}({\cal H})$ is sometimes called supersymmetric Fock space. But such
a notation is misleading. Supersymmetry is an additional property of
functionals or operators defined on ${\cal S}({\cal H});$ it is a relation
between the values of such a functional (operator) on ${\cal S}({\cal H}_{%
\overline{0}})$ with those on ${\cal S}({\cal H}_{\overline{1}}).$ Let us
assume that the spaces ${\cal H}_{\overline{0}}$ and ${\cal H}_{\overline{1}%
} $ are isomorphic spaces with an isometric isomorphism $\Theta :{\cal H}_{%
\overline{0}}\rightarrow {\cal H}_{\overline{1}}.$ We define an operator $S:%
{\cal H}\rightarrow {\cal H}$, which maps ${\cal H}_{\overline{0}%
}\rightarrow {\cal H}_{\overline{1}}$ and ${\cal H}_{\overline{1}%
}\rightarrow {\cal H}_{\overline{0}}$ (i.e. $S$ is an odd operator) by $%
Sf=-\Theta f_0+\Theta ^{-1}f_1$ if $f=f_0+f_1,f_k\in {\cal H}_{\overline{k}%
},k=0,1.$ Then the Gaussian functional $L(F)$ satisfies the supersymmetry
relation $L(D_SF)=0$ for all $F\in {\cal S}_{fin}({\cal H}),$ where $D_S$ is
the antiderivation of the algebra ${\cal S}_{fin}({\cal H})$, which is
generated by the odd operator $S$, i.e. $D_Sf=Sf$ if $f\in {\cal H}$, see 
\cite{Haba/Kupsch:1995}. The restriction to tensors of rank 2 yields the
identity $\omega (Sf,g)+(-1)^{\pi (f)}\omega (f,Sg)=0,$ if $f\in {\cal H}_{%
\overline{0}}{\cal \cup H}_{\overline{1}}$ and $g\in {\cal H},$ for the
bilinear form (\ref{h.7}) or (\ref{Gauss1}). This form is therefore in the
strict sense a supersymmetric form (with respect to the operator $S$).

\section{Superspace and analytic functions\label{super}}

\subsection{Superspaces\label{Sspace}}

Let ${\cal H}={\cal H}_{\overline{0}}\oplus {\cal H}_{\overline{1}}$ be a $%
{\Bbb Z}_2$-graded Hilbert space as introduced in Sect.\ref{Hilbert} and $%
\Lambda $ be an infinite dimensional Grassmann algebra considered as
superalgebra $\Lambda =\Lambda _{\overline{0}}\oplus \Lambda _{\overline{1}%
}, $ see Sect.\ref{Term}. Then the $\Lambda $-module ${\cal H}^\Lambda
=\Lambda \otimes {\cal H}$ is a subspace of the algebra ${\cal S}%
_{fin}^\Lambda ({\cal H}):=\Lambda \otimes {\cal S}_{fin}{\cal (H)}$. The
product of ${\cal S}_{fin}^\Lambda ({\cal H})$ is generated by\footnote{%
The tensor product is the algebraic tensor product. The definition used here
corresponds to the ``Grassmann envelope of the first kind'' in \cite
{Berezin:1987}, p.92.} $\zeta _1\cdot \zeta _2=\lambda _1\lambda _2\otimes
(f_1\odot f_2)$ for elements $\zeta _i=\lambda _i\otimes f_i\in \Lambda
\otimes {\cal H},$ $i=1,2.$ If the factors $\zeta _i$ are confined to the
space $(\Lambda _{\overline{0}}\otimes {\cal H}_{\overline{0}}){\cal \oplus (%
}\Lambda _{\overline{1}}\otimes {\cal H}_{\overline{1}})\subset \Lambda
\otimes {\cal H},$ this product is commutative. The general rule for the
product of two decomposable elements of the algebra ${\cal S}_{fin}^\Lambda (%
{\cal H})$ is $\Xi _i=\lambda _i\otimes F_i\in \Lambda \otimes {\cal S}%
_{fin}({\cal H)}={\cal S}_{fin}^\Lambda ({\cal H}),$ $i=1,2,$%
\begin{equation}
\Xi _1\cdot \Xi _2=\lambda _1\lambda _2\otimes (F_1\odot F_2),  \label{la.1}
\end{equation}
if $\Xi _i=\lambda _i\otimes F_i\in \Lambda \otimes {\cal S}_{fin}({\cal H)}=%
{\cal S}_{fin}^\Lambda ({\cal H}),$ $i=1,2.$ In the following we shall
assume that also $\Lambda $ has a parity conserving involution $\lambda
\rightarrow \lambda ^{*}.$ Then 
\begin{equation}
\left( \lambda \otimes F\right) ^{*}=\lambda ^{*}\otimes F^{*}  \label{la.2}
\end{equation}
induces an involution on ${\cal S}_{fin}^\Lambda ({\cal H})$. The bilinear
form (\ref{alg11}) has a unique $\Lambda $-bilinear extension $\Xi _1,\Xi
_2\in {\cal S}_{fin}^\Lambda ({\cal H})\rightarrow \left\langle \Xi _1\mid
\Xi _2\right\rangle \in \Lambda $ such that 
\begin{equation}
\left\langle \Xi _1\mid \Xi _2\right\rangle =\lambda _1\lambda
_2\left\langle F_1\mid F_2\right\rangle  \label{la.5}
\end{equation}
if $\Xi _i=\lambda _i\otimes F_i$ with $\lambda _i\in \Lambda ,F_i\in {\cal S%
}_{fin}({\cal H}),i=1,2$. This form satisfies the symmetry relation $%
\left\langle \Xi _1\mid \Xi _2\right\rangle ^{*}=\left\langle \Xi _2^{*}\mid
\Xi _1^{*}\right\rangle .$ Correspondingly, the inner product of ${\cal S}%
_{fin}({\cal H})$ has an extension to a $\Lambda $-sesquilinear form on $%
{\cal S}_{fin}^\Lambda ({\cal H})\times {\cal S}_{fin}^\Lambda ({\cal H})$ 
\begin{equation}
\left( \Xi _1\mid \Xi _2\right) :=\left\langle \Xi _1^{*}\mid \Xi
_2\right\rangle  \label{la.4}
\end{equation}
if $\Xi _i\in {\cal S}_{fin}^\Lambda ({\cal H}),i=1,2$. With the
identification $F\in {\cal S}_{fin}{\cal (H)}\Longrightarrow \kappa
_0\otimes F$ $\in {\cal S}_{fin}^\Lambda {\cal (H)}$, the algebra ${\cal S}%
_{fin}{\cal (H)}$ has a natural injection in ${\cal S}_{fin}^\Lambda {\cal %
(H)},$ which is consistent with the definitions (\ref{la.5}) and (\ref{la.4}%
). The factorization (\ref{alg12}) has an extension to the bilinear form (%
\ref{la.5}). We only state the following result. If $F_{1,2}\in {\cal S}(%
{\cal E})$ and $\Xi _{1,2}\in {\cal S}^\Lambda ({\cal E})$ then 
\begin{equation}
\left\langle \Xi _1^{*}\Xi _2\mid F_1^{*}\odot F_2\right\rangle
=\left\langle \Xi _1^{*}\mid F_2\right\rangle \left\langle \Xi _2\mid
F_1^{*}\right\rangle .  \label{la.5a}
\end{equation}

For the superalgebra $\Lambda $ we choose the Grassmann algebra $\Lambda
=\oplus _{p\geq 0}\Lambda _p,$ with a norm $\left\| \lambda \right\|
_\Lambda ^2=\sum_{n=0}^\infty (p!)^{-2}\left\| \lambda _p\right\| _p^2$ if $%
\lambda =\sum_{p=0}^\infty \lambda _p,\lambda _p\in \Lambda _p$. Here $%
\Lambda _p$ is the subspace of $\Lambda $ generated by the products of $p$
linearly independent generators. Then the norm of the unit $\kappa _0$ is $%
\left\| \kappa _0\right\| _\Lambda =1,$ and the antisymmetric tensor product 
$\lambda _1,\lambda _2\in \Lambda \rightarrow \lambda _1\lambda _2\in
\Lambda $ is continuous with the estimate, see App.\ref{NormEst}, 
\begin{equation}
\left\| \lambda _1\lambda _2\right\| _\Lambda \leq c\left\| \lambda
_1\right\| _\Lambda \left\| \lambda _2\right\| _\Lambda  \label{sfu.2}
\end{equation}
where the constant is $c=\sqrt{3}$. A smaller value of the constant $c$
might be possible, but $c\geq \sqrt{\frac 43}$ necessarily holds for a
Hilbert norm.\footnote{%
Instead of the Grassman algebra with the Hilbert norm we could have chosen
an infinite dimensional Banach-Grassmann algebra as defined in \cite
{Rogers:1980}. In that case the product is continuous with a constant $c=1$
in (\ref{sfu.2}), but we have to use the projective tensor product of Banach
spaces to define $\Lambda \widehat{\otimes }{\cal S}({\cal H}).$ Then the
norm estimates become more involved, but the final results remain unchanged.}
The inner product of $\Lambda $ is denoted by $\left( .\mid .\right)
_\Lambda .$ We introduce the unique Hilbert cross norms $\left\| \Xi
\right\| _p$ on the spaces ${\cal S}_p^\Lambda ({\cal H})=\Lambda \widehat{%
\otimes }{\cal S}_p({\cal H})$ which satisfy $\left\| \Xi \right\|
_p=\left\| \lambda \right\| _\Lambda \left\| F\right\| _p$ if $\Xi =\lambda
\otimes F$ with $\lambda \in \Lambda $ and $F\in {\cal S}_p({\cal H})$ for $%
p\in {\Bbb N.}.$ The algebraic tensor product ${\cal S}_{fin}^\Lambda ({\cal %
H}):=\Lambda \otimes {\cal S}_{fin}{\cal (H)}$ is then provided with the
norms $\left\| \Xi \right\| _{(\gamma )}\in {\Bbb R}_{{\bf +}},\gamma \in 
{\Bbb R},$ defined by $\left\| \Xi \right\| _{(\gamma )}^2=\sum_{p=0}^\infty
(n!)^\gamma \left\| \Xi _p\right\| _p^2$ if $\Xi =\sum_{p=0}^\infty \Xi _p$
with $\Xi _p\in {\cal S}_p^\Lambda ({\cal H})$. The completion of ${\cal S}%
_{fin}^\Lambda ({\cal H})$ with this norm is denoted as ${\cal S}_{(\gamma
)}^\Lambda ({\cal H})=\Lambda \widehat{\otimes }{\cal S}_{(\gamma )}({\cal H}%
)$. The space ${\cal S}_{(0)}^\Lambda ({\cal H})$ will simply be denoted by $%
{\cal S}^\Lambda ({\cal H}).$ With the identification $F\in {\cal S}%
_{(\gamma )}{\cal (H)}\Longrightarrow \kappa _0\otimes F\in {\cal S}%
_{(\gamma )}^\Lambda ({\cal H})$ the space ${\cal S}_{(\gamma )}{\cal (H)}$
has a natural isometric injection into ${\cal S}_{(\gamma )}^\Lambda ({\cal H%
})$.

The completed $\Lambda $-module $\Lambda \widehat{\otimes }{\cal H}$ will be
denoted by ${\cal H}^\Lambda .$ For the representation of the Fock space $%
{\cal S}({\cal H})$ by functions the even part of the module $\Lambda 
\widehat{\otimes }{\cal H}$ is the fundamental space. We denote as {\it %
superspace }the completed space ${\cal H}_\Lambda =(\Lambda _{\overline{0}}%
\widehat{\otimes }{\cal H}_{\overline{0}}){\cal \oplus (}\Lambda _{\overline{%
1}}\widehat{\otimes }{\cal H}_{\overline{1}})\subset {\cal H}^\Lambda .$
This space is a $\Lambda _{\overline{0}}$-module, and it can by obtained as
the $\Lambda _{\overline{0}}$-extension of the {\it restricted superspace }$%
{\cal H}_\Lambda ^{res}=\kappa _0\widehat{\otimes }{\cal H}_{\overline{0}}%
{\cal \oplus (}\Lambda _1\widehat{\otimes }{\cal H}_{\overline{1}})\cong 
{\cal H}_{\overline{0}}{\cal \oplus (}\Lambda _1\widehat{\otimes }{\cal H}_{%
\overline{1}}),$ where $\Lambda _1$ is the generating Hilbert space of the
Grassmann algebra $\Lambda ,$%
\begin{equation}
{\cal H}_\Lambda =\Lambda _{\overline{0}}\widehat{\otimes }{\cal H}_\Lambda
^{res}.  \label{la.6a}
\end{equation}

The bilinear form (\ref{h.7}) or (\ref{Gauss1}) has a unique $\Lambda $%
-bilinear and continuous extensions to ${\cal H}^\Lambda \times {\cal H}%
^\Lambda ,$ which is again denoted by $\omega .$ Restricted to the
superspace 
\begin{equation}
(\xi ,\eta )\in {\cal H}_\Lambda \times {\cal H}_\Lambda \rightarrow \omega
(\xi ,\eta )=\left\langle \Omega \mid \xi \eta \right\rangle \in \Lambda
\label{la.6}
\end{equation}
is a symmetric form, $\omega (\xi ,\eta )=\omega (\eta ,\xi ).$

All ${\Bbb C}$-linear (bounded) operators on ${\cal H}$ or ${\cal S}({\cal H}%
)$ have a unique $\Lambda $-linear extension to ${\cal H}^\Lambda =\Lambda 
\widehat{\otimes }{\cal H}$ or ${\cal S}^\Lambda ({\cal H})=\Lambda \widehat{%
\otimes }{\cal S}({\cal H}).$ If $T:{\cal S}({\cal H})\rightarrow {\cal S}(%
{\cal H})$ then the $\Lambda $-extension {\rm T}$:{\cal S}^\Lambda ({\cal H}%
)\rightarrow {\cal S}^\Lambda ({\cal H})$ satisfies ${\rm T}(\lambda \otimes
F)=\lambda \otimes TF$ for $\lambda \in \Lambda $ and $F\in {\cal S}({\cal H}%
)$. The operator $T^{\prime }$ is called the transposed operator of $T,$ if $%
\left\langle T^{\prime }F\mid G\right\rangle =\left\langle F\mid
TG\right\rangle $ holds for all $F,G\in {\cal S}({\cal H}).$ The $\Lambda $%
-extension ${\rm T}^{\prime }$ of $T^{\prime }$ satisfies 
\begin{equation}
\left\langle {\rm T}^{\prime }\Xi \mid H\right\rangle =\left\langle \Xi \mid 
{\rm T}H\right\rangle  \label{la.8}
\end{equation}
for all $\Xi ,H\in {\cal S}^\Lambda ({\cal H}).$

Given an orthonormal basis $\left\{ e_a\in {\cal H},a\in {\Bbb N}\right\} $
of ${\cal H},$ any $\zeta \in {\cal H}^\Lambda $ has the representation $%
\zeta =\sum_{a=1}^\infty \lambda _a\otimes e_a$ with $\lambda _a\in \Lambda
,\sum_a\left\| \lambda _a\right\| _\Lambda ^2<\infty $. The norm of $\zeta $
is calculated as $\left\| \zeta \right\| ^2=\left\| \zeta \right\|
_1^2=\sum_a\left\| \lambda _a\right\| _\Lambda ^2.$ The norm of a product $%
\zeta _1\zeta _{2...}\zeta _p$ of elements $\zeta _n\in {\cal H}^\Lambda
,n=1,...,p\geq 2,$ is estimated with the help of (\ref{sfu.2}) 
\begin{equation}
\left\| \zeta _1\zeta _{2...}\zeta _p\right\| _p\leq c^{p-1}\sqrt{p!}%
\prod_{n=1}^p\left\| \zeta _n\right\| .  \label{sfu.4}
\end{equation}

Given an orthonormal basis $\left\{ E_n\in {\cal S}({\cal H}),\ n\in {\Bbb N}%
\right\} $ of the Fock space ${\cal S}({\cal H})$ any $\Xi \in {\cal S}%
^\Lambda ({\cal H})$ can be decomposed as $\Xi =\sum_n\lambda _n\otimes E_n$
with $\lambda _n\in \Lambda $. The norm is $\left\| \Xi \right\|
^2=\sum_n\left\| \lambda _n\right\| _\Lambda ^2.$ The $\Lambda $%
-sesquilinear form (\ref{la.4}) of $\Xi _1=\sum_n\lambda _{1,n}\otimes E_n$
and $\Xi _2=\sum_n\lambda _{2,n}\otimes E_n$ is calculated as $\left( \Xi
_1\mid \Xi _2\right) =\sum_n\lambda _{1,n}^{*}\lambda _{2,n}\in \Lambda .$
The estimate (\ref{sfu.2}) yields the bound $\left\| \left( \Xi _1\mid \Xi
_2\right) \right\| _\Lambda \leq c\left\| \Xi _1\right\| \left\| \Xi
_2\right\| $ and, since the involution is isometric, the $\Lambda $-bilinear
form (\ref{la.5}) is majorized by the same bound 
\begin{equation}
\left\| \left\langle \Xi _1\mid \Xi _2\right\rangle \right\| _\Lambda \leq
c\left\| \Xi _1\right\| \left\| \Xi _2\right\| .  \label{la.4a}
\end{equation}

The bilinear form (\ref{la.5}) ${\cal S}_{fin}^\Lambda ({\cal H})\times 
{\cal S}_{fin}^\Lambda ({\cal H})\rightarrow \Lambda $ can also be extended
to a bilinear continuous pairing ${\cal S}_{(-\gamma )}^\Lambda ({\cal H}%
)\times {\cal S}_{(\gamma )}^\Lambda ({\cal H})\rightarrow \Lambda .$

\subsection{Coherent states}

The use of coherent states for the representation of the bosonic Fock space 
\cite{Klauder/Skagerstam:1985} can be generalized to superalgebras. But it
is only the $\Lambda $-extended superalgebra ${\cal S}^\Lambda ({\cal H})$
for which the simple definition by the exponential series is possible 
\begin{equation}
\zeta \in {\cal H}_\Lambda \rightarrow \exp \zeta =\sum_{p=0}^\infty \frac
1{p!}\zeta ^p\in {\cal S}^\Lambda ({\cal H})  \label{sfu.5}
\end{equation}
with the product (\ref{la.1}). The absolute convergence of this series
follows from the estimate (\ref{sfu.4}),\\$\left\| \exp \zeta \right\|
^2\leq \sum_{p=0}^\infty \left( p!\right) ^{-2}\left\| \zeta ^p\right\|
^2\leq \sum_{p=0}^\infty (p!)^{-1}c^{2p}\left\| \zeta \right\| ^{2p}<\infty
. $ Actually the series converges in any space ${\cal S}_{(\gamma )}^\Lambda
({\cal H})$ with $\gamma <1.$ The series (\ref{sfu.5}) satisfies the usual
functional relation \\$\exp (\zeta _1+\zeta _2)=(\exp \zeta _1)(\exp \zeta
_2)$ for $\zeta _1,\zeta _2\in {\cal H}_\Lambda .$ Since $\left\| \exp \zeta
-1-\zeta \right\| \leq const\left\| \zeta \right\| ^2$ if $\left\| \zeta
\right\| \leq 1,$ the mapping (\ref{sfu.5}) is Fr\'echet differentiable at $%
\zeta =0,$ and as a consequence of the functional relation it is Fr\'echet
differentiable everywhere. Moreover, as Fr\'echet differentiable mapping
between complex Banach spaces it is analytic. More explicitly, for $\zeta
=\sum_{a\in {\bf A}}\lambda _a\otimes f_a$ the series expansion (\ref{sfu.5}%
) is given by $\exp \zeta =\sum_{p=0}^\infty \frac
1{p!}\sum_{(a_1,...,a_p)\in {\bf A}^p}\lambda _{a_1}\cdot \cdot \cdot
\lambda _{a_p}\otimes f_{a_1}\odot ...\odot f_{a_p}.$ If $\zeta =\kappa
_0\otimes f\in \Lambda _{\overline{0}}\otimes {\cal H}_{\overline{0}}$ we
obtain $\exp \zeta =\kappa _0\otimes \exp f$ with the usual coherent states $%
\exp f$ of the bosonic Fock space ${\cal T}^{+}({\cal H}_{\overline{0}})$.

The ${\Bbb C}$-linear span of the coherent states (\ref{sfu.5}) will be
denoted by ${\cal C}({\cal H}_\Lambda )$ and the $\Lambda $-linear span is
called ${\cal C}^\Lambda ({\cal H}_\Lambda ).$ The ${\Bbb C}$-linear span of
the coherent states of the bosonic Fock space is a dense set of this Fock
space. But neither ${\cal C}({\cal H})$ nor ${\cal C}^\Lambda ({\cal H}%
_\Lambda )$ are dense in ${\cal S}^\Lambda ({\cal H}).$ E.g. the space $%
\kappa _0\widehat{\otimes }{\cal H}_{\overline{1}}$ is orthogonal to all
elements of ${\cal C}^\Lambda ({\cal H}_\Lambda )$. Nevertheless, many
calculations of superanalysis can be reduced to calculations on the set $%
{\cal C}({\cal H}_\Lambda ).$ Given a bounded linear mapping $T:{\cal S}(%
{\cal H})\rightarrow {\cal B},$ where ${\cal B}$ is a complex Hilbert space,
e.g. ${\Bbb C}$ or ${\cal S}({\cal H}),$ this operator has a unique
extension to a bounded $\Lambda $-linear mapping ${\rm T}=\kappa _0\otimes T:%
{\cal S}^\Lambda ({\cal H})\rightarrow \Lambda \widehat{\otimes }{\cal B}$.
The image of ${\rm T}$ on the set of coherent states completely determines $%
{\rm T}$ and consequently $T.$

\begin{lemma}
\label{coherent}The linear operator $T:{\cal S}({\cal H})\rightarrow {\cal B}
$ is uniquely determined by ${\rm T}\Xi $ with $\Xi \in {\cal C}({\cal H}%
_\Lambda ).$
\end{lemma}

\noindent {\bf Proof }The restriction $T\mid _{{\cal S}({\cal H}_{\overline{0%
}})}$ can be obtained from ${\rm T}\exp (\kappa _0\otimes f)=\kappa
_0\otimes (T\exp f)$ with $f\in {\cal H}_{\overline{0}}$. The mapping $%
f^{}\in {\cal H}_{\overline{0}}\rightarrow {\rm T}\exp (\kappa _0\otimes
f)\in \Lambda \widehat{\otimes }{\cal B}$ is analytic, and we can calculate $%
T(f_1\vee ...\vee f_n),f_a\in {\cal H}_{\overline{0}},$ from the derivative $%
\frac{\partial ^n}{\partial z_1...\partial z_n}{\rm T}\left( \exp \kappa
_0\otimes \sum_{a=1}^nz_af_a\right) ,z_a\in {\Bbb C},$ at $z_1=...=z_n=0.$

For fermionic arguments $\zeta \in \Lambda _{\overline{1}}\widehat{\otimes }%
{\cal H}_{\overline{1}}$ we may choose a finite sum $\zeta
=\sum_{a=1,...,n}\kappa _a\otimes g_a$ with orthonormal elements $\kappa
_a\in \Lambda _1$ and arbitrary vectors $g_a\in {\cal H}_{\overline{1}}.$
Then $\left( \kappa _1...\kappa _n\mid {\rm T}\exp \zeta \right) _\Lambda
=(n!)^{-3}T(g_1\wedge ...\wedge g_n)$, and $T$ is determined on ${\cal S}%
_{fin}({\cal H}_{\overline{1}}).$

The image of $T(F)$ with $F=(f_1\vee ...\vee f_m)\odot (g_1\wedge ...\wedge
g_n),$ $f_a\in {\cal H}_{\overline{0}},g_b\in {\cal H}_{\overline{1}%
},a=1,...,m,$ $b=1,...,n,$ can be calculated from ${\rm T}\exp \zeta $ with $%
\zeta =\sum_{a=1}^m\kappa _0\otimes f_a+\sum_{b=1}^n\kappa _b\otimes g_b.$
Then linearity and continuity determine $T$ uniquely. \hfill $\square $ 
\smallskip 

Lemma \ref{coherent} has an obvious generalization to closable unbounded
operators, if the $\Lambda $-linear span of ${\cal S}_{fin}^\Lambda ({\cal H}%
)\cup {\cal C}({\cal H}_\Lambda )$ belongs to the domain of ${\rm T.}$

\subsection{Analytic functions\label{Analytic}}

We define the topological space ${\cal A(H}_\Lambda {\cal )}$ of analytic
functions{\it \ }on the{\it \ }superspace ${\cal H}_\Lambda $ as the linear
space of all entire analytic functions $\zeta \in {\cal H}_\Lambda
\rightarrow \varphi (\zeta )\in \Lambda $ with the topology of uniform
convergence on bounded sets. We define a mapping ${\cal S}^\Lambda ({\cal H}%
)\rightarrow {\cal A(H}_\Lambda {\cal )}$ by 
\begin{equation}
\Xi \in {\cal S}^\Lambda ({\cal H})\rightarrow \varphi _\Xi (\zeta
)=\left\langle \exp \zeta \mid \Xi \right\rangle \in \Lambda ,  \label{sfu.6}
\end{equation}
where the pairing $\left\langle .\mid .\right\rangle $ refers only to ${\cal %
S}({\cal H})$.

\begin{lemma}
\label{entire}For $\Xi \in {\cal S}_{(\gamma )}^\Lambda ({\cal H})$ with $%
\gamma >-1$ the function (\ref{sfu.6}) is an entire analytic function ${\cal %
H}_\Lambda \rightarrow \Lambda .$ If $\gamma =0$ it satisfies the uniform
bound $\left\| \varphi _\Xi (\zeta )\right\| _\Lambda \leq \left\| \Xi
\right\| \exp \left( 2^{-1}c^2\left\| \zeta \right\| ^2\right) .$
\end{lemma}

\noindent {\bf Proof} The function (\ref{sfu.6}) has the series
representation 
\begin{equation}
\varphi _\Xi (\zeta )=\sum_{p=0}^\infty \frac 1{p!}\left\langle \zeta ^p\mid
\Xi _p\right\rangle  \label{sfu.7}
\end{equation}
where $\Xi _p$ is the projection $\Xi _p\in \Lambda \widehat{\otimes }{\cal S%
}_p{\cal (H)}$ of $\Xi =\sum_{p=0}^\infty \Xi _p.$ The homogeneous
polynomial $\zeta \rightarrow \left\langle \zeta ^p\mid \Xi _p\right\rangle
\in \Lambda $ is holomorphic because it is holomorphic on all finite
dimensional subspaces and it is continuous with the bound $\left\|
\left\langle \zeta ^p\mid \Xi _p\right\rangle \right\| _\Lambda \leq \sqrt{p!%
}c^p\left\| \Xi _p\right\| _p\left\| \zeta \right\| ^p$, see (\ref{sfu.4})
and (\ref{la.4a}). The series (\ref{sfu.7}) converges uniformly with the
bound 
\begin{equation}
\left\| \varphi _\Xi (\zeta )\right\| _\Lambda \leq \sum_p\frac{c^p}{\sqrt{p!%
}}\left\| \Xi _p\right\| _p\left\| \zeta \right\| ^p\leq \sqrt{\left(
\sum_p(p!)^\gamma \left\| \Xi _p\right\| ^2\right) \left( \sum_q\frac{c^{2q}%
}{(q!)^{1+\gamma }}\left\| \zeta \right\| ^{2q}\right) }\leq \left\| \Xi
\right\| _{(\gamma )}f_\gamma (\left\| \zeta \right\| )  \label{sfu.8}
\end{equation}
where the function $f_\gamma (t)=\sqrt{\sum_q\frac{c^{2q}}{(q!)^{1+\gamma }}%
t^{2q}}$ is locally bounded for all $t\geq 0$ if $\gamma >-1.$ For $\gamma
=0 $ this function is $f_0(t)=\exp \frac{c^2t^2}2$. \hfill $\square $ 
\smallskip 

If $\zeta =\stackunder{a\in {\bf A}}{\sum }\lambda _a\otimes x_a\in {\cal H}%
_\Lambda $ with $a\in {\bf A}\subset {\Bbb N}$ the expansion (\ref{sfu.7})
has the explicit form 
\begin{equation}
\begin{array}{ll}
\varphi _\Xi (\zeta ) & =\sum_{p=0}^\infty \frac 1{p!}\sum_{(a_1,...,a_p)\in 
{\bf A}^p}\lambda _{a_1}\cdot \cdot \cdot \lambda _{a_p}\left\langle \Xi
_p\mid x_{a_1}\odot \cdot \cdot \cdot \odot x_{a_p}\right\rangle \\ 
& =\sum_{p=0}^\infty \frac 1{\sqrt{p!}}\sum_{(a_1,...,a_p)\in {\bf A}%
^p}\lambda _{a_1}\cdot \cdot \cdot \lambda _{a_p}\left\langle \Xi _p\mid
x_{a_1}\otimes \cdot \cdot \cdot \otimes x_{a_p}\right\rangle
\end{array}
\label{sfu.9}
\end{equation}
The correct symmetrization of the series in the second line follows from the
correlated parities of the factors $\lambda _a.$ For $\Xi \in {\cal S}%
_{fin}^\Lambda {\cal (H)}$ and a finite set ${\bf A}$ this series has only a
finite number of terms.

The Fr\'echet derivative of (\ref{sfu.6}) with increment $\zeta _1\in {\cal H%
}_\Lambda $ is $\varphi _\Xi ^{^{\prime }}(\zeta )(\zeta _1)=\\%
\lim_{t\rightarrow +0}t^{-1}\left( \varphi _\Xi (\zeta +t\zeta _1)-\varphi
_\Xi (\zeta )\right) =\sum_{p=0}^\infty \frac 1{p!}\left\langle \zeta
^p\zeta _1\mid \Xi _{p+1}\right\rangle =\sum_{p=0}^\infty \frac
1{p!}\left\langle \zeta _1\zeta ^p\mid \Xi _{p+1}\right\rangle ,$ where $%
\zeta _1$ appears as left factor. More generally the n-th Fr\'echet
derivative of (\ref{sfu.6}) with increments $\zeta _1,...,\zeta _n\in {\cal H%
}_\Lambda $ is $\varphi _\Xi ^{(n)}(\zeta )(\zeta _1,...,\zeta
_n)=\sum_{p=0}^\infty \frac 1{p!}\left\langle \zeta _1\cdot \cdot \cdot
\zeta _n\zeta ^p\mid \Xi _{p+n}\right\rangle .$ This derivative has the
uniform bound, see (\ref{sfu.8}), 
\begin{equation}
\left\| \varphi _\Xi ^{(n)}(\zeta )(\zeta _1..,\zeta _n)\right\| _\Lambda
\leq \sqrt{n!}2^nc^n\left\| \Xi \right\| _{(\gamma )}\left\| \zeta
_1\right\| \cdot \cdot \cdot \left\| \zeta _n\right\| f_\gamma (\left\|
\zeta \right\| )  \label{sfu.10}
\end{equation}
for all $\zeta \in {\cal H}_\Lambda $ and any $\gamma >-1.$

The restriction of (\ref{sfu.6}) to $\Xi =\kappa _{\overline{0}}\otimes
F,F\in {\cal S}({\cal H}),$ will be denoted by $\varphi _F(\zeta ).$ The
function space ${\cal A}({\cal H}_\Lambda )$ is an algebra with respect to
multiplication of the functions. An essential point of the theory of these
analytic functions is

\begin{theorem}
\label{homomorph}The mapping $F\in {\cal S}_{fin}{\cal (H)}\rightarrow
\varphi _F(\zeta )\in {\cal A}({\cal H}_\Lambda )$ is an homomorphism of the
algebras, i.e. $\varphi _{F\odot G}(\zeta )=\varphi _F(\zeta )\varphi
_G(\zeta )$ if $F,G\in {\cal S}_{fin}({\cal H}).$
\end{theorem}

\noindent {\bf Proof }The definitions (\ref{alg4}) and (\ref{sfu.6}) yield
for $F\in {\cal S}_p({\cal H})$ , $G\in {\cal S}_q({\cal H})$ and $\zeta =%
\stackunder{a\in {\bf A}}{\sum }\lambda _a\otimes x_a\in {\cal H}_\Lambda $%
\[
\begin{array}{lll}
\varphi _{F\odot G}(\zeta ) & = & \frac 1{\left( p+q\right)
!}\sum_{(a_1,...,a_{p+q})\in {\bf A}^{p+q}}\lambda _{a_1}\cdot \cdot \cdot
\lambda _{a_{p+q}}\left\langle F\odot G\mid x_{a_1}\odot \cdot \cdot \cdot
\odot x_{a_{p+q}}\right\rangle \\ 
& = & \frac 1{\left( p+q\right) !\sqrt{p!q!}}\sum_{(a_1,...,a_{p+q})\in {\bf %
A}^{p+q}}\lambda _{a_1}\cdot \cdot \cdot \lambda _{a_{p+q}}\times \\ 
&  & \sum_\sigma \chi _\sigma (x_{a_1},...,x_{a_{p+q}})\left\langle F\mid
x_{\sigma (a_1)}\otimes \cdot \cdot \otimes x_{\sigma (a_p)}\right\rangle
\left\langle G\mid x_{\sigma (a+1)}\otimes \cdot \cdot \cdot \otimes
x_{\sigma (a_{p+q})}\right\rangle
\end{array}
\]

\noindent where $\sum_\sigma $ extends over all permutations $\sigma $ of $%
\left\{ 1,...,p+q\right\} .$ Since $\lambda _a$ has the same parity as $x_a,$
the last expression leads to 
\[
\varphi _{F\odot G}(\zeta )=\frac 1{\sqrt{p!q!}}\sum_{(a_1,...,a_{p+q})\in 
{\bf A}^{p+q}}\lambda _{a_1}\cdot \cdot \cdot \lambda _{a_{p+q}}\left\langle
F\mid x_{a_1}\otimes \cdot \cdot \otimes x_{a_p}\right\rangle \left\langle
G\mid x_{a+1}\otimes \cdot \cdot \cdot \otimes x_{a_{p+q}}\right\rangle 
\]
Following the second identity (\ref{sfu.9}) this function is $\varphi
_F(\zeta )\varphi _G(\zeta ),$ and linearity in $F$ and $G$ completes the
proof. \hfill $\square $ \smallskip

The analytic functions (\ref{sfu.6}) provide a representation of the $%
\Lambda $-extended Fock space ${\cal S}^\Lambda ({\cal H})$.

\begin{lemma}
The mapping $\Xi \in {\cal S}^\Lambda ({\cal H})\rightarrow \varphi (\zeta
)\in {\cal A}({\cal H}_\Lambda )$ is continuous and injective.
\end{lemma}

\noindent {\bf Proof} The continuity follows from the estimate (\ref{sfu.8}%
). The injectivity is a consequence of the stronger Lemma \ref{injection}. %
\hfill $\square $ \smallskip 

The image of the mapping (\ref{sfu.6}) is denoted as ${\cal F}^\Lambda (%
{\cal H}_\Lambda ).$ The analytic functions of this space can be restricted
to arguments $\zeta $ in the Hilbert space ${\cal H}_\Lambda ^{res}=\kappa _0%
\widehat{\otimes }{\cal H}_{\overline{0}}\oplus \Lambda _1\widehat{\otimes }%
{\cal H}_{\overline{1}}$ and they are entire analytic functions in the
variable $\zeta \in {\cal H}_\Lambda ^{res}.$ Knowing the functions (\ref
{sfu.6}) on ${\cal H}_\Lambda ^{res}$ the values on ${\cal H}_\Lambda $ can
be obtained by analytic continuation.

\begin{corollary}
The functions $\varphi \in {\cal F}^\Lambda ({\cal H}_\Lambda )$ are
uniquely determined by their restriction to ${\cal H}_\Lambda ^{res}.$
\end{corollary}

The proof is included in the proof of Lemma \ref{injection}.

In the following sections we also use restrictions of the analytic functions
to the diagonal real subspace. Let ${\cal H}$ be a Hilbert space with the
structure ${\cal H}={\cal E}\oplus {\cal E}^{*}$ as introduced in Sect. \ref
{Hilbert}. If $\xi =\sum_{a\in {\bf A}}\lambda _a\otimes x_a\in \Lambda 
\widehat{\otimes }{\cal E}$ then the involution (\ref{la.2}) leads to $\xi
^{*}=\sum_{a\in {\bf A}}\lambda _a^{*}\otimes x_a^{*}\in \Lambda \otimes 
{\cal E}^{*}$ and $\xi +\xi ^{*}\in \Lambda \widehat{\otimes }{\cal H}$. On
the other hand each $\zeta \in \Lambda \widehat{\otimes }{\cal H}$ can be
uniquely decomposed into $\zeta =\xi _1+\xi _2^{*}$ with $\xi _{1,2}\in
\Lambda \widehat{\otimes }{\cal E}$. Therefore, if $\zeta \in \Lambda 
\widehat{\otimes }{\cal H}$ satisfies the reality condition $\zeta
^{*}=\zeta ,$ it has the unique decomposition $\zeta =\xi +\xi ^{*}$ with $%
\xi \in \Lambda \widehat{\otimes }{\cal E}.$ If $\zeta ^{*}=\zeta \in {\cal H%
}_\Lambda $ then $\xi \in {\cal E}_\Lambda =\Lambda _{\overline{0}}\widehat{%
\otimes }{\cal E}_{\overline{0}}\oplus \Lambda _{\overline{1}}\widehat{%
\otimes }{\cal E}_{\overline{1}}.$ In agreement with (\ref{h.4}) we define
the real subspace 
\begin{equation}
{\cal H}_\Lambda ^D=\left\{ \zeta \mid \zeta ^{*}=\zeta ,\zeta \in {\cal H}%
_\Lambda \right\} .  \label{sdiff.2}
\end{equation}
The restriction of the function (\ref{sfu.6}) $\varphi _\Xi (\zeta )$ to
this real space 
\begin{equation}
\zeta \in {\cal H}_\Lambda ^D\rightarrow \varphi _\Xi (\zeta )=\left\langle
\exp \zeta \mid \Xi \right\rangle =\left( \exp \zeta \mid \Xi \right) \in
\Lambda  \label{sdiff.3}
\end{equation}
is an infinite Fr{\'e}chet differentiable function for all $\Xi \in {\cal S}%
^\Lambda ({\cal H})$. We denote by ${\cal A}({\cal H}_\Lambda ^D)$ the space
of all functions which are restrictions of the analytic functions $\varphi
(\zeta )\in {\cal A}({\cal H}_\Lambda )$ to arguments $\zeta \in {\cal H}%
_\Lambda ^D.$ We provide ${\cal A}({\cal H}_\Lambda ^D)$ with the topology
of uniform convergence for each Fr{\'e}chet derivative on open bounded sets.
The estimates (\ref{sfu.8}) and (\ref{sfu.10}) remain valid for $\zeta \in 
{\cal H}_\Lambda ^D.$ The functions (\ref{sdiff.3}) satisfy the simple
reality condition $\varphi _{\Xi ^{*}}(\zeta )=\left( \varphi _\Xi (\zeta
)\right) ^{*}$ only if $\Xi \in \Lambda _{\overline{0}}\widehat{\otimes }%
{\cal S}({\cal H}).$

\begin{lemma}
\label{injection}For $\gamma >-1$ the mapping 
\begin{equation}
\Xi \in {\cal S}_{(\gamma )}^\Lambda {\cal (H)}\rightarrow \varphi _\Xi
(\zeta )\in {\cal A}({\cal H}_\Lambda ^D)  \label{sdiff.4}
\end{equation}
is continuous and injective.
\end{lemma}

\noindent {\bf Proof} The continuity of (\ref{sdiff.4}) follows from the
estimates (\ref{sfu.8}) and (\ref{sfu.10}). To derive the injectivity it is
sufficient to prove that $\varphi _\Xi (\zeta )=0$ implies $\Xi =0$, because 
${\cal S}^\Lambda ({\cal H})$ and ${\cal A}({\cal H}_\Lambda ^D)$ are linear
spaces. The function $t\in {\Bbb R}{\bf \rightarrow }\varphi _\Xi (t\zeta
)=\left\langle \exp t\zeta \mid \Xi \right\rangle =\sum_{n=0}^\infty \frac{%
t^n}{n!}\left\langle \zeta ^n\mid \Xi _n\right\rangle ,$ with $\zeta \in 
{\cal H}_\Lambda ^D$ fixed, is infinite differentiable, and the derivative
of order $n$ evaluated at $t=0$ is the polynomial $\xi \in {\cal E}_\Lambda
\rightarrow \left\langle (\xi +\xi ^{*})^n\mid \Xi _n\right\rangle \in
\Lambda .$ We choose $\xi =\alpha \otimes x+\sum_{b=1,...,n}\vartheta
_b\otimes y_b$ with $\alpha \in \Lambda _{\overline{0}},x\in {\cal E}_{%
\overline{0}},$ and $\vartheta _a\in \Lambda _{\overline{1}},y_a\in {\cal E}%
_{\overline{1}},$ for $a=1,...,n$. The polynomial $\left\langle (\xi +\xi
^{*})^n\mid \Xi _n\right\rangle $ is then a polynomial in the variables $%
{\bf \lambda }=(\alpha ,\vartheta _1,...,\vartheta _n)\in \Lambda ^{1,n}$ as
considered in App.\ref{Pol}. Assume $\varphi _\Xi (\zeta )=\left\langle \exp
\zeta \mid \Xi \right\rangle =0$ for $\zeta \in {\cal H}_\Lambda ^D$, then
the n-th derivative of $\varphi _\Xi (t\zeta )$ satisfies $\left\langle
\zeta ^n\mid \Xi _n\right\rangle =0$ for all $\zeta \in {\cal H}_\Lambda ^D$%
, and consequently all polynomials in the variables ${\bf \lambda }=(\alpha
,\vartheta _1,...,\vartheta _n)\in \Lambda ^{1,n}$ are identical zero. As a
consequence of Lemma \ref{polynom} in App.\ref{Pol} we obtain $\left\langle
x^{\odot p}\odot x^{*\odot q}\odot y_{a_1}\odot ...\odot y_{a_r}\odot
y_{b_1}^{*}\odot ...\odot y_{b_s}^{*}\mid \Xi _n\right\rangle =0$ for all
index sets $1\leq a_1<...<a_r\leq n$, $1\leq b_1<...<b_s<n$ and all numbers $%
p,q,r,s\geq 0$ with $p+q+r+s=n$. Since the vectors $x\in {\cal E}_{\overline{%
0}}$ and $y_a\in {\cal E}_{\overline{1}},$ $a=1,...,n,$ are arbitrary, the
tensor $\Xi _n\in \Lambda \widehat{\otimes }{\cal S}_n({\cal H})$ vanishes.
This argument applies to all tensors $\Xi _n,n=0,1,...,$ and $\Xi
=\sum_{n=0}^\infty \Xi _n$ has to vanish.

Following Corollary \ref{restricted} in App. \ref{Pol} we can take $\xi
=\alpha \otimes x+\sum_{b=1,...,n}\vartheta _b\otimes y_b$ with $\alpha \in 
{\Bbb C},x\in {\cal E}_{\overline{0}},$ and $\vartheta _a\in \Lambda
_1,y_a\in {\cal E}_{\overline{1}},a=1,...,n$, for the proof given above.
Hence already the values of $\varphi _\Xi (\zeta )$ for $\zeta \in {\cal H}%
_\Lambda ^{res}$ determine $\Xi $ uniquely. \hfill $\square $ \smallskip

The image of the mapping (\ref{sdiff.4}) is denoted as ${\cal F}_{(\gamma
)}^\Lambda ({\cal H}_\Lambda ^D)$. If $\Xi $ is restricted to $\Xi =\kappa
_0\otimes F$ with $F\in {\cal S}_{(\gamma )}({\cal H})$ the image is called $%
{\cal F}_{(\gamma )}({\cal H}_\Lambda ^D),$ such that ${\cal F}_{(\gamma
)}^\Lambda ({\cal H}_\Lambda ^D)=\Lambda \widehat{\otimes }{\cal F}_{(\gamma
)}{\cal (H}_\Lambda ^D{\cal )}.$ For $\gamma =0$ the subscript $(\gamma )$
is omitted in agreement with preceding notations.

For the investigations in Sects.\ref{Smeasure}-\ref{WS} it is convenient to
consider the functions (\ref{sdiff.3}) as functions on the space ${\cal E}%
_\Lambda $ choosing the argument $\zeta =\xi +\xi ^{*}$ with $\xi \in {\cal E%
}_\Lambda .$ We shall therefore write 
\begin{equation}
\varphi _\Xi (\xi ,\xi ^{*})=\left\langle \exp (\xi +\xi ^{*})\mid \Xi
\right\rangle ,\xi \in {\cal E}_\Lambda  \label{sdiff.4a}
\end{equation}
and we denote the corresponding function spaces by ${\cal F}_{(\gamma
)}^\Lambda ({\cal E}_{\Lambda {\Bbb R}}).$ The reason for this notation
comes from the fact that the arguments are actually taken from ${\cal E}%
_{\Lambda {\Bbb R}},$ the underlying real space of ${\cal E}_\Lambda .$ It
is convenient to introduce for the functions (\ref{sdiff.4a}) derivatives
along vectors belonging to the subspaces ${\cal E}_\Lambda $ and ${\cal E}%
_\Lambda ^{*},$ similar to the partial derivatives $\frac \partial {\partial
z_a}$ or $\frac \partial {\partial \overline{z}_a}$ for functions defined on 
${\Bbb C}^n,$ $\frac \partial {\partial z_a}=\frac 12\left( \frac \partial
{\partial x_a}-i\frac \partial {\partial y_a}\right) $ and $\frac \partial
{\partial \overline{z}_a}=\frac 12\left( \frac \partial {\partial
x_a}+i\frac \partial {\partial y_a}\right) $ if $z_a=x_a+iy_a$ with $%
x_a,y_a\in {\Bbb R},a=1,...,n.$ We define the derivative of a function $%
\varphi (\xi ,\xi ^{*})$ along $\eta \in {\cal E}_\Lambda $ or $\eta ^{*}\in 
{\cal E}_\Lambda ^{*}$ by 
\begin{equation}
\begin{array}{l}
\varphi ^{\prime }(\xi ,\xi ^{*})(\eta )=\frac 12\lim_{t\rightarrow
+0}t^{-1}\left( \varphi (\xi +t\eta ,\xi ^{*}+t\eta ^{*})-i\varphi (\xi
+it\eta ,\xi ^{*}-it\eta ^{*})-(1-i)\varphi (\xi ,\xi ^{*})\right) , \\ 
\varphi ^{\prime }(\xi ,\xi ^{*})(\eta ^{*})=\frac 12\lim_{t\rightarrow
+0}t^{-1}\left( \varphi (\xi +t\eta ,\xi ^{*}+t\eta ^{*})+i\varphi (\xi
+it\eta ,\xi ^{*}-it\eta ^{*})-(1+i)\varphi (\xi ,\xi ^{*})\right) .
\end{array}
\label{sdiff.7}
\end{equation}
For this definition only arguments of the real subspace have been used. If
the function $\varphi (\xi ,\xi ^{*})$ can be analytically continued to
arguments $(\xi ,\eta ^{*})\in {\cal E}_\Lambda \times {\cal E}_\Lambda
^{*}, $ an assumption which will be satisfied in almost all our
applications, these definitions agree with\\$\varphi ^{\prime }(\xi ,\xi
^{*})(\eta )=\lim_{t\rightarrow +0}t^{-1}\left( \varphi (\xi +t\eta ,\xi
^{*})-\varphi (\xi ,\xi ^{*})\right) $ and \\$\varphi ^{\prime }(\xi ,\xi
^{*})(\eta ^{*})=\lim_{t\rightarrow +0}t^{-1}\left( \varphi (\xi ,\xi
^{*}+t\eta ^{*})-\varphi (\xi ,\xi ^{*})\right) .$

The spaces ${\cal F}_{(\gamma )}^\Lambda ({\cal H}_\Lambda ^D)$ or ${\cal F}%
_{(\gamma )}^\Lambda ({\cal E}_{\Lambda {\Bbb R}})$ are again Hilbert spaces
with the norm induced by the mapping (\ref{sdiff.4}). This mapping transfers
also the $\Lambda $-bilinear form (\ref{la.5}) to a $\Lambda $-bilinear form 
${\cal F}_{(-\gamma )}^\Lambda ({\cal E}_{\Lambda {\Bbb R}})\times {\cal F}%
_{(\gamma )}^\Lambda ({\cal E}_{\Lambda {\Bbb R}})\rightarrow \Lambda $%
\begin{equation}
\left\langle \varphi _\Xi \parallel \varphi _H\right\rangle :=\left\langle
\Xi \mid H\right\rangle .  \label{sdiff.8}
\end{equation}
The space ${\cal F}^\Lambda ({\cal H}_\Lambda ^D)\cong {\cal F}^\Lambda (%
{\cal E}_{\Lambda {\Bbb R}})$ is a reproducing kernel space and will be
investigated in more detail in Sect.\ref{RKS}.

\section{Gaussian functionals and supermeasures\label{Smeasure}}

\subsection{Superdistributions\label{Sdistrib}}

In this section we introduce a notion of the supermeasure in a form adapted
to the investigation of Gaussian functionals. As in the last part of Sect.%
\ref{Analytic} we consider a topological space of $\Lambda -$valued
functions $f(\zeta )$ defined on the real diagonal subspace (\ref{sdiff.2}) $%
{\cal H}_\Lambda ^D,$ or, equivalently, on ${\cal E}_\Lambda $ (or more
precisely on ${\cal E}_{\Lambda {\Bbb R}}).$ We assume that these functions
are restriction of $\Lambda _{\overline{0}}-$analytic functions on ${\cal H}%
_\Lambda $. The space of all these functions is a $\Lambda -$module and will
be denoted by ${\cal Z}.$ (In the concrete case to be considered in Sect.\ref
{Gfunct} the space ${\cal Z}$ is a subspace of ${\cal F}^\Lambda ({\cal E}%
_{\Lambda {\Bbb R}})$.) For each $\eta \in {\cal E}_\Lambda $ let $\varphi
_{[\eta ]}$ be the function on ${\cal E}_{\Lambda {\Bbb R}}$ defined as $%
\varphi _{[\eta ]}(\xi ,\xi ^{*})=\exp i\left\langle \xi +\xi ^{*}\mid \eta
+\eta ^{*}\right\rangle .$ Here we adopt the notations of Sect.\ref{Analytic}
for functions defined on ${\cal E}_{\Lambda {\Bbb R}}.$ We assume that $%
\varphi _{[\eta ]}(\xi ,\xi ^{*})\in {\cal Z}$ for all $\eta \in {\cal E}%
_\Lambda $ and that the function $\eta \longmapsto \varphi _{[\eta ]}:{\cal E%
}_{\Lambda {\Bbb R}}\rightarrow {\cal Z}$ is G\^ateau differentiable.
Finally, let ${\cal Z}^{\prime }$ be the topological $\Lambda -$module of
the continuous $\Lambda -$linear mappings of ${\cal Z}$ into $\Lambda .$ The
elements of ${\cal Z}^{\prime }$ will be called ${\cal Z}$-distributions%
\footnote{%
For a general theory of distributions on infinite dimensional spaces see 
\cite{Dalecky/Fomin:1991}.} or also superdistribution on ${\cal E}_{\Lambda 
{\Bbb R}}.$ If $v\in {\cal Z}^{\prime }$ then the Fourier transform $\tilde
v $ of $v$ is the function on ${\cal E}_{\Lambda {\Bbb R}}$ taking values in 
$\Lambda $ and defined by 
\begin{equation}
\tilde v(\eta ,\eta ^{*})=\left\langle v\parallel \varphi _{[\eta
]}\right\rangle =\left\langle v\parallel \exp i\left\langle .\mid \eta +\eta
^{*}\right\rangle \right\rangle ,  \label{sdist.1}
\end{equation}
where the $\Lambda -$bilinear pairing between ${\cal Z}^{\prime }$ and $%
{\cal Z}$ is denoted by $\left\langle .\parallel .\right\rangle .$ We will
assume that the Fourier transform of $v$ defines $v$ uniquely.

A ${\cal Z}$-distribution $v$ is called the canonical Gaussian distribution
on ${\cal E}_{\Lambda {\Bbb R}}$, if $\widetilde{v}(\eta ,\eta ^{*})=\\\exp
(-\frac 12\omega (\eta +\eta ^{*},\eta +\eta ^{*}))$, where $\omega $ is the
restriction of the bilinear form (\ref{la.6}) to ${\cal H}_\Lambda ^D\times 
{\cal H}_\Lambda ^D;$ in this context the form $\omega $ is called
correlation functional of $v$. Since $\frac 12\omega (\eta +\eta ^{*},\eta
+\eta ^{*})=\left\langle \eta ^{*}\mid \eta \right\rangle =(\eta \mid \eta
), $ it is the restriction of the sesquilinear $\Lambda -$extension (\ref
{la.4}) of the positive definite inner product of ${\cal H}$ to ${\cal E}%
_{\Lambda {\Bbb R}}\times {\cal E}_{\Lambda {\Bbb R}}$. The canonical
Gaussian distribution on ${\cal E}_{\Lambda {\Bbb R}}$ has therefore the
Fourier transform 
\begin{equation}
\tilde v(\eta ,\eta ^{*})=\exp \left( -\frac 12\omega (\eta +\eta ^{*},\eta
+\eta ^{*})\right) =e^{-\left\langle \eta ^{*}\mid \eta \right\rangle }.
\label{sdist.2}
\end{equation}
The general Gaussian distribution is characterized by the Fourier transform $%
e^{-b(\eta ,\eta ^{*})}$ with a quadratic correlation functional $b(\eta
,\eta ^{*}):{\cal E}_{\Lambda {\Bbb R}}$ $\rightarrow \Lambda _{\overline{0}%
} $ which is derived from the $\Lambda -$extension of some positive
(semi)definite form on ${\cal H}\times {\cal H}$.

\subsection{The Gaussian functional on the Fock space\label{Gfunct}}

The linear functional (\ref{Gauss2}) has a unique $\Lambda $-linear
extension to ${\cal S}_{fin}^\Lambda ({\cal H})$%
\begin{equation}
\Xi \in {\cal S}_{fin}^\Lambda ({\cal H})\rightarrow L(\Xi )=\left\langle
\exp \Omega \mid \Xi \right\rangle \in \Lambda .  \label{la.7}
\end{equation}
For the following investigations we choose as representation of ${\cal S}%
^\Lambda ({\cal H})$ the function space ${\cal F}^\Lambda ({\cal E}_{\Lambda 
{\Bbb R}})$ with the functions (\ref{sdiff.4a}) $\varphi _\Xi (\xi ,\xi
^{*})=\left\langle \exp (\xi +\xi ^{*})\mid \Xi \right\rangle ,\xi \in {\cal %
E}_\Lambda $. The linear functional (\ref{la.7}) is well defined for $\Xi
\in {\cal S}_{fin}^\Lambda ({\cal H}),$ but for a theory of integration we
would like to apply it also to tensors which do not have finite rank. For
that purpose it is convenient to use a triplet of Hilbert spaces ${\cal H}%
^{+}\subset {\cal H}\subset {\cal H}^{-}$ with Hilbert-Schmidt embeddings,
already introduced in Sect.\ref{Gauss}. The spaces ${\cal E}^{+}$ and ${\cal %
E}^{-*}$ (or ${\cal E}^{-}$ and ${\cal E}^{+*}$) are dual spaces with
respect to the bilinear pairing (\ref{h.9}). The tensor $\Omega $ is an
element of ${\cal S}_2({\cal H}^{-})$ and $\exp \Omega \in {\cal S}%
_{(-\gamma )}({\cal H}^{-})$ with $0<\gamma <1.$ The space ${\cal S}%
_{(-\gamma )}({\cal H}^{-})$ can be embedded into ${\cal S}_{(-\gamma
)}^\Lambda ({\cal H}^{-})=\Lambda \widehat{\otimes }{\cal S}_{(-\gamma )}(%
{\cal H}^{-}),$ and the functional (\ref{la.7}) is a continuous functional
on ${\cal S}_{(\gamma )}^\Lambda ({\cal H}^{+})=\Lambda \widehat{\otimes }%
{\cal S}_{(\gamma )}({\cal H}^{+}).$ Any tensor $\Xi \in {\cal S}_{(\gamma
)}^\Lambda ({\cal H}^{+})\subset {\cal S}_{(\gamma )}^\Lambda ({\cal H})$
can be represented by a differentiable functions in the space ${\cal F}%
_{(\gamma )}^\Lambda ({\cal E}_{\Lambda {\Bbb R}}^{-})\subset {\cal F}%
_{(\gamma )}^\Lambda ({\cal E}_{\Lambda {\Bbb R}})$ following the
constructions of Sect.\ref{Analytic}. Here the inclusion ${\cal F}_{(\gamma
)}^\Lambda ({\cal E}_{\Lambda {\Bbb R}}^{-})\subset {\cal F}_{(\gamma
)}^\Lambda ({\cal E}_{\Lambda {\Bbb R}})$ means that the restriction of a
function $\varphi \in {\cal F}_{(\gamma )}^\Lambda ({\cal E}_{\Lambda {\Bbb R%
}}^{-})$ to arguments $\xi \in {\cal E}_\Lambda $ is an element of ${\cal F}%
_{(\gamma )}^\Lambda ({\cal E}_{\Lambda {\Bbb R}}).$ The tensor $\exp \Omega
\in {\cal S}_{(-\gamma )}({\cal H}^{-})$ is represented by the function $%
v(\xi ,\xi ^{*})=\left\langle \exp (\xi +\xi ^{*})\mid \exp \Omega
\right\rangle =e^{\omega \left( \xi ,\xi ^{*}\right) },$ which is a
differentiable function in the variable $\xi \in {\cal E}_{\Lambda {\Bbb R}}$%
, but it is not an element of ${\cal F}_{(-\gamma )}^\Lambda ({\cal E}%
_{\Lambda {\Bbb R}})$, we only have $v(\xi ,\xi ^{*})\in {\cal F}_{(-\gamma
)}^\Lambda ({\cal E}_{\Lambda {\Bbb R}}^{+}).$ Using the bilinear pairing (%
\ref{sdiff.8}) the functional (\ref{la.7}) is a linear functional on ${\cal F%
}_{fin}^\Lambda ({\cal E}_{\Lambda {\Bbb R}})=\left\{ \varphi _\Xi (\xi ,\xi
^{*})\mid \Xi \in {\cal S}_{fin}^\Lambda ({\cal H})\right\} $%
\begin{equation}
\left\langle v\parallel \varphi _\Xi \right\rangle =\left\langle \exp \Omega
\mid \Xi \right\rangle   \label{sdiff.9}
\end{equation}
and it is can be extended to a continuous functional on the space ${\cal F}%
_{(\gamma )}^\Lambda ({\cal E}_{\Lambda {\Bbb R}}^{-})=\\ \left\{ \varphi
_\Xi (\xi ,\xi ^{*})\mid \Xi \in {\cal S}_{(\gamma )}^\Lambda ({\cal H}%
^{+})\right\} $. Following (\ref{Gauss3}) we also know that the functional
is well defined on the linear span of all functions $\varphi _{\Xi
^{*}H}(\xi ,\xi ^{*})$ with $\Xi ,H\in {\cal S}^\Lambda ({\cal E})$.The
functional (\ref{sdiff.9}) will be written as integral 
\begin{equation}
\left\langle \exp \Omega \mid \Xi \right\rangle =\int \varphi _\Xi (\xi ,\xi
^{*})dv(\xi ,\xi ^{*})  \label{sdiff.10}
\end{equation}
and $dv(\xi ,\xi ^{*})$ is called the Gaussian {\it supermeasure} related to
the quadratic form $\omega \left( \xi ,\xi ^{*}\right) $. The identity (\ref
{sdiff.10}) is (formally) equivalent to $\int \exp (\xi +\xi ^{*})dv(\xi
,\xi ^{*})=\exp \Omega .$ The Fourier transform (\ref{sdist.1}) of the
distribution $v$ is 
\begin{equation}
\int e^{i\left\langle \xi +\xi ^{*}\mid \eta +\eta ^{*}\right\rangle }dv(\xi
,\xi ^{*})=\left\langle \exp \Omega \mid \exp i(\eta +\eta
^{*})\right\rangle =e^{-\frac 12\omega (\eta +\eta ^{*},\eta +\eta
^{*})}=e^{-\left\langle \eta ^{*}\mid \eta \right\rangle }  \label{sdiff.101}
\end{equation}
with $\eta \in {\cal E}_\Lambda $. Here the first identity follows from the
definition (\ref{sdiff.10}) and the calculation of the functional is
performed in App.\ref{CohStates}. The result (\ref{sdiff.102}) generalizes
the statements for numerical Gaussian (pro)measures. Moreover $\exp \Omega $
or, more precisely, $v(\xi ,\xi ^{*})=\left\langle \exp (\xi +\xi ^{*})\mid
\exp \Omega \right\rangle $ is the canonical Gaussian distribution as
defined by the Fourier transform (\ref{sdist.2}). The analytic continuation
of (\ref{sdiff.101}) leads to the Fourier-Laplace transform 
\begin{equation}
\int e^{\left\langle \xi +\xi ^{*}\mid \varsigma +\tau ^{*}\right\rangle
}dv(\xi ,\xi ^{*})=\left\langle \exp \Omega \mid \exp (\varsigma +\tau
^{*})\right\rangle =e^{\frac 12\omega (\varsigma +\tau ^{*},\varsigma +\tau
^{*})}=e^{\left\langle \tau ^{*}\mid \varsigma \right\rangle }
\label{sdiff.102}
\end{equation}
with $\varsigma ,\tau \in {\cal E}_\Lambda .$

The functional (\ref{la.7}) is originally defined on ${\cal S}_{fin}^\Lambda
({\cal H}),$ and the supermeasure is a promeasure on the space of all
finitely based functions ${\cal F}_{fin}^\Lambda ({\cal E}_{\Lambda {\Bbb R}%
})=\left\{ \varphi _\Xi (\xi ,\xi ^{*})\mid \Xi \in {\cal S}_{fin}^\Lambda (%
{\cal H})\right\} .$ But since the functional is defined on ${\cal S}%
_{(\gamma )}^\Lambda ({\cal H}^{+}),\gamma >0,$ this promeasure can be
extended to distributions on ${\cal E}_{\Lambda {\Bbb R}}^{-},$ the space on
which all test functions $\varphi _\Xi (\xi ,\xi ^{*}),\Xi \in {\cal S}%
_{(\gamma )}^\Lambda ({\cal H}^{+}),$ are continuous functions. The class of
integrable functions includes the spaces ${\cal F}_{(\gamma )}^\Lambda (%
{\cal E}_{\Lambda {\Bbb R}}^{-})=\left\{ \varphi _\Xi (\xi ,\xi ^{*}),\Xi
\in {\cal S}_{(\gamma )}^\Lambda ({\cal H}^{+})\right\} $ for any $\gamma >0.
$ For these functions we derive the following statement

\begin{lemma}
\label{IntFunct} If $\varphi (\xi ,\xi ^{*})\in {\cal F}_{(\gamma )}^\Lambda
({\cal E}_{\Lambda {\Bbb R}}^{-})$ for some $\gamma >0$, then also $\varphi
(\xi +\varsigma ,\xi ^{*}+\upsilon ^{*})$ with $\varsigma ,\upsilon \in 
{\cal E}_\Lambda ^{-}$ and $\varphi (z_1\xi ,z_2\xi ^{*})$ with $z_{1,2}\in 
{\Bbb C}$ are integrable functions.
\end{lemma}

\noindent {\bf Proof }We assume $0<\gamma <1.$ From the definition of the
function follows $\varphi (\xi ,\xi ^{*})=\varphi _\Xi (\xi ,\xi ^{*})$ with
a tensor $\Xi \in {\cal S}_{(\gamma )}^\Lambda ({\cal H}^{+}).$ and
therefore $\varphi (\xi +\varsigma ,\xi ^{*}+\upsilon ^{*})=\left\langle
\exp (\xi +\xi ^{*})\mid \exp (\varsigma +\tau ^{*})\cont \Xi \right\rangle
. $ For $\varsigma ,\upsilon \in {\cal E}_\Lambda ^{-}$ we have $\exp
(\varsigma +\tau ^{*})\in {\cal S}_{(\alpha )}^\Lambda ({\cal H}^{-}),\alpha
<1,$ and Lemma \ref{Contraction} implies $\exp (\varsigma +\tau ^{*})\cont %
\Xi \in {\cal S}_{(\beta )}^\Lambda ({\cal H}^{+})$ with $0<\beta <\gamma .$
Hence $\left\langle \exp (\xi +\xi ^{*})\mid \exp (\varsigma +\tau ^{*})%
\cont \Xi \right\rangle \in {\cal F}_{(\beta )}^\Lambda ({\cal E}_{\Lambda 
{\Bbb R}}^{-})$ is integrable.

For $z_{1,2}\in {\Bbb C}$ the mapping $\zeta =\xi +\eta ^{*}\rightarrow {\rm %
A}[z_1,z_2]\zeta :=z_1\xi +z_2\eta ^{*}$ for $\xi ,\eta \in \Lambda \widehat{%
\otimes }{\cal E}^{\pm }$ is a bounded operator on $\Lambda \widehat{\otimes 
}{\cal H}^{\pm },$ and $\Gamma ({\rm A})$ is a bounded extension on ${\cal S}%
_{(\alpha )}^\Lambda ({\cal H}^{\pm }),\alpha \in {\Bbb R.}$ From $%
\left\langle {\rm A}[z_1,z_2]\zeta \mid \varsigma \right\rangle
=\left\langle \zeta \mid {\rm A}[z_2,z_1]\varsigma \right\rangle $ for $%
\zeta \in \Lambda \widehat{\otimes }{\cal H}^{-}$ and $\varsigma \in \Lambda 
\widehat{\otimes }{\cal H}^{+}$ we obtain $\varphi (z_1\xi ,z_2\xi
^{*})=\left\langle \exp (\xi +\xi ^{*})\mid \Gamma ({\rm A}[z_2,z_1])\Xi
\right\rangle .$ Since $\Gamma ({\rm A}[z_2,z_1])\Xi \in {\cal S}_{(\gamma
)}^\Lambda ({\cal H}^{+})$ if $\Xi \in {\cal S}_{(\gamma )}^\Lambda ({\cal H}%
^{+})$ the proof is complete. \hfill $\square$ \smallskip 

The Gaussian functional (\ref{la.7}) is the tensor product of the functional 
$\left\langle \exp \Omega _0\mid .\right\rangle $ which operates
non-trivially only on the bosonic tensors of ${\cal S}_{fin}^\Lambda ({\cal H%
}_{\overline{0}}),$ and of the functional $\left\langle \exp \Omega _1\mid
.\right\rangle $ which operates on the fermionic tensors of ${\cal S}%
_{fin}^\Lambda ({\cal H}_{\overline{1}}).$ The bosonic functional can be
calculated with the help of the canonical Gaussian promeasure $\mu _0(dx)$
on ${\cal E}_{\overline{0}}\cong \kappa _0\otimes {\cal E}_{\overline{0}%
}\subset \Lambda _{\overline{0}}\widehat{\otimes }{\cal E}_{\overline{0}},$
see App.\ref{Boson}. Hence the functional (\ref{la.7}) can be evaluated with
the vector valued Gaussian (pro)measure $\mu _0(dx)\exp \Omega _1$ defined
on ${\cal E}_{\overline{0}{\Bbb R}}$ (or ${\cal E}_{\overline{0}{\Bbb R}%
}^{-})$ with values in ${\cal S}_{(-\gamma )}({\cal H}_{\overline{1}}^{-}).$
Equivalently, the Gaussian superdistribution $v$ with Fourier transform (\ref
{sdist.2}) has the interpretation of the vector valued promeasure $\mu
_0(dx)v_1:{\cal E}_{\overline{0}{\Bbb R}}\rightarrow {\cal Z}_{\overline{1}%
}^{\prime }$ where ${\cal Z}_{\overline{1}}={\cal F}_{(\gamma )}^\Lambda
(\Lambda _{\overline{1}}\widehat{\otimes }{\cal E}_{\overline{1}%
}^{-}),0<\gamma <1.$ The original definition of the supermeasure in \cite
{Smolyanov/Shavgulidze:1988} is related to this interpretation.

There is another method to calculate the functional (\ref{Gauss2}) by 
Gaussian integration. In App.\ref{Integration} we refer to a representation
of the tensor algebra ${\cal T}_{fin}^{-}({\cal H})$ by an algebra of
functions with the help of an ordering prescription without $\Lambda $%
-extension. Then the Gaussian functional (\ref{Gauss2}), including the
fermionic part, can be evaluated with a positive Gaussian measure on ${\cal E%
}_{{\Bbb R}}^{-}={\cal E}_{\overline{0}{\Bbb R}}^{-}\oplus {\cal E}_{%
\overline{1}{\Bbb R}}^{-}$. For more details see App.\ref{Integration}.

\section{Reproducing kernel spaces and Bargmann-Fock representation for
superalgebras\label{RKS-BF}}

The representation of the bosonic Fock space as a Hilbert space of
(anti)holomorphic functions (Bargmann-Fock representation or holomorphic
representation) is an effective tool in quantum mechanics \cite
{Bargmann:1961}\cite{Berezin:1971} and in quantum field theory \cite
{Segal:1962}\cite{Segal:1963}\cite{Kree/Raczka:1978}. This Hilbert space of
functions is a reproducing kernel Hilbert space, and a great part of the
calculations done within this representation only refer to this property. In
the reproducing kernel space technique the inner product can be defined in a
rather abstract way, and one has a great flexibility in choosing the domain
on which the functions are defined (exploiting the fact that an analytic
function is already uniquely determined on rather small sets), see e.g. \cite
{Kupsch:1991}, where these techniques have been applied to the fermionic
Fock space. In this section we generalize these constructions to analytic
functions on superspaces. In Sect.\ref{RKS} we restrict the domain of the
analytic functions $\varphi _F$ to the real subspace ${\cal H}_\Lambda ^D.$
Actually a restriction to even smaller but densely embedded subspaces is
possible. In Sect.\ref{BF} we use the full analyticity of the functions and
investigate the Bargmann-Fock representation. The resulting function spaces
are isomorphic reproducing kernel spaces, which exhibit many features known
from the analysis of the bosonic Fock space. In the Bargmann-Fock
representation one has an integral representation for the inner product,
where the ''measure'' is the supermeasure{\it ,} defined in Sect.\ref
{Smeasure}.

\subsection{Reproducing kernel spaces\label{RKS}}

We define a sesquilinear form on ${\cal F}^\Lambda ({\cal H}_\Lambda ^D)$ by 
\begin{equation}
\left( \varphi _\Xi \parallel \varphi _H\right) :=\left( \Xi \mid H\right) ,
\label{sdiff.6}
\end{equation}
if $\Xi ,H\in {\cal S}^\Lambda ({\cal H}).$ This form is positive definite
if $\Xi =id\otimes F$ and $H=id\otimes G$ with $F,G\in {\cal S}({\cal H})$,
and it defines an inner product on ${\cal F}({\cal H}_\Lambda ^D).$ The
mapping (\ref{sdiff.4}) is therefore an isomorphism between the Hilbert
spaces ${\cal S}({\cal H})$ and ${\cal F}({\cal H}_\Lambda ^D).$ If $\Xi
=id\otimes F$ we simply write $\varphi _F$ instead of $\varphi _{id\otimes
F} $. We define a kernel function ${\cal H}_\Lambda ^D\times {\cal H}%
_\Lambda ^D\rightarrow \Lambda _{\overline{0}}$ by 
\begin{equation}
K(\zeta _1,\zeta _2):=\left( \exp \zeta _1\mid \exp \zeta _2\right)
=e^{\left( \zeta _1\mid \zeta _2\right) }=e^{\left\langle \zeta _1\mid \zeta
_2\right\rangle }.  \label{ker.1}
\end{equation}
The identity used on the right side of this definition is calculated in App.%
\ref{CohStates}. If $\eta \in {\cal H}_\Lambda ^D$ then $\exp \eta \in {\cal %
S}^\Lambda ({\cal H}),$ and the function $\zeta \rightarrow K(\zeta ,\eta )$
is an element of ${\cal F}^\Lambda ({\cal H}_\Lambda ^D)$. The importance of
this kernel function follows from

\begin{lemma}
\label{Kernel}The functions $\varphi (\zeta )\in {\cal F}^\Lambda {\cal (H}%
_\Lambda ^D{\cal )}$ satisfy the identity 
\begin{equation}
\left( K(.,\eta )\parallel \varphi \right) =\varphi (\eta )  \label{ker.2}
\end{equation}
with the kernel (\ref{ker.1}).
\end{lemma}

\noindent {\bf Proof} Assume $\varphi (\zeta )\in {\cal F}^\Lambda {\cal (H}%
_\Lambda ^D{\cal )}.$ Then Lemma \ref{injection} and the definition of $%
{\cal F}^\Lambda {\cal (H}_\Lambda ^D{\cal )}$ imply that there exists a
unique element $\Xi \in {\cal S}^\Lambda ({\cal H})$ such that $\varphi
(\zeta )=\varphi _\Xi (\zeta ).$ On the other hand for $\eta \in {\cal H}%
_\Lambda ^D$ we have $G:=\exp \eta \in {\cal S}^\Lambda ({\cal H})$ and $%
\varphi _G(\zeta )=\left( \exp \zeta \mid \exp \eta \right) =K(\zeta ,\eta
). $ Hence (\ref{sdiff.6}) implies $\left( K(.,\eta )\parallel \varphi _\Xi
\right) =\left( \exp \eta \mid \Xi \right) =\varphi _\Xi (\eta )$. \hfill $%
\square $ \smallskip

The space ${\cal F}^\Lambda {\cal (H}_\Lambda ^D{\cal )}$ has therefore $%
K(\zeta _1,\zeta _2)$ as reproducing kernel, and ${\cal F(H}_\Lambda ^D{\cal %
)}\subset {\cal F}^\Lambda {\cal (H}_\Lambda ^D{\cal )}$ is a Hilbert space
with reproducing kernel. But since the functions in ${\cal F(H}_\Lambda ^D%
{\cal )}$ have values in a non-commutative algebra, ${\cal F(H}_\Lambda ^D%
{\cal )}$ differs from the usual function spaces, and the kernel function $%
K(\zeta _1,\zeta _2)$ does not satisfy all of the properties of a
reproducing kernel as given in \cite{Aronszajn:1950}. E.g. the function $%
\zeta \rightarrow K(\zeta ,\eta ),\eta \in \left( \Lambda _{\overline{1}}%
\widehat{\otimes }{\cal H}_{\overline{1}}\right) \cap {\cal H}_\Lambda ^D$,
is not an element of the Hilbert space ${\cal F(H}_\Lambda ^D{\cal )}.$

\subsection{Bargmann-Fock representation\label{BF}}

The functions (\ref{sdiff.3}) have a unique analytic extension to arguments $%
\zeta =\xi +\eta ^{*},$with $\xi \in {\cal E}_\Lambda $ and $\eta ^{*}\in 
{\cal E}_\Lambda ^{*}$, see Lemma \ref{entire}. We shall denote the
analytically continued functions as 
\begin{equation}
\varphi _\Xi (\xi ,\eta ^{*})=\left\langle \exp (\xi +\eta ^{*})\mid \Xi
\right\rangle =\left( \exp (\xi ^{*}+\eta )\mid \Xi \right)  \label{sdiff.12}
\end{equation}
with independent arguments $\xi \in {\cal E}_\Lambda $ and $\eta ^{*}\in 
{\cal E}_\Lambda ^{*}$.\footnote{%
More precisely, $\varphi $ depends on variables $\left( \xi _{{\bf R}},\eta
_{{\bf R}}\right) \in {\cal E}_{\Lambda {\bf R}}^{(1)}\times {\cal E}%
_{\Lambda {\bf R}}^{(2)},$ where ${\cal E}_{\Lambda {\bf R}}^{(1,2)}$ are
two copies of ${\cal E}_{\Lambda {\bf R}},$ the underlying real space of $%
{\cal E}_\Lambda .$} The space of these functions with $\Xi \in {\cal S}%
^\Lambda ({\cal H})$ is denoted as ${\cal F}^\Lambda \left( {\cal E}_\Lambda
\times {\cal E}_\Lambda ^{*}\right) .$ If $\Xi $ is restricted to $\Xi
=id\otimes F$ with $F\in {\cal S}({\cal H}),$ the space is called ${\cal F}%
\left( {\cal E}_\Lambda \times {\cal E}_\Lambda ^{*}\right) ,$ and these
functions satisfy the identities 
\begin{equation}
\left\{ 
\begin{array}{l}
\left( \varphi _F(\xi ,\eta ^{*})\right) ^{*}=\varphi _{F^{*}}(\xi ^{*},\eta
), \\ 
\varphi _F(\xi ,\eta ^{*})\varphi _G(\xi ,\eta ^{*})=\varphi _{F\odot G}(\xi
,\eta ^{*}),
\end{array}
\right.  \label{sdiff.13}
\end{equation}
for the second identity see Lemma \ref{homomorph}. The space ${\cal F}%
^\Lambda \left( {\cal E}_\Lambda \times {\cal E}_\Lambda ^{*}\right) $ is of
course isomorphic to the space ${\cal F}^\Lambda ({\cal H}_\Lambda ^D),$ and 
${\cal F}\left( {\cal E}_\Lambda \times {\cal E}_\Lambda ^{*}\right) $ is
still a Hilbert space with a reproducing kernel. Using the variables $\xi
\in {\cal E}_\Lambda $ and $\eta ^{*}\in {\cal E}_\Lambda ^{*}$ the kernel (%
\ref{ker.1}) has the analytic continuation 
\begin{equation}
K(\xi _1,\eta _1^{*};\xi _2,\eta _2^{*})=e^{\left\langle \xi _1\mid \xi
_2^{*}\right\rangle +\left\langle \eta _1^{*}\mid \eta _2\right\rangle },
\label{ker.3}
\end{equation}
and all arguments of Sect.\ref{RKS} can be repeated.

As an advantage of the analytic continuation of the argument $\zeta \in 
{\cal H}_\Lambda ^D$ to $\xi +\eta ^{*}\in {\cal E}_\Lambda \oplus {\cal E}%
_\Lambda ^{*}$ we can write the form (\ref{sdiff.6}) as integral.

\begin{lemma}
\label{Bargmann}For $\Xi ,H\in {\cal S}_{fin}^\Lambda ({\cal H)}$ the
sesquilinear form (\ref{sdiff.6}) has the integral representation\footnote{%
Here the integration is extended over two independent copies of ${\cal E}%
_{\Lambda {\bf R}}$, see the foregoing footnote. The operator {\rm j }is the%
{\rm \ }$\Lambda $-linear extension of (\ref{h.11}). The measure is
considered as promeasure on ${\cal E}_{\Lambda {\bf R}}.$} 
\begin{equation}
\left( \Xi \mid H\right) =\int \left( \varphi _\Xi (\xi ,{\rm j}\eta
^{*})\right) ^{*}\varphi _H(\xi ,\eta ^{*})dv(\xi ,\xi ^{*})dv(\eta ,\eta
^{*}).  \label{sdiff.14}
\end{equation}
\end{lemma}

\noindent {\bf Proof} We first assume $\Xi =\kappa _0\otimes F$ and $%
H=\kappa _0\otimes G$ with $F,G\in {\cal S}_{fin}({\cal H)}.$ To prove (\ref
{sdiff.14}) in this case it is sufficient to consider tensors $%
F=F_1^{*}\odot F_2$ and $G=G_1^{*}\odot G_2$ with $F_{1,2}\in {\cal S}_{fin}(%
{\cal E)}$ and $G_{1,2}\in {\cal S}_{fin}({\cal E)}$. Then $\varphi _F(\xi
,\eta ^{*})=\left\langle \exp \eta ^{*}\mid F_2\right\rangle \left\langle
\exp \xi \mid F_1^{*}\right\rangle $ and \\$\varphi _G(\xi ,\eta
^{*})=\left\langle \exp \eta ^{*}\mid G_2\right\rangle \left\langle \exp \xi
\mid G_1^{*}\right\rangle ,$ see (\ref{la.5a}). The integrand of (\ref
{sdiff.14}) is therefore \\$\left\langle \exp \xi ^{*}\mid F_1\right\rangle
\left\langle \exp \eta \mid JF_2^{*}\right\rangle \left\langle \exp \eta
^{*}\mid G_2\right\rangle \left\langle \exp \xi \mid G_1^{*}\right\rangle =$ %
\\$\left\langle \exp \xi ^{*}\mid F_1\right\rangle \left\langle \exp (\eta
+\eta ^{*})\mid G_2\odot JF_2^{*}\right\rangle \left\langle \exp \xi \mid
G_1^{*}\right\rangle =$ \\$(-1)^{\pi (F_1)(\pi (G_2)+\pi (F_2))}\left\langle
\exp (\eta +\eta ^{*})\mid G_2\odot JF_2^{*}\right\rangle \left\langle \exp
(\xi +\xi ^{*})\mid G_1^{*}\odot F_1\right\rangle .$ Here $J$ is the
operator (\ref{Gauss6}). Only terms with $\pi (G_2)=\pi (F_2),$ for which
the sign function is $+1,$ contribute to the integration (\ref{sdiff.10})
with $dv(\eta ,\eta ^{*})$. Moreover $\left\langle \exp \Omega \mid G_2\odot
JF_2^{*}\right\rangle =\left\langle \exp \Omega \mid F_2^{*}\odot
G_2\right\rangle =\left( F_2\mid G_2\right) ,$ and 
\[
\int \left( \varphi _F(\xi ,{\rm j}\eta ^{*})\right) ^{*}\varphi _G(\xi
,\eta ^{*})dv(\xi ,\xi ^{*})dv(\eta ,\eta ^{*})=\left( F\mid G\right) 
\]
follows for all $F$ and $G\in {\cal S}_{fin}({\cal H}).$ The $\Lambda $%
-extension of this identity to $\Xi =\lambda \otimes F$ and $H=\mu \otimes G$
with $\lambda ,\mu \in \Lambda $ and $F,G\in {\cal S}_{fin}({\cal H})$ is
calculated for the left side using the definitions (\ref{la.5}) (\ref
{sdiff.12}), and for the right side with (\ref{la.4}). It yields for both
sides an additional left factor $\lambda ^{*}$ and an additional right
factor $\mu $. The identity (\ref{sdiff.14}) then follows by $\Lambda $%
-linearity. \hfill $\square $ \smallskip

If $\Xi ,H$ are restricted to $\Xi =\kappa _0\otimes F$ and $H=\kappa
_0\otimes G$ with $F,G\in {\cal S}_{fin}({\cal H}),$ the integral (\ref
{sdiff.14}) yields the inner product of the Fock space ${\cal S}({\cal H}).$
The antilinear mapping 
\begin{equation}
\varphi _F(\xi ,\eta ^{*})\rightarrow \left( \varphi _F(\xi ,{\rm j}\eta
^{*})\right) ^{*}=\left\langle JF^{*}\mid \exp (\xi ^{*}+\eta )\right\rangle
,  \label{sdiff.15}
\end{equation}
which is needed to calculate the integrand, is not an involution, see the
discussion of the conjugation (\ref{Gauss7}) in Sect.\ref{Gauss}. The
problem comes from the inversion $j$ which contributes to the fermionic
terms $F\in {\cal S}({\cal E}_{\overline{1}}^{*}).$ The inner product of the
fermionic Fock space is usually written only for the Fock space ${\cal T}%
^{-}({\cal E}_{\overline{1}})$ of just one species of fermions, see e.g.
Sect.2.4 of \cite{Faddeev/Slavnov:1980}. If $F,G\in {\cal S}({\cal H}_{%
\overline{0}}\oplus {\cal E}_{\overline{1}})$ then (\ref{sdiff.14})
simplifies to $\int \left( \varphi _F(\xi )\right) ^{*}\varphi _G(\xi
)dv(\xi ,\xi ^{*})=\left( F\mid G\right) $ without the phase inversion $j$.

As in the case of the classical Bargmann-Fock space \cite{Kondratiev:1991}%
\cite{Yokoi:1995} we can extend Lemma \ref{Bargmann} to all functions $%
\varphi _\Xi $ with $\Xi \in {\cal S}^\Lambda ({\cal H}).$ But then the
functions should be extended to arguments $\left( \xi ,\eta ^{*}\right) \in 
{\cal E}_\Lambda ^{-}\times {\cal E}_\Lambda ^{-*},$ and the supermeasure
has to be taken as distribution on ${\cal E}_\Lambda ^{-}.$ The
corresponding space of functions is then more appropriately denoted by $%
{\cal F}^\Lambda ({\cal E}_\Lambda ^{-}\times {\cal E}_\Lambda ^{-*}).$ The
extended functions are not necessarily analytic on the space ${\cal E}%
_\Lambda ^{-}\times {\cal E}_\Lambda ^{-*}.$

\subsection{Kernels and symbols of operators\label{Opkernel}}

In this subsection we give a short introduction to integral operators in the
Bargmann-Fock representation. The class of linear operators which allows
such a representation includes all bounded operators, and also differential
operators. Actually it is more convenient to use the techniques of the
reproducing kernel spaces of Sect.\ref{RKS} for these constructions. The
kernels of the Bargmann-Fock representation can then be recovered by
analytic continuation in the arguments.

The sesquilinear form (\ref{sdiff.6}) is related to the bilinear form (\ref
{sdiff.8}) by $\left( \varphi _\Xi \parallel \varphi _H\right) =\left\langle
\varphi _{\Xi ^{*}}\parallel \varphi _H\right\rangle .$ Since $\eta
^{*}=\eta $ for $\eta \in {\cal H}_\Lambda ^D$ the identity (\ref{ker.2})
can also be written as $\left\langle K(.,\eta )\parallel \varphi
\right\rangle =\varphi (\eta ).$ Let $\Upsilon $ be the mapping of ${\cal S}%
^\Lambda ({\cal H})$ onto ${\cal F}^\Lambda ({\cal H}_\Lambda ^D)\subset 
{\cal A}({\cal H}_\Lambda ^D)$ defined in Lemma \ref{injection}. Given an
operator $T$ on ${\cal S}({\cal H})$ with $\Lambda $-extension ${\rm T}$ on $%
{\cal S}^\Lambda ({\cal H})$ we denote by ${\rm T}_{{\cal F}}$ its
representation on ${\cal F}^\Lambda ({\cal H}_\Lambda ^D),$ i.e. ${\rm T}_{%
{\cal F}}$ is defined by ${\rm T}_{{\cal F}}\Upsilon =\Upsilon {\rm T}$.

\begin{proposition}
For any $\varphi \in {\cal F}^\Lambda ({\cal H}_\Lambda ^D)$ and any $\eta
\in {\cal H}_\Lambda ^D$ 
\begin{equation}
\left( {\rm T}_{{\cal F}}\varphi \right) (\eta )=\left\langle K_T(.,\eta
)\parallel \varphi \right\rangle  \label{ker.4}
\end{equation}
where the kernel $K_T:{\cal H}_\Lambda ^D\times {\cal H}_\Lambda
^D\rightarrow \Lambda $ is given by 
\begin{equation}
K_T(\zeta _1,\zeta _2):=\left\langle {\rm T}\exp \zeta _1\mid \exp \zeta
_2\right\rangle =\left\langle \exp \zeta _1\mid {\rm T}^{\prime }\exp \zeta
_2\right\rangle =\left( \exp \zeta _1\mid {\rm T}^{\prime }\exp \zeta
_2\right) .  \label{ker.5}
\end{equation}
\end{proposition}

\noindent {\bf Proof }The proof is quite similar to that of Lemma \ref
{Kernel}. If $\varphi \in {\cal F}^\Lambda ({\cal H}_\Lambda ^D)$ then there
exists a unique element $\Xi \in {\cal S}^\Lambda ({\cal H})$ for which $%
\varphi =\varphi _\Xi .$ For any $\eta \in {\cal H}_\Lambda ^D$ we have $%
\exp \eta \in {\cal S}^\Lambda ({\cal H}),$ and hence, if $\eta _1,\eta
_1\in {\cal H}_\Lambda ^D,$ and if $K_T$ is defined by (\ref{ker.5}) then \\$%
\left\langle K_T(.,\eta _2)\parallel \varphi \right\rangle =\left\langle
\left\langle \exp .\mid {\rm T}^{\prime }\exp \eta _2\right\rangle \parallel
\varphi _\Xi \right\rangle =\left\langle {\rm T}^{\prime }\exp \eta _2\mid
\Xi \right\rangle =\left\langle \exp \eta _2\mid {\rm T}\Xi \right\rangle =%
{\rm T}_{{\cal F}}\varphi .$ Here ${\rm T}^{\prime }$ is the transposed
operator (\ref{la.8}). \hfill $\square $ \smallskip

Take as example a (bounded) operator $A:{\cal H}\rightarrow {\cal H},$ and
let $T$ be its second quantization $T=\Gamma (A):{\cal S}({\cal H}%
)\rightarrow {\cal S}({\cal H}).$ This operator is represented on ${\cal F}%
^\Lambda ({\cal H}_\Lambda ^D)$ as $\varphi (\zeta )\rightarrow \left( {\rm T%
}_{{\cal F}}\varphi \right) (\zeta )=\varphi (A\zeta ),$ and the kernel of $%
\Gamma (A)$ is $K_{\Gamma (A)}(\zeta _1,\zeta _2)=\left\langle \Gamma ({\rm A%
})\exp \zeta _1\mid \exp \zeta _2\right\rangle =\left\langle \exp ({\rm A}%
\zeta _1)\mid \exp \zeta _2\right\rangle =e^{\left\langle {\rm A}\zeta
_1\mid \zeta _2\right\rangle }.$ If $A=I$ then $\Gamma (I)$ is the identity
operator on ${\cal S}({\cal H}),$ and we obtain the reproducing kernel (\ref
{ker.1}).

As further examples we consider two unbounded mappings, which are $\Lambda $%
-extensions of the creation and the annihilation operators. The mapping $\Xi
\in {\cal S}^\Lambda ({\cal H})\rightarrow {\rm T}_\varpi \Xi :=\varpi \Xi
\in {\cal S}^\Lambda ({\cal H})$ with an element $\varpi \in {\cal H}%
_\Lambda $ is an extension of the creation operator on ${\cal S}({\cal H}).$
On the space ${\cal F}^\Lambda ({\cal H}_\Lambda ^D)$ it is a multiplication
operator: $\left\langle \exp \zeta \mid \varpi \Xi \right\rangle
=\left\langle \zeta \mid \varpi \right\rangle \left\langle \exp \zeta \mid
\Xi \right\rangle .$ The kernel of this operator is $K_T(\zeta _1,\zeta
_2)=\left\langle \varpi \exp \zeta _1\mid \exp \zeta _2\right\rangle
=\left\langle \varpi \mid \zeta _2\right\rangle e^{\left\langle \zeta _1\mid
\zeta _2\right\rangle }.$\\The mapping $\Xi \in {\cal S}^\Lambda ({\cal H}%
)\rightarrow {\rm C}_\varpi \Xi :=\varpi \cont \Xi \in {\cal S}^\Lambda (%
{\cal H})$ with an element $\varpi \in {\cal H}_\Lambda $ is an extension of
the annihilation operator on ${\cal S}({\cal H}).$ The kernel of this
operator is \\$K_C(\zeta _1,\zeta _2)=\left\langle \varpi \cont \exp \zeta
_1\mid \exp \zeta _2\right\rangle =\left\langle \exp \zeta _1\mid \varpi
\exp \zeta _2\right\rangle =\left\langle \zeta _1\mid \varpi \right\rangle
e^{\left\langle \zeta _1\mid \zeta _2\right\rangle }.$\\If $\varpi \in {\cal %
H}_\Lambda ^D$ the operator ${\rm C}_\varpi $ is mapped by $\Upsilon $ onto
a differential operator on the space of functions ${\cal F}^\Lambda ({\cal H}%
_\Lambda ^D)$: $\left\langle \exp \zeta \mid \varpi \cont \Xi \right\rangle
=\left\langle \varpi \exp \zeta \mid \Xi \right\rangle =\varphi _\Xi
^{(1)}(\zeta )(\varpi )$, the Fr\'echet derivative with increment $\varpi .$

As in the usual analysis one can define the {\it Wick symbol} of an operator 
$T$ by, see e.g. \cite{Berezin:1971}\cite{Kree/Raczka:1978} \cite
{Berezin:1980}, 
\begin{equation}
w_T(\zeta _1,\zeta _2):=e^{-\left\langle \zeta _1\mid \zeta _2\right\rangle
}K_T(\zeta _1,\zeta _2)=K_T(\zeta _1,\zeta _2)e^{-\left\langle \zeta _1\mid
\zeta _2\right\rangle }\in \Lambda _{\overline{0}}  \label{ker.6}
\end{equation}
such that the identity operator has the symbol $1,$ and the symbols of
differential operators are polynomials. The operator symbols offer therefore
a simple prescription to relate classical differential operators to
operators on the Fock space ${\cal S}({\cal H})$. Another application is
their use in the evaluation of Feynman path integrals, see the (formal)
calculations in \cite{Berezin:1980} or \cite{Faddeev/Slavnov:1980}.

\section{Wiener-Segal representation\label{WS}}

\subsection{The Wiener-Segal representation and Gauss transform\label{Gt}}

The Wiener-Segal representation \cite{Wiener:1930}\cite{Ito:1951}\cite
{Segal:1956} is based on Wick ordering which has a well defined meaning also
for superalgebras. For $\Xi \in {\cal S}_{fin}^\Lambda ({\cal H})$ we
calculate the Wick ordered tensor ${\rm W}\Xi \in {\cal S}_{fin}^\Lambda (%
{\cal H})$ following the rule (\ref{Gauss4}): ${\rm W}(\lambda \otimes
F)=\lambda \otimes WF$ for $\lambda \in \Lambda $ and $F\in {\cal S}_{fin}(%
{\cal H}).$ Wick ordered polynomials are then defined as elements of ${\cal F%
}^\Lambda ({\cal E}_{\Lambda {\Bbb R}})$ by 
\begin{equation}
\Phi _\Xi (\xi ,\xi ^{*}):=\left\langle \exp \left( \xi +\xi ^{*}\right)
\mid {\rm W}\Xi \right\rangle =\left\langle \exp (\xi +\xi ^{*}-\Omega )\mid
\Xi \right\rangle .  \label{sdiff.16}
\end{equation}
For $\Xi =\kappa _0\otimes F$ with $F\in {\cal S}_{fin}({\cal H})$ we simply
write $\Phi _F(\xi ,\xi ^{*}).$ These polynomials form again an algebra
under pointwise multiplication which is isomorphic to the superalgebra $%
{\cal S}_{fin}({\cal H}).$ But in contrast to Theorem \ref{homomorph} the
rule is now 
\begin{equation}
\Phi _F(\xi ,\xi ^{*})\Phi _G(\xi ,\xi ^{*})=\Phi _{F\bigtriangleup G}(\xi
,\xi ^{*}),
\end{equation}
where $F\bigtriangleup G$ is Le Jan's Wiener-Grassmann product, which is
generated as associative product by $f\bigtriangleup g=f\odot g+\omega (f,g)$
for $f,g\in {\cal H}$, see \cite{LeJan:1988}\cite{Kupsch:1990}\cite
{Meyer:1993}. The polynomials (\ref{sdiff.16}) are continuous functions of $%
\xi \in {\cal E}_\Lambda $. The functions (\ref{sdiff.16}) can be
''integrated'' with respect to the supermeasure 
\begin{equation}
\int \Phi _\Xi (\xi ,\xi ^{*})d\upsilon (\xi ,\xi ^{*})=\left\langle \exp
\Omega \mid {\rm W}\Xi \right\rangle =\left\langle 1\mid \Xi \right\rangle .
\label{sdiff.17}
\end{equation}
Motivated by (\ref{sdiff.15}), we define the conjugation 
\begin{equation}
\Phi _\Xi ^{\dagger }(\xi ,\xi ^{*}):=\left( \Phi _{{\rm J}\Xi }(\xi ,\xi
^{*})\right) ^{*}  \label{sdiff.18}
\end{equation}
where ${\rm J}$ is the $\Lambda $-linear extension of (\ref{Gauss6}).

\begin{lemma}
The $\Lambda $-sesquilinear form (\ref{la.4}) can be calculated as the
integral 
\begin{equation}
\int \Phi _\Xi ^{\dagger }(\xi ,\xi ^{*})\Phi _H(\xi ,\xi ^{*})dv(\xi ,\xi
^{*})=\left( \Xi \mid H\right)  \label{sdiff.19}
\end{equation}
for all $\Xi ,H\in {\cal S}_{fin}^\Lambda ({\cal H}).$
\end{lemma}

\noindent {\bf Proof} We first restrict the tensors to $\Xi =\kappa
_0\otimes F$ and $H=\kappa _0\otimes G$ with $F,G\in {\cal S}_{fin}({\cal H}%
).$ From the definition (\ref{sdiff.16}) and Lemma \ref{homomorph} follows
that also the product of the functions $\Phi _F$ and $\Phi _G$ can be
integrated 
\begin{equation}
\int \Phi _F(\xi ,\xi ^{*})\Phi _G(\xi ,\xi ^{*})dv(\xi ,\xi
^{*})=\left\langle \exp \Omega \mid WF\odot WG\right\rangle \stackrel{(\ref
{Gauss10})}{=}\left\langle F\mid JG\right\rangle .  \label{sdiff.20}
\end{equation}
For $\Xi =\kappa _0\otimes F$ the antilinear mapping (\ref{sdiff.18}) is
related to (\ref{Gauss7}) by \\$\left( \Phi _{JF}(\xi ,\xi ^{*})\right)
^{*}=\left\langle WJF\mid \exp (\xi +\xi ^{*})\right\rangle
^{*}=\left\langle (WF)^{\dagger }\mid \exp (\xi +\xi ^{*})\right\rangle $.
The identity (\ref{Gauss9}) implies then that the sesquilinear form $\int
\Phi _F^{\dagger }(\xi ,\xi ^{*})\Phi _G(\xi ,\xi ^{*})dv(\xi ,\xi ^{*})$ is
the inner product of ${\cal S}({\cal H})$%
\begin{equation}
\int \Phi _F^{\dagger }(\xi ,\xi ^{*})\Phi _G(\xi ,\xi ^{*})dv(\xi ,\xi
^{*})=\left( F\mid G\right) .  \label{sdiff.21}
\end{equation}
Both sides of the identity (\ref{sdiff.20}) have a unique extension to
tensors $\Xi =\lambda \otimes F$ and $H=\mu \otimes G$ with $\lambda ,\mu
\in \Lambda $ and $F,G\in {\cal S}_{fin}({\cal H}).$ The definitions (\ref
{la.5}) (\ref{la.4}) (\ref{sdiff.16}) and (\ref{sdiff.18}) lead to an
additional left factor $\lambda ^{*}$ and an additional right factor $\mu $,
see the proof for Lemma \ref{Bargmann}. Then $\Lambda $-linearity finally
yields (\ref{sdiff.19}). \hfill $\square $ \smallskip

So far the functions (\ref{sdiff.16}) have been introduced only for $\Xi \in 
{\cal S}_{fin}^\Lambda ({\cal H})$. But the forms (\ref{sdiff.20}) and (\ref
{sdiff.21}) are well defined for the whole Fock space ${\cal S}({\cal H})$,
and they have their continuous extensions to ${\cal S}^\Lambda ({\cal H})$.
We can therefore define the completion of the space of functions (\ref
{sdiff.16}) to $\left\{ \Phi _\Xi (\xi ,\xi ^{*})\mid \Xi \in {\cal S}%
^\Lambda ({\cal H})\right\} $. Since Wick ordering is not continuous within $%
{\cal S}({\cal H}),$ the resulting functions have to be considered as
generalized functions. We choose a triplet of Hilbert spaces ${\cal H}%
^{+}\subset {\cal H}\subset {\cal H}^{-}$ with Hilbert-Schmidt embeddings.
Then the Wick ordered tensor $WF$ is defined as element of ${\cal S}({\cal H}%
)$ for all tensors $F$ in a dense subset ${\cal S}_{(\gamma )}({\cal H}%
^{+})\subset {\cal S}({\cal H}),\gamma >0,$ see Corollary \ref{Wick} in App.%
\ref{NormEst}. Hence, if $\Xi \in \Lambda \widehat{\otimes }{\cal S}%
_{(\gamma )}({\cal H}^{+})$ with $\gamma >0,$ the functions (\ref{sdiff.16})
are differentiable functions, which can be analytically continued to
functions in ${\cal F}\left( {\cal E}_\Lambda ^{-}\times {\cal E}_\Lambda
^{-*}\right) .$ Since ${\cal S}_{(\gamma )}({\cal H}^{+})$ has a Hilbert
topology, we can use the bilinear form (\ref{sdiff.20}) to define
generalized functions $\Phi _H$ for all $H$ in the (topological) dual space $%
\Lambda \widehat{\otimes }{\cal S}_{(-\gamma )}({\cal H}^{-}).$ This space
of generalized functions includes $\left\{ \Phi _\Xi (\xi ,\xi ^{*})\mid \Xi
\in {\cal S}^\Lambda ({\cal H})\right\} .$

The space of functions $\left\{ \Phi _F(\xi ,\xi ^{*})\mid F\in {\cal S}(%
{\cal H})\right\} $ is by construction a Hilbert space with the (formal $%
{\cal L}^2$-) inner product (\ref{sdiff.21}). This space of functions is
denoted as ${\cal W}({\cal E}_{\Lambda {\Bbb R}}^{-})$, see the discussion
of the domain of the supermeasure in Sect.\ref{Gfunct}. As images of the
Fock space ${\cal S}({\cal H})$ the spaces ${\cal F}\left( {\cal E}_{\Lambda 
{\Bbb R}}\right) $ or ${\cal F}\left( {\cal E}_\Lambda ^{-}\times {\cal E}%
_\Lambda ^{-*}\right) $ and ${\cal W}({\cal E}_{\Lambda {\Bbb R}}^{-})$ are
isomorphic. The $\Lambda $-extension is again denoted by ${\cal W}^\Lambda (%
{\cal E}_{\Lambda {\Bbb R}}^{-}).$ If $\Xi \in {\cal S}_{(\gamma )}^\Lambda (%
{\cal H}^{+})=\Lambda \widehat{\otimes }{\cal S}_{(\gamma )}({\cal H}^{+})$
with $\gamma >0$ both the functions $\Phi _\Xi (\xi ,\xi ^{*})\in {\cal W}%
^\Lambda ({\cal E}_{\Lambda {\Bbb R}}^{-})$ and $\varphi _\Xi \left( \xi
,\eta ^{*}\right) \in {\cal F}^\Lambda \left( {\cal E}_\Lambda ^{-}\times 
{\cal E}_\Lambda ^{-*}\right) $ are differentiable functions, moreover $%
\varphi _\Xi $ and $\Phi _\Xi $ can be analytically continued to functions
on ${\cal E}_\Lambda ^{-}\times {\cal E}_\Lambda ^{-*},$ such that $\varphi
_\Xi $ and $\Phi _\Xi \in {\cal F}^\Lambda \left( {\cal E}_\Lambda
^{-}\times {\cal E}_\Lambda ^{-*}\right) $. As in the analysis of the
bosonic Fock space \cite{Kondratiev:1991}\cite{Yokoi:1995}, these functions
can be related by the Gauss transform.

\begin{lemma}
\label{Gausstrans}For $\Xi \in {\cal S}_{(\gamma )}^\Lambda ({\cal H}%
^{+}),0<\gamma <1,$ the functions (\ref{sdiff.12}) and (\ref{sdiff.16}) are
related by the integral transforms 
\begin{equation}
\Phi _\Xi (\xi ,\xi ^{*})=\int \varphi _\Xi \left( \xi +i\zeta ,\xi
^{*}+i\zeta ^{*}\right) dv(\zeta ,\zeta ^{*})  \label{sdiff.22}
\end{equation}
and 
\begin{equation}
\varphi _\Xi \left( \xi ,\eta ^{*}\right) =\int \Phi _\Xi (\zeta +\xi ,\eta
^{*}+\xi ^{*})dv(\zeta ,\zeta ^{*}).  \label{sdiff.23}
\end{equation}
\end{lemma}

\noindent {\bf Proof }For $\Xi \in {\cal S}_{(\gamma )}^\Lambda ({\cal H}%
^{+}),0<\gamma <1,$ the functions $\varphi _\Xi $ and $\Phi _\Xi $ are
elements of ${\cal F}_{(\alpha )}^\Lambda ({\cal E}_{\Lambda {\Bbb R}}^{-})$
with $0<\alpha <\gamma $ and Lemma \ref{IntFunct} implies that the functions 
$\varphi _\Xi \left( \xi +i\zeta ,\xi ^{*}+i\zeta ^{*}\right) $ and $\Phi
_\Xi (\zeta +\xi ,\eta ^{*}+\xi ^{*})$ are integrable. Following Lemma \ref
{coherent} it is sufficient to verify the identities (\ref{sdiff.22}) and (%
\ref{sdiff.23}) for coherent states. If $\varsigma ,\tau \in {\cal H}^{+}$
then $\Xi =\exp (\varsigma +\tau ^{*})\in {\cal S}_{(\gamma )}^\Lambda (%
{\cal H}^{+})\subset {\cal S}^\Lambda ({\cal H})$ has the Bargmann-Fock
representation $\varphi _\Xi (\xi ,\eta ^{*})=\left\langle \exp (\xi +\eta
^{*})\mid \exp (\varsigma +\tau ^{*})\right\rangle =e^{\left\langle \xi \mid
\tau ^{*}\right\rangle +\left\langle \eta ^{*}\mid \varsigma \right\rangle
}, $ see (\ref{coh.4}) in App.\ref{CohStates}. The Wick ordered form of $\Xi 
$ is ${\rm W}\exp (\varsigma +\tau ^{*})=\exp (\varsigma +\tau
^{*})e^{-\left\langle \tau ^{*}\mid \varsigma \right\rangle }\in {\cal S}%
_{(\gamma )}^\Lambda ({\cal H}^{+}),$ see (\ref{coh.7}), and it is
represented by the function 
\begin{equation}
\Phi _\Xi (\zeta ,\zeta ^{*})=\left\langle \exp (\zeta +\zeta ^{*})\mid \exp
(\varsigma +\tau ^{*})\right\rangle e^{-\left\langle \tau ^{*}\mid \varsigma
\right\rangle }=e^{\left\langle \zeta \mid \tau ^{*}\right\rangle
+\left\langle \zeta ^{*}\mid \varsigma \right\rangle -\left\langle \tau
^{*}\mid \varsigma \right\rangle }  \label{sdiff.24}
\end{equation}
in the Wiener-Segal representation. The analytic continuation of this
function is\\$\Phi _\Xi (\xi ,\eta ^{*})=e^{\left\langle \xi \mid \tau
^{*}\right\rangle +\left\langle \eta ^{*}\mid \varsigma \right\rangle
-\left\langle \tau ^{*}\mid \varsigma \right\rangle }\in {\cal F}\left( 
{\cal E}_\Lambda ^{-}\times {\cal E}_\Lambda ^{-*}\right) .$ The identities (%
\ref{sdiff.22}) and (\ref{sdiff.23}) can now easily be derived from the
Laplace transform (\ref{sdiff.102}) of the supermeasure. The integral on the
right side of (\ref{sdiff.22}) is calculated as $\int e^{\left\langle \zeta
+i\eta \mid \tau ^{*}\right\rangle +\left\langle \zeta ^{*}+i\eta ^{*}\mid
\varsigma \right\rangle }dv(\eta ,\eta ^{*})=e^{\left\langle \xi \mid \tau
^{*}\right\rangle +\left\langle \eta ^{*}\mid \varsigma \right\rangle
}e^{-\left\langle \tau ^{*}\mid \varsigma \right\rangle }=\Phi _\Xi (\zeta
,\zeta ^{*}),$ and (\ref{sdiff.23}) follows from $\int e^{\left\langle \zeta
+\xi \mid \tau ^{*}\right\rangle +\left\langle \zeta ^{*}+\eta ^{*}\mid
\varsigma \right\rangle -\left\langle \tau ^{*}\mid \varsigma \right\rangle
}dv(\zeta ,\zeta ^{*})=e^{\left\langle \xi \mid \tau ^{*}\right\rangle
+\left\langle \eta ^{*}\mid \varsigma \right\rangle }=\varphi _\Xi \left(
\xi ,\eta ^{*}\right) .$ \hfill $\square $ \smallskip 

We can restrict the argument in (\ref{sdiff.23}) to the real subspace with $%
\xi =\eta .$ Then we do not need the analytic continuation of $\Phi _\Xi ,$
and (\ref{sdiff.23}) is an integral transform from a dense subspace of $%
{\cal W}^\Lambda ({\cal E}_{\Lambda {\Bbb R}}^{-})$ into the reproducing
kernel space ${\cal F}^\Lambda \left( {\cal E}_{\Lambda {\Bbb R}}\right) .$

\subsection{The Ornstein-Uhlenbeck semigroup}

With help of the Wiener-Segal representation many results of classical
analysis can be transferred to superanalysis. As example we consider the
Ornstein-Uhlenbeck semigroup on the Fock space ${\cal S}({\cal H}),$ and its
representation by the Mehler formula. For the bosonic case this semigroup
and the Mehler formula have been investigated in detail by Meyer \cite
{Meyer:1982} and by Kusuoka \cite{Kusuoka:1991}. The role of the
Ornstein-Uhlenbeck semigroup in the quantum field theory of bosons can
clearly be seen in \cite{Simon:1974}. So far results for the fermionic Fock
space have only been obtained using an ordering prescription, which is
unstable against perturbations. \cite{Kupsch:1991}

To avoid technical complications we assume that $A$ is a strictly
positive-definite selfadjoint operator on ${\cal H}$ with a pure point
spectrum satisfying the additional properties\\i) The operator $A$ is even,
i.e. $A{\cal H}_{\overline{k}}\subset {\cal H}_{\overline{k}},k=0,1.$\\ii)
The operator $A$ is real, i.e. $Af^{*}=(Af)^{*}$ for $f\in {\cal H}.$\\Then $%
e^{-At},t\geq 0,$ is a contraction semigroup on ${\cal H},$ and $\Gamma
(e^{-At}),t\geq 0,$ is by definition the Ornstein-Uhlenbeck semigroup with
generator $d\Gamma (A)\;$on ${\cal S}({\cal H}).$

On the dense subset $\left\{ \Phi _\Xi (\xi ,\xi ^{*})\mid \Xi \in {\cal S}%
_{(\gamma )}^\Lambda ({\cal H}^{+}),0<\gamma <1\right\} \subset {\cal W}(%
{\cal E}_{\Lambda {\Bbb R}}^{-})$ we define for $t\geq 0$ the
transformations 
\begin{equation}
\left( P_t\Phi \right) (\zeta ,\zeta ^{*})=\int \Phi (e^{-{\rm A}t}\zeta -%
\sqrt{I-e^{-2{\rm A}t}}\eta ,e^{-{\rm A}t}\zeta ^{*}-\sqrt{I-e^{-2{\rm A}t}}%
\eta ^{*})dv(\eta ,\eta ^{*}).  \label{ou.1}
\end{equation}
For the classical Wiener-Segal representation this is the Mehler formula of
the Ornstein-\\Uhlenbeck semigroup, see e.g. \cite{Meyer:1982}\cite
{Kusuoka:1991}. But also in superanalysis we have

\begin{lemma}
\label{OU}The Ornstein-Uhlenbeck semigroup $\Gamma (e^{-At}),t\geq 0,$ is
represented by the family of transformations (\ref{ou.1}).
\end{lemma}

\noindent {\bf Proof} It is sufficient to give the proof for coherent
states. The Ornstein-Uhlenbeck semigroup has a unique extension $\Gamma (e^{-%
{\rm A}t}),t\geq 0,$ to the space ${\cal S}^\Lambda ({\cal H}).$ These
transformations map the coherent state $\Xi =\exp (\varsigma +\tau ^{*})\in 
{\cal S}^\Lambda ({\cal H})$ onto $\Gamma (e^{-{\rm A}t})\Xi =\exp (e^{-{\rm %
A}t}\varsigma +e^{-{\rm A}t}\tau ^{*}).$ The Wick ordered form is {\rm W}$%
\Gamma (e^{-{\rm A}t})\Xi =\exp (e^{-{\rm A}t}\varsigma +e^{-{\rm A}t}\tau
^{*})e^{-\left\langle e^{-2{\rm A}t}\tau ^{*}\mid \varsigma \right\rangle }.$

The integral (\ref{ou.1}) can easily be calculated for the function (\ref
{sdiff.24}). The Fourier-Laplace transform (\ref{sdiff.102}) yields 
\[
\begin{array}{ll}
\left( P_t\Phi _\Xi \right) (\zeta ,\zeta ^{*}) & =e^{\left\langle e^{-{\rm A%
}t}\zeta \mid \tau ^{*}\right\rangle +\left\langle e^{-{\rm A}t}\zeta
^{*}\mid \varsigma \right\rangle -\left\langle e^{-2{\rm A}t}\tau ^{*}\mid
\varsigma \right\rangle }=e^{\left\langle \zeta \mid e^{-{\rm A}t}\tau
^{*}\right\rangle +\left\langle \zeta ^{*}\mid e^{-{\rm A}t}\varsigma
\right\rangle -\left\langle e^{-2{\rm A}t}\tau ^{*}\mid \varsigma
\right\rangle } \\ 
& =\left\langle \exp (\zeta +\zeta ^{*})\mid \exp (e^{-{\rm A}t}\varsigma
+e^{-{\rm A}t}\tau ^{*})\right\rangle e^{-\left\langle e^{-2{\rm A}t}\tau
^{*}\mid \varsigma \right\rangle } \\ 
& =\left\langle \exp (\zeta +\zeta ^{*})\mid {\rm W}\Gamma (e^{-{\rm A}%
t})\Xi \right\rangle .
\end{array}
\]
Hence (\ref{ou.1}) is a representation of the Ornstein-Uhlenbeck semigroup. %
\hfill $\square $

\bigskip\ 

\begin{center}
{\bf Acknowledgment}
\end{center}

\noindent O. G. Smolyanov would like to thank the Deutsche
Forschungsgemeinschaft (DFG) for financial support. \appendix

\section{Norm estimates for tensor algebras\label{NormEst}}

In this section we present Hilbert norm estimates for rather general ${\Bbb Z%
}$-graded algebras ${\cal A}$ over the field ${\Bbb K}={\Bbb R}$ or ${\Bbb C}
$.. We assume a structure ${\cal A}=\oplus _{n=0}^\infty {\cal A}_n$ as
considered for the tensor algebras in Sect.\ref{alg}. Thereby the one
dimensional space ${\cal A}_0={\Bbb K}$ is spanned by the unit $e,$ and the
product $F\circ G$ maps ${\cal A}_p\times {\cal A}_q$ into ${\cal A}_{p+q}$
for all $p,q\in \left\{ 0,1,...\right\} .$ The algebra is provided with the
Hilbert norm 
\begin{equation}
\left\| F\right\| ^2=\sum_{n=0}^\infty w(n)\left\| F_n\right\| _n^2\text{ if 
}F=\sum_{n=0}^\infty F_n\text{ with }F_n\in {\cal A}_n.  \label{a1.1}
\end{equation}
Here $\left\| .\right\| _n$ is a Hilbert norm of ${\cal A}_n,$ and the
factors $w(n)$ are positive weights with the normalization $w(0)=1.$ We
assume a norm estimate of the product of homogeneous elements as given in (%
\ref{alg8})\footnote{%
Here we keep the normalizations (\ref{alg8}) of the superalgebra in Sect.\ref
{alg}. The factor $\frac{(p+q)!}{p!q!}$ can be easily absorbed by a
redefinitionn of the norms $\left\| .\right\| _p.$} 
\begin{equation}
\left\| F_p\circ G_q\right\| _{p+q}^2\leq \frac{(p+q)!}{p!q!}\left\|
F_p\right\| _p^2\left\| G_q\right\| _q^2  \label{a1.2}
\end{equation}
if $F_p\in {\cal A}_p$ and $G_q\in {\cal A}_q.$

\begin{proposition}
\label{Estim1}If the norm is defined with the weights $w(n)=\left( n!\right)
^\gamma ,\gamma \leq -2,$ then the product of the algebra is continuous with
the uniform norm estimate 
\begin{equation}
\left\| F\circ G\right\| \leq \sqrt{3}\left\| F\right\| \left\| G\right\| .
\label{a1.3}
\end{equation}
\end{proposition}

\noindent {\bf Proof} The norm of $F\circ G$ for $F=\sum_{n=0}^\infty F_n$
and $G=\sum_{n=0}^\infty G_n$ with $F_n,G_n\in {\cal A}_n$ is given by 
\[
\begin{array}{l}
\left\| F\circ G\right\| ^2=\left\| \sum_{m,n}F_m\circ G_n\right\| ^2 \\ 
\leq \left| F_0G_0\right| ^2+3\left( \left| F_0\right| ^2\left\|
\sum_{n=1}^\infty G_n\right\| ^2+\left\| \sum_{m=1}^\infty F_m\right\|
^2\left| G_0\right| ^2+\left\| \sum_{m\geq 1,n\geq 1}F_m\circ G_n\right\|
^2\right) \\ 
\leq \left| F_0G_0\right| ^2+3\sum_{n\geq 1}w(n)\left( \left| F_0\right|
^2\left\| G_n\right\| _n^2+\left\| F_n\right\| _n^2\left| G_0\right|
^2+\left\| \sum_{p+q=n}^{^{\prime }}F_p\circ G_q\right\| _n^2\right)
\end{array}
\]
The symbol $\sum^{^{\prime }}$ means summation with the constraint $p\geq
1,q\geq 1$. The sum \\$\sum_{p+q=n,p\geq 1,q\geq
1}...=\sum_{p+q=n}^{^{\prime }}...$ has $n-1$ terms, hence

$\left\| \sum_{p+q=n}^{^{\prime }}F_p\circ G_q\right\| _n^2\leq
(n-1)\sum_{p+q=n}^{^{\prime }}\left\| F_p\circ G_q\right\| _n^2\stackrel{(%
\ref{a1.2})}{\leq }(n-1)\sum_{p+q=n}^{^{\prime }}\left( 
\begin{array}{l}
n \\ 
p
\end{array}
\right) \left\| F_p\right\| _p^2\left\| G_q\right\| _q^2$. \\If $w(n)$ is
chosen such that 
\begin{equation}
(p+q-1)\frac{(p+q)!}{p!q!}w(p+q)\leq w(p)w(q)\text{ for all }p,q\geq 1,
\label{a1.4}
\end{equation}
we finally obtain $\sum_{n\geq 1}w(n)\left\| \sum_{p+q=n}^{^{\prime
}}F_p\circ G_q\right\| _n^2\leq \left( \sum_{p\geq 1}w(p)\left\| F_p\right\|
_p^2\right) \left( \sum_{q\geq 1}w(q)\left\| G_q\right\| _q^2\right) ,$
hence 
\[
\begin{array}{l}
\left\| F\circ G\right\| ^2 \\ 
\leq \left| F_0G_0\right| ^2+3\left( \left| F_0\right| ^2\left\|
\sum_{n=1}^\infty G_n\right\| ^2+\left\| \sum_{m=1}^\infty F_m\right\|
^2\left| G_0\right| ^2+\left\| \sum_{m=1}^\infty F_m\right\| ^2\left\|
\sum_{n=1}^\infty G_n\right\| ^2\right) \\ 
\leq 3\left\| F\right\| ^2\left\| G\right\| ^2.
\end{array}
\]
With $w(n)=\frac 1{n!}\alpha (n)$ the inequalities (\ref{a1.4}) are
equivalent to $(p+q-1)\alpha (p+q)\leq \alpha (p)\alpha (q)$ for all $%
p,q\geq 1.$ A function which satisfies these constraints is $\alpha
(n)=(n!)^{-\gamma },\gamma \geq 1.$ \hfill $\square $ \smallskip 

It might be possible to derive an estimate (\ref{a1.3}) with a constant $c<%
\sqrt{3},$ but the value $c=1$ is definitely excluded due to the following

\begin{lemma}
Let ${\cal A}$ be an algebra over the field ${\Bbb K}={\Bbb R}$ or ${\Bbb C}$
with dimension $\dim {\cal A}\geq 2.$ If this algebra satisfies the
properties\\i)${\cal A}$ is provided with a Hilbert inner product $\left(
.\mid .\right) $ normalized at the unit $e$, $\left\| e\right\| ^2=\left(
e\mid e\right) =1$,\\ii)there exists at least one element $f\in {\cal A}%
,f\neq 0$, such that each two of the elements $e,f$ and $f^2=f\circ f$ are
orthogonal, \\then the norm estimate $\left\| F\circ G\right\| \leq c\left\|
F\right\| \left\| G\right\| $ is not valid for some $F,G\in {\cal A},$ if $c<%
\sqrt{\frac 43}.$
\end{lemma}

\noindent {\bf Proof }Since{\bf \ }$f\neq 0$ we can normalize it and assume $%
\left\| f\right\| =1$. Take $F=e+\alpha f$ with $\alpha \in {\Bbb R..}$ Then 
$F^2=e+2\alpha f+\alpha ^2f^2$ and $\left\| F^2\right\| ^2=1+4\alpha
^2+\alpha ^4\left\| f^2\right\| ^2.$ On the other hand $\left\| F\right\|
^2=1+\alpha ^2,$ and $\left\| F^2\right\| ^2\leq c^2\left\| F^2\right\| ^4$
implies $1+4\alpha ^2+\alpha ^4\left\| f^2\right\| \leq c^2(1+\alpha ^2)^2.$
But this inequality is true for all $\alpha \geq 0$ only if $c^2\geq
\sup_{\alpha \geq 0}\frac{1+4\alpha ^2}{(1+\alpha ^2)^2}=\frac 43$. \hfill $%
\square $ \smallskip

This Lemma applies to ${\Bbb Z}$-graded algebras as considered above with
any(!) choice of the weights $w(n)>0,w(0)=1.$ For these algebras we can
simply choose $f$ from the generating space ${\cal A}_1$. The Lemma is true
for any algebra with unit, which has two linearly independent nilpotent
elements $f_1$ and $f_2$. In that case there always exists a nilpotent
element $\alpha f_1+\beta f_2\neq 0,$ which is orthogonal to the unit
element, such that the second condition is satisfied.

The Grassmann superalgebra $\Lambda $ of the main text is exactly chosen as
the antisymmetric tensor algebra with the product defined according to (\ref
{alg4}) with projection onto antisymmetric tensors, and a norm (\ref{a1.1})
with $w(n)=(n!)^{-2}.$

By a slight modification of the estimates given for the proof of Proposition 
\ref{Estim1} one can obtain another interesting result. Let $w_\gamma
(n),\gamma \in {\Bbb R},$ be a one parameter family of weights with $%
w_\alpha (n)\leq w_\beta (n)$ if $\alpha \leq \beta $ for all $n\in {\Bbb N}$%
. The corresponding norms (\ref{a1.1}) are denoted by $\left\| F\right\|
_{(\gamma )}.$ If these weights satisfy\footnote{%
For this estimate the one dimensional space generated by the unit has not to
be separated, and the sum $\sum_{p+q=n}...$ contains $n+1$ terms.} 
\begin{equation}
(p+q+1)\frac{(p+q)!}{p!q!}w_\gamma (p+q)\leq c\cdot w_\alpha (p)w_\beta (q)%
\text{ for }p,q\geq 0,  \label{a1.4a}
\end{equation}
for some $\alpha ,\beta ,\gamma \in {\Bbb R},$ then the product of ${\cal A}$
is estimated by $\left\| F\circ G\right\| _{(\gamma )}\leq \sqrt{c}\cdot
\left\| F\right\| _{(\alpha )}\left\| G\right\| _{(\beta )}$ with exactly
that constant which appears in (\ref{a1.4a}). If we now choose $w_\gamma
(n)=(n!)^\gamma ,$ the estimate (\ref{a1.4a}) is satisfied if $\gamma <\min
(\alpha ,\beta ),$ because $(p+q)!\leq 2^{p+q}p!q!,$ and $p\in \left\{
0,1,...\right\} \rightarrow (p+1)2^p(p!)^{-1-\varepsilon }$ is a bounded
function for any fixed $\varepsilon >0.$ Hence we have derived

\begin{proposition}
\label{Estim2}If the algebra ${\cal A}$ is equipped with the family of norms 
$\left\| .\right\| _{(\gamma )}$ induced by the weight functions $w_\gamma
(n)=(n!)^\gamma ,\gamma \in {\Bbb R},$ the product satisfies the norm
estimates 
\[
\left\| F\circ G\right\| _{(\gamma )}\leq c\cdot \left\| F\right\| _{(\alpha
)}\left\| G\right\| _{(\beta )} 
\]
if $\gamma <\min (\alpha ,\beta )$ with a constant $c,$ which depends on the
parameters $\alpha ,\beta $ and $\gamma .$
\end{proposition}

The norm estimates of this appendix can be applied to the superalgebra $%
{\cal S}_{fin}({\cal H})$ of Sect. \ref{alg}. If we chose a norm (\ref{a1.1}%
) with a weight $w(n)=(n!)^\gamma ,\gamma \in {\Bbb R},$ the completion of $%
{\cal S}_{fin}({\cal H})$ is denoted as ${\cal S}_{(\gamma )}({\cal H}).$
These spaces satisfy the inclusions ${\cal S}_{(\alpha )}({\cal H})\subset 
{\cal S}_{(\beta )}({\cal H})$ if $\alpha \leq \beta .$ The Fock space of
Sect \ref{alg} is ${\cal S}_{(0)}({\cal H})={\cal S}({\cal H}),$ and the
spaces ${\cal S}_{(-\gamma )}({\cal H})$ and ${\cal S}_{(\gamma )}({\cal H})$
are dual with respect to the pairing (\ref{alg11}). Following Proposition 
\ref{Estim2} the graded tensor product (\ref{alg4}) can be extended from $%
{\cal S}_{fin}({\cal H})$ to any space ${\cal S}_{(\gamma )}({\cal H}%
)\subset {\cal S}({\cal H})$ with $\gamma <0,$ such that $F,G\in {\cal S}%
_{(\gamma )}({\cal H})\rightarrow F\odot G\in {\cal S}({\cal H}).$

For the constructions presented in Sects. \ref{Gauss}-\ref{WS} the
convergence of the exponential series of vectors (coherent states) and of
tensors of rank 2 (Gaussian functionals) are important. The bound (\ref{a1.2}%
) (or the equivalent estimate (\ref{alg8})) imply 
\begin{equation}
\left\| f^{\odot n}\right\| _n\leq \sqrt{n!}\left\| f\right\| ^n\text{ if }%
f\in {\cal H}\text{ and }\left\| F^{\odot n}\right\| _{2n}^2\leq \frac{(2n)!%
}{2^{2n}}\left\| F\right\| _2^{2n}\text{ if }F\in {\cal H}^{\odot 2}.
\label{a1.5}
\end{equation}
A simple estimate then leads to 
\begin{equation}
\exp f\in {\cal S}_{(\gamma )}({\cal H})\text{ if }\gamma <1,\text{ and }%
\exp F\in {\cal S}_{(\gamma )}({\cal H})\text{ if }\gamma <0.  \label{a1.6}
\end{equation}
The convergence of $\exp F$ within ${\cal S}_{(0)}({\cal H})$ can only be
derived from (\ref{a1.5}), if the norm of $F$ is small.

The tensor $\Omega $ related to the bilinear continuous form (\ref{Gauss1})
is not an element of ${\cal S}_2({\cal H})$. We have to choose a triplet of
Hilbert spaces ${\cal H}^{+}\subset {\cal H}\subset {\cal H}^{-}$ with
Hilbert-Schmidt embeddings\footnote{%
These embeddings should induce the corresponding triplets for the subspaces $%
{\cal E}_{\overline{0}},{\cal E}_{\overline{1}},{\cal E}_{\overline{0}}^{*}$
and ${\cal E}_{\overline{1}}^{*},$ such that also ${\cal H}^{-}$ and ${\cal H%
}^{+}$ are Hilbert spaces with the structure presented in Chap.\ref{Hilbert}.%
}, then $\Omega \in {\cal S}_2({\cal H}^{-}).$ Following (\ref{a1.6}) the
tensor $\exp \Omega $ is an element of ${\cal S}_{(-\alpha )}({\cal H}^{-})$
with arbitrarily small $\alpha >0$, and the functional (\ref{Gauss2}) is
continuous on the dual space ${\cal S}_{(\alpha )}({\cal H}^{+}).$ To obtain
a domain on which Wick ordering is defined we first derive

\begin{proposition}
\label{Contraction}If $Y\in {\cal S}_{(-\alpha )}({\cal H}^{-})$ and $F\in 
{\cal S}_{(\gamma )}({\cal H}^{+}),0<\alpha <\gamma ,$ then $Y\cont F\in 
{\cal S}_{(\alpha )}({\cal H}^{+})\subset {\cal S}({\cal H}).$
\end{proposition}

\noindent {\bf Proof }Let $H\in {\cal S}_{(-\alpha )}({\cal H}^{-}).$
Following Proposition \ref{Estim2} the tensor $H\odot Y$ is an element of $%
{\cal S}_{(-\gamma )}({\cal H}^{-}),$ and $F\in {\cal S}_{(\gamma )}({\cal H}%
^{+})\rightarrow \left\langle H\odot Y\mid F\right\rangle $ is a continuous
functional, which satisfies \\$\left| \left\langle H\odot Y\mid
F\right\rangle \right| \leq const\left\| H\right\| _{(-\alpha )}\left\|
F\right\| _{(\gamma )}$ with the norms of ${\cal S}_{(-\alpha )}({\cal H}%
^{-})$ and ${\cal S}_{(\gamma )}({\cal H}^{+})$, respectively. Since $%
\left\langle H\odot Y\mid F\right\rangle =\left\langle H\mid Y\cont
F\right\rangle $ by definition of the contraction, the tensor $Y\cont F$ is
an element of ${\cal S}_{(\alpha )}({\cal H}^{+})\subset {\cal S}({\cal H}).$
\hfill $\square $

\begin{corollary}
\label{Wick}If $F\in {\cal S}_{(\gamma )}({\cal H}^{+}),\gamma >0,$ then $%
WF=\exp (-\Omega )\cont F\in {\cal S}_{(\alpha )}({\cal H}^{+})\subset {\cal %
S}({\cal H})$ with $0<\alpha <\gamma .$
\end{corollary}

\section{Polynomials on Grassmann Algebras\label{Pol}}

In this appendix we derive statements about polynomials defined on an
infinite dimensional Grassmann algebra $\Lambda $ with involution. The
Grassmann algebra $\Lambda $ is the antisymmetric tensor algebra over a
complex Hilbert space ${\cal F}$. It has the direct sum structure $\Lambda
=\oplus _{p=0}^\infty \Lambda _p$ with $\Lambda _p={\cal T}_p^{-}({\cal F}),$
and it is provided with a Hilbert norm (\ref{a1.1}) such that the product is
continuous. For the proofs it is convenient to use a real orthonormal basis $%
\kappa _a=\kappa _a^{*}\in {\cal F}=\Lambda _1,a\in {\bf N.}$ Then $\kappa _{%
{\bf A}}=(-i)\kappa _{a_1}...\kappa _{a_n},{\bf A}=\left\{
a_1<...<a_n\right\} \subset {\Bbb N}$ forms a real basis of $\Lambda _n.$
Any element $\rho \in \Lambda $ has the representation $\rho =\sum_{{\bf %
A\subset }{\it P}({\Bbb N})}h({\bf A})e_{{\bf A}}$ with complex coefficients 
$h({\bf A}),$ where ${\it P}({\Bbb N})$ is the power set of ${\Bbb N}$, i.e.
the set of all finite subsets of ${\Bbb N}.$

\begin{proposition}
\label{p1} If an element $\rho $ of the Grassmann algebra $\Lambda $
satisfies one of the following conditions 
\[
\left\{ 
\begin{array}{l}
\text{i) }\rho \lambda =0\text{ for all }\lambda \in \Lambda _1{\bf ,}\text{
or} \\ 
\text{ii) }\rho \lambda \lambda ^{*}=0\text{ for all }\lambda \in \Lambda _1%
{\bf ,}
\end{array}
\right. 
\]
then $\rho =0.$
\end{proposition}

\noindent {\bf Proof} Any element of the Grassmann algebra $\Lambda $ has
the series representation \\$\rho =\sum_{{\bf A\in }{\it P}({\Bbb N})}h({\bf %
A)}\kappa _{{\bf A}}.$ To derive the consequences of assumption i) we choose
for $\lambda $ a basis element $\lambda =\kappa _b,b\in {\Bbb N}{\bf .}$
Then $0=\rho \kappa _b=\sum_{{\bf A\in }{\it P}({\Bbb N}),{\bf A\cap }%
\left\{ b\right\} {\bf =\emptyset }}h({\bf A)}e_{{\bf A\cup }\left\{
b\right\} }$ implies $h({\bf A})=0$ for all ${\bf A\in }{\it P}({\Bbb N})$
with ${\bf A\cap }\left\{ b\right\} =\emptyset .$ Any set ${\bf A\in }{\it P}%
({\Bbb N})$ is a finite set, and for any ${\bf A\in }{\it P}({\Bbb N})$
there exist $b\in {\Bbb N}{\bf ,}$ such that ${\bf A\cap }\left\{ b\right\}
=\emptyset $ is satisfied. Hence assumption i) implies $h({\bf A)=}0$ for
all ${\bf A\in }{\it P}({\Bbb N})$.

To derive the consequences of assumption ii) we choose $\lambda =\kappa
_b+i\kappa _{b+1},b\in {\Bbb N}{\bf .}$ Then assumption ii) implies $h({\bf A%
})=0$ for all ${\bf A\in }{\it P}({\Bbb N})$ with ${\bf A\cap }\left\{
b,b+1\right\} =\emptyset .$ Since $b\in {\Bbb N}$ is arbitrary, we obtain
again $h({\bf A})=0$ for all ${\bf A\in }{\it P}({\Bbb N})$ without
restriction. \hfill $\square $ \medskip\ 

In Sect.\ref{RKS} we have to evaluate continuous polynomials on the
underlying real space of $\Lambda ^{1,n}=\Lambda _{\overline{0}}\times
\left( \Lambda _{\overline{1}}\right) ^n$. These polynomials have the
structure 
\[
\begin{array}{l}
{\bf \lambda }=(\alpha ,\vartheta _1,...,\vartheta _n)\in \Lambda _{{\Bbb R}%
}^{1,n}\rightarrow \\ 
{\rm pol}({\bf \lambda )=}\sum_{(p,q,{\bf A},{\bf B})}\rho (p,q,{\bf A},{\bf %
B})\alpha ^p\alpha ^{*q}\vartheta _{{\bf A}}\vartheta _{{\bf B}}^{*}\in
\Lambda ,
\end{array}
\]
where the sum extends over all numbers $p=0,1,...,m$ and $q=0,1,...,m$ and
all ordered subsets ${\bf A,B}\subset \left\{ 1,...,n\right\} $. The factors 
$\vartheta _{{\bf A}}$ and $\vartheta _{{\bf B}}^{*}$ are the fermionic
monomials $\vartheta _{{\bf A}}=\vartheta _{a_1}...\vartheta _{a_r}$ and $%
\vartheta _{{\bf B}}^{*}=\vartheta _{b_1}^{*}...\vartheta _{b_s}^{*}$ if $%
{\bf A}=\left\{ a_1<a_2<...<a_r\right\} $ and ${\bf B}=\left\{
b_1<b_2<...<b_s\right\} $. The number of elements of these sets is $\left| 
{\bf A}\right| =r\leq n$ and $\left| {\bf B}\right| =s\leq n.$ The
coefficients $\rho (p,q,{\bf A},{\bf B})$ are arbitrary elements of $\Lambda 
$. The polynomial is homogeneous of degree $N$ if only terms with $p+q+r+s=N$
contribute.

\begin{lemma}
\label{polynom}The polynomial ${\rm pol}{\cal (}{\bf \lambda )}$ is
identical zero on $\Lambda _{{\Bbb R}}^{1,n}$ if and only if all
coefficients $\rho $ vanish.
\end{lemma}

\noindent {\bf Proof} If all $\rho (p,q,{\bf A,B})=0$ then ${\rm pol}({\bf %
\lambda )=}0$ follows obviously for ${\bf \lambda }\in \Lambda _{{\Bbb R}%
}^{1,n}$. To derive the inverse statement we substitute the arguments $%
\alpha \in \Lambda _{\overline{0}}$ by $z\alpha \in \Lambda _{\overline{0}}$
and $\vartheta _a\in \Lambda _{\overline{1}}$ by $z_a\vartheta _a\in \Lambda
_{\overline{1}}$\\with complex numbers $z=x+iy,z_a=x_a+iy_a$ where $x,x_a\in 
{\Bbb R}$ and $y{\bf ,}y_a\in {\Bbb R}{\bf ,}a=1,...,n.$ The function ${\rm %
pol}(z\alpha ,...{\bf )}$ is then a polynomial in the real variables $x,x_a$
and $y,y_b,$ or, equivalently, a polynomial in the variables $z,z_a$ and $%
\overline{z},\overline{z}_b$, $a,b=1,...,n.$ Since the multiplication is
continuous within $\Lambda ,$ this polynomial is a differentiable function
in $z,z_a$ and $\overline{z},\overline{z_b},$ and the partial derivative $%
\left( \frac \partial {\partial z}\right) ^p\left( \frac \partial {\partial
z_{a_1}}\right) \cdot \cdot \cdot \left( \frac \partial {\partial
z_{a_r}}\right) \left( \frac \partial {\partial \overline{z}}\right)
^q\left( \frac \partial {\partial \overline{z}_{b_1}}\right) .\cdot \cdot
\left( \frac \partial {\partial \overline{z}_{b_s}}\right) {\rm pol}(z\alpha
,...{\bf )}$ at $z=z_1=...=\overline{z}_n$is exactly $\rho (p,q,{\bf A,B}%
)p!q!\alpha ^p\alpha ^{*q}\vartheta _{{\bf A}}\vartheta _{{\bf B}}^{*}$. If $%
{\rm pol}({\bf \lambda )}$ is identical zero, all derivatives vanish, and we
have $\rho (p,q,{\bf A,B})\alpha ^p\alpha ^{*q}\vartheta _{{\bf A}}\vartheta
_{{\bf B}}^{*}=0$ for all $p,q=0,1,...,m$ and all subsets ${\bf A,B\subset }%
\left\{ 1,...,n\right\} $ with arbitrary $\left( \alpha ,\vartheta
_1,...,\vartheta _n\right) \in \Lambda ^{1,n}$. For the bosonic arguments we
can choose $\alpha =\alpha ^{*}=id.$ The remaining identities with fermionic
factors $\vartheta _a\in \Lambda _1,a=1,...,n,$ can then be reduced by
induction with the help of Proposition \ref{p1} to $\rho (p,q,{\bf A,B})=0$. 
\hfill
$\square $ \smallskip

Since the proof has only used arguments $\alpha \in {\Bbb C}$ and $\vartheta
_a\in \Lambda _1\subset \Lambda _{\overline{1}}$ we have even derived

\begin{corollary}
\label{restricted}The polynomial ${\rm pol}{\cal (}{\bf \lambda )}$ is
uniquely determined by arguments ${\bf \lambda }=(\alpha ,\vartheta
_1,...,\vartheta _n)$ with $\alpha \in {\Bbb C}\subset \Lambda _{\overline{0}%
}$ and $\vartheta _a\in \Lambda _1\subset \Lambda _{\overline{1}},a=1,...,n$.
\end{corollary}

\noindent Lemma and Corollary have an obvious generalization to polynomials
on the spaces \\$\Lambda ^{m,n}=\left( \Lambda _{\overline{0}}\right)
^m\times \left( \Lambda _{\overline{1}}\right) ^n,$ with $m,n\in {\Bbb N}.$

\section{Calculations for coherent states\label{CohStates}}

\subsection{Basic identities}

The $\Lambda $-bilinear forms $(\xi ,\eta )\in {\cal H}_\Lambda \times {\cal %
H}_\Lambda \rightarrow \left\langle \xi \mid \eta \right\rangle \in \Lambda $
and $\omega (\xi ,\eta )\in \Lambda $ are uniquely defined by linear
extension to the algebraic superspace ${\cal H}_\Lambda ^{_{alg}}=\Lambda _{%
\overline{0}}\otimes {\cal H}_{\overline{0}}\oplus \Lambda _{\overline{1}%
}\otimes {\cal H}_{\overline{1}}$ and by closure. The form $\omega $ is
symmetric on ${\cal H}_\Lambda \times {\cal H}_\Lambda ,$ and we have 
\begin{equation}
\omega (\xi ,\eta )=\left\langle \xi \mid {\rm j}\eta \right\rangle
\label{coh.1}
\end{equation}
for all $\xi ,\eta \in {\cal H}_\Lambda ,$ where ${\rm j}$ is the $\Lambda $%
-extension of (\ref{h.11}).If $\xi \in {\cal E}_\Lambda ^{*}$ and/or $\eta
\in {\cal E}_\Lambda =\Lambda _{\overline{0}}\otimes {\cal E}_{\overline{0}%
}\oplus \Lambda _{\overline{1}}\otimes {\cal E}_{\overline{1}},$ this
identity implies $\left\langle \xi \mid \eta \right\rangle =\omega (\xi
,\eta ).$ The bilinear form (\ref{la.5}) has the factorization 
\begin{equation}
\left\langle \xi _m\cdot \cdot \cdot \xi _1\mid \eta _1\cdot \cdot \cdot
\eta _n\right\rangle =\delta _{mn}\sum_\sigma \left\langle \xi _1\mid \eta
_{\sigma (1)}\right\rangle ...\left\langle \xi _n\mid \eta _{\sigma
(n)}\right\rangle  \label{coh.2}
\end{equation}
for $\xi _a,\eta _b\in {\cal H}_\Lambda ,a=1,...,m$ and $b=1,...,n,$ see (%
\ref{alg11a}). The summation extends over all permutations $\sigma $ of $%
\left\{ 1,...,n\right\} .$ Using the identity (\ref{coh.1}) we finally
obtain with the $\Lambda $-extension of (\ref{Gauss6}), denoted by ${\rm J},$%
\begin{equation}
\left\langle \xi _m\cdot \cdot \cdot \xi _1\mid {\rm J}\eta _1\cdot \cdot
\cdot \eta _n\right\rangle =\delta _{mn}\sum_\sigma \omega (\xi _1,\eta
_{\sigma (1)})...\omega (\xi _n,\eta _{\sigma (n)}).  \label{coh.3}
\end{equation}

\subsection{Bilinear forms and contractions of coherent states}

The bilinear pairing (\ref{coh.2}) of two coherent states (\ref{sfu.5})
yields 
\begin{equation}
\left\langle \exp \xi \mid \exp \eta \right\rangle =\sum_{p=0}^\infty \left(
\frac 1{p!}\right) ^2p!\left\langle \xi \mid \eta \right\rangle
^p=e^{\left\langle \xi \mid \eta \right\rangle }\in \Lambda _{\overline{0}}.
\label{coh.4}
\end{equation}
Since $(\exp \xi )^{*}=\exp \xi ^{*},$ this identity implies $\left( \exp
\xi \mid \exp \eta \right) =e^{\left( \xi ,\eta \right) }.$ The contraction (%
\ref{alg13}) has a $\Lambda $-extension which we denote by the same symbol $%
\cont $ . Since $\exp \eta \in {\cal S}_{(\gamma )}^\Lambda ({\cal H})$ for
any $\gamma <1$ the contraction $\exp \eta \cont \Xi $ is a well defined
element of ${\cal S}^\Lambda ({\cal H})$ if $\Xi \in {\cal S}_{(\alpha
)}^\Lambda ({\cal H})$ with $\alpha >0.$ The proof of this statement follows
by $\Lambda -$extension from Proposition \ref{Contraction} in App. \ref
{NormEst}. As simple application we calculate 
\begin{equation}
\exp \eta \cont \exp \xi =e^{\left\langle \xi \mid \eta \right\rangle }\exp
\xi  \label{coh.5}
\end{equation}
if $\xi ,\eta \in {\cal H}_\Lambda .$ This identity follows from \\$%
\left\langle \exp \zeta \mid \exp \eta \cont \exp \xi \right\rangle
=\left\langle \exp (\eta +\zeta )\mid \exp \xi \right\rangle \stackrel{(\ref
{coh.4})}{=}e^{\left\langle \eta +\zeta \mid \xi \right\rangle
}=e^{\left\langle \xi \mid \eta \right\rangle }\left\langle \exp \zeta \mid
\exp \xi \right\rangle ,$ and Lemma \ref{injection}. The result is also true
for arguments $\xi \in {\cal H}_\Lambda ^{+},\eta \in {\cal H}_\Lambda ^{-}.$

\subsection{Gaussian functionals and Wick ordering}

For $\xi ,\eta \in {\cal E}_\Lambda $ we have $\exp (\xi +\eta ^{*})=\left(
\exp \eta \right) ^{*}\cdot \left( \exp \xi \right) \in {\cal S}({\cal E}%
_\Lambda ^{*})\cdot {\cal S}({\cal E}_\Lambda ),$ and \\$\left\langle \exp
\Omega \mid \exp (\xi +\eta ^{*})\right\rangle =\left( \exp \eta \mid \exp
\xi \right) =e^{\left( \eta \mid \xi \right) }=e^{\frac 12\omega (\xi +\eta
^{*},\xi +\eta ^{*})}$ follows from (\ref{Gauss3}) and (\ref{coh.4}). Hence 
\begin{equation}
\left\langle \exp \Omega \mid \exp \zeta \right\rangle =e^{\frac 12\omega
(\zeta ,\zeta )}  \label{coh.6}
\end{equation}
follows for all $\zeta \in {\cal H}_\Lambda .$

For Wick ordering we first take a $\xi \in {\cal H}_\Lambda ^{+}$ then $\exp
\xi \in {\cal S}_{(\gamma )}^\Lambda ({\cal H}^{+})$ with $0<\gamma <1$.
Following Proposition \ref{Contraction} in App.\ref{NormEst} the
contractions $\exp (\pm \Omega )\cont \exp \xi $ are defined, and we
calculate $\left\langle \exp \eta \mid {\rm W}^{\pm 1}\exp \xi \right\rangle
=\left\langle \exp \eta \cdot \exp (\mp \Omega )\mid \exp \xi \right\rangle
=\left\langle \exp (\mp \Omega )\mid \exp \eta \cont \exp \xi \right\rangle 
\stackrel{(\ref{coh.5})}{=}\\e^{\left\langle \eta \mid \xi \right\rangle
}\left\langle \exp (\mp \Omega )\mid \exp \xi \right\rangle =e^{\left\langle
\eta \mid \xi \right\rangle \mp \frac 12\omega (\xi ,\xi )}$ for arbitrary $%
\eta \in {\cal H}_\Lambda ^{-}.$ Then (\ref{coh.4}) and Lemma \ref{injection}
yield 
\begin{equation}
{\rm W}^{\pm 1}\exp \xi =e^{\mp \frac 12\omega (\xi ,\xi )}\exp \xi .
\label{coh.7}
\end{equation}
This result has a unique continuous extension to $\xi \in {\cal H}_\Lambda $.

For the proof of the fundamental identity (\ref{Gauss5}) we first derive 
\begin{equation}
\left\langle \exp \Omega \mid \left( \exp \xi \right) \left( \exp \eta
\right) \right\rangle =\left\langle {\rm W}^{-1}\exp \xi \mid {\rm JW}%
^{-1}\exp \eta \right\rangle  \label{coh.8}
\end{equation}
with $\xi ,\eta \in {\cal H}_\Lambda .$ Following (\ref{coh.4}) the left
side of (\ref{coh.8}) is $e^{\frac 12\omega (\xi +\eta ,\xi +\eta )}.$ The
right side is calculated with the help of (\ref{coh.3}) and (\ref{coh.7}) as 
$e^{\frac 12\omega (\xi ,\xi )}e^{\frac 12\omega (\eta ,\eta )}\left\langle
\exp \xi \mid {\rm J}\exp \eta \right\rangle =e^{\frac 12\omega (\xi ,\xi
)+\frac 12\omega (\eta ,\eta )+\omega (\xi ,\eta )},$ and (\ref{coh.8}) is
valid for all $\xi ,\eta \in {\cal H}_\Lambda .$ The bilinear mapping ($%
F,G)\in {\cal S}_{fin}({\cal H})\times {\cal S}_{fin}({\cal H})\rightarrow
\left\langle \exp \Omega \mid F\odot G\right\rangle $ has a unique $\Lambda $%
-extension. The identity (\ref{coh.8}) and the arguments used for the
derivation of Lemma \ref{coherent} imply that 
\begin{equation}
\left\langle \exp \Omega \mid F\odot G\right\rangle =\left\langle
W^{-1}F\mid JW^{-1}G\right\rangle  \label{coh.9}
\end{equation}
is true for all $F,G\in {\cal S}_{fin}({\cal H}).$ Since $%
(W^{-1}F)^{*}=W^{-1}F^{*},$ this identity is equivalent to (\ref{Gauss5}).

\section{Gaussian integrals\label{Integration}}

In this appendix we indicate a method which allows to calculate the Gaussian
functional (\ref{Gauss2}) on the superalgebra ${\cal S}_{fin}({\cal H})$ by
Gaussian integration without use of superanalysis.

\subsection{The bosonic functional\label{Boson}}

The tensors $F\in {\cal S}({\cal H}_{\overline{0}})={\cal T}^{+}({\cal H}_{%
\overline{0}})$ of the bosonic Fock space can be represented by the
functions 
\begin{equation}
x\in {\cal E}_{\overline{0}}\rightarrow \varphi _F(x,x^{*})=\left\langle
F\mid \exp (x+x^{*})\right\rangle .  \label{int.1}
\end{equation}
These functions are restrictions of the functions (\ref{sdiff.4a}) to $\Xi
=id\otimes F$ and to arguments $\zeta =id\otimes x,x\in {\cal E}_{\overline{0%
}}.$

The bosonic part of the Gaussian functional (\ref{Gauss2}) can be calculated
by integration with respect to the canonical Gaussian promeasure $\mu _0$ on 
${\cal E}_{\overline{0}}$ (more precisely on the underlying real space $%
{\cal E}_{\overline{0}{\Bbb R}}$). This promeasure is characterized by the
Laplace transform 
\begin{equation}
\int_{{\cal E}_{\overline{0}{\Bbb R}}}e^{\left\langle f^{*}\mid
x\right\rangle +\left\langle g\mid x^{*}\right\rangle }\mu
_0(dx,dx^{*})=e^{\left\langle f^{*}\mid g\right\rangle }\text{ if }f,g\in 
{\cal E}_{\overline{0}}.  \label{int.2}
\end{equation}

\begin{lemma}
The Gaussian functional (\ref{Gauss2}) restricted to $F\in {\cal S}_{alg}(%
{\cal H}_{\overline{0}})={\cal T}_{alg}^{+}({\cal H}_{\overline{0}})$ has
the integral representation 
\begin{equation}
\left\langle \exp \Omega _0\mid F\right\rangle =\int_{{\cal E}_{\overline{0}%
{\Bbb R}}}\varphi _F(x,x^{*})\mu _0(dx,dx^{*}).  \label{int.3}
\end{equation}
\end{lemma}

{\bf Proof} The proof follows from a simple calculation for coherent states $%
F=\exp f,f\in {\cal H}_{\overline{0}}.$ Then (\ref{Gauss2a}) implies $%
\left\langle \exp \Omega _0\mid \exp (f^{*}+g)\right\rangle =e^{\omega
(f^{*},g)}=e^{\left\langle f^{*}\mid g\right\rangle }.$ A comparison with (%
\ref{int.2}) yields the result. \hfill $\square $ \smallskip

\subsection{The fermionic functional\label{Fermion}}

There is no canonical counterpart to the functions (\ref{int.1}) in the
fermionic case. But using the ordering prescription on the Hilbert space $%
{\cal H}_{\overline{1}}$ one can transfer a great part of the constructions
given for the symmetric tensor algebra in Sect.\ref{Boson} to the
antisymmetric tensor algebra. These techniques have been introduced in
quantum field theory by K. O. Friedrichs \cite{Friedrichs:1953}, and they
have been further developed by Klauder\cite{Klauder:1960} and by
Garbaczewski and Rzewuski\cite{Garb/Rzew:1974}, for the following
calculations see \cite{Kupsch:1990}\cite{Kupsch:1991}. In this appendix we
only use the ordering of a basis. The cited literature gives more general
ordering prescriptions. The disadvantage of all these methods is their
explicit dependence on the chosen ordering prescription, which is not
invariant against general unitary transformations. On the other hand, these
constructions do not need an extension to a superspace.

Let $\left\{ e_a\mid a\in {\Bbb N}\right\} $ be an orthonormal basis of $%
{\cal E}_{\overline{1}}.$ If ${\bf A}$ is the finite ordered subset\\${\bf A}%
=\left\{ a_1<a_2<...<a_p\right\} \subset {\Bbb N}$ of the integers, the
tensor $e_{{\bf A}}=e_{a_1}\wedge ...\wedge e_{a_p}$ is an element of an ON
basis of ${\cal T}_p^{-}({\cal E}_{\overline{1}})$ and $e_{{\bf A}%
}^{*}=e_{a_p}^{*}\wedge ...\wedge e_{a_1}^{*}$ is an element of a basis of $%
{\cal T}_p^{-}({\cal E}_{\overline{1}}^{*}).$ Moreover, $\left\{ e_{{\bf A}%
}^{*}\wedge e_{{\bf B}}\mid {\bf A\in }{\it P}({\Bbb N}),{\bf B}\in {\it P}(%
{\Bbb N})\right\} ,$ where ${\it P}({\Bbb N})$ is the power set of ${\Bbb N}%
, $ is an orthonormal basis of the Fock space ${\cal S}({\cal H}_{\overline{1%
}})={\cal T}^{-}({\cal H}_{\overline{1}}).$ Any $F\in {\cal S}({\cal H}_{%
\overline{1}})={\cal T}^{-}({\cal H}_{\overline{1}})$ has the decomposition $%
F=\sum_{{\bf A},{\bf B}}f({\bf A},{\bf B})e_{{\bf A}}^{*}\wedge e_{{\bf B}},$
where the sum extends over ${\bf A\in }{\it P}({\Bbb N})$ and ${\bf B}\in
{\it P}({\Bbb N}).$ The fermionic part of the tensor (\ref{Gauss1}) $\Omega $
has the representation $\Omega _1=\sum_{a=1}^\infty e_a^{*}\wedge e_a$ (with
respect to any orthonormal basis of ${\cal E}_{\overline{1}}),$ and the
exponential is calculated as $\exp \Omega _1=\sum_{{\bf A\in }{\it P}({\Bbb N%
})}e_{{\bf A}}^{*}\wedge e_{{\bf A}}.$ The Gaussian functional has therefore
the representation 
\begin{equation}
\left\langle \exp \Omega _1\mid F\right\rangle =\sum_{{\bf A\in }{\it P}(%
{\Bbb N})}f({\bf A},{\bf A})  \label{int.5}
\end{equation}
for all $F\in {\cal T}_{fin}^{-}({\cal H}_{\overline{1}}).$ To define
functions like (\ref{int.1}) also for tensors $F\in {\cal T}_{fin}^{-}({\cal %
H}_{\overline{1}})$ we need a nontrivial substitute for the coherent states.
We first introduce the numerical monomials $x\in {\cal H}_{\overline{1}%
}\rightarrow x({\bf A}):=\prod_{a\in {\bf A}}\left( e_a,x\right) \in {\Bbb C.%
}$. The substitutes for powers $"\frac 1{n!}x^n"$ are then the tensors $%
\sum_{{\bf A},\left| {\bf A}\right| =n}x({\bf A})e_{{\bf A}}\in {\cal T}%
_n^{-}({\cal E}_{\overline{1}}),$ and the substitute for a coherent state is
the absolutely converging series $E(x,x^{*})=\sum_{{\bf A},{\bf B}}x({\bf A}%
)x({\bf B})e_{{\bf A}}^{*}\wedge e_{{\bf B}}\in {\cal T}^{-}({\cal H}_{%
\overline{1}}),$ where the summation extends over ${\bf A\in }{\it P}({\Bbb N%
})$ and ${\bf B}\in {\it P}({\Bbb N}).$ It is exactly this definition where
the ordering prescription enters. Fermionic functions can then be defined in
the same way as the bosonic functions (\ref{int.1})

\begin{equation}
\psi _F(x,x^{*}):=\left\langle E(x,x^{*})\mid F\right\rangle =\sum_{{\bf A,B}%
}f({\bf B},{\bf A})\overline{x({\bf A})}x({\bf B})  \label{int.8}
\end{equation}
We integrate the numerical function (\ref{int.8}) with the canonical
Gaussian promeasure $\mu _1$ of the space ${\cal E}_{\overline{1}{\Bbb R}},$
see (\ref{int.2}) for the corresponding measure on ${\cal E}_{\overline{0}%
{\Bbb R}},$ and we obtain exactly the Gaussian functional (\ref{int.5}) 
\begin{equation}
\int_{{\cal E}_{\overline{1}{\Bbb R}}}\psi _F(x,x^{*})\mu
_1(dx,dx^{*})=\sum_{{\bf A\in }{\it P}({\Bbb N})}f({\bf A},{\bf A}).
\label{int.9}
\end{equation}
The promeasures $\mu _0$ and $\mu _1$ have unique extensions to $\sigma -$%
additive measures on ${\cal E}_{\overline{0}{\Bbb R}}^{-},$ or ${\cal E}_{%
\overline{1}{\Bbb R}}^{-},$ respectively. The functions $\left\{ \varphi
_F(x,x^{*})\mid F\in {\cal S}({\cal H}_{\overline{0}})\right\} $ (or $%
\left\{ \psi _G(x,x^{*})\mid G\in {\cal S}({\cal H}_{\overline{1}})\right\} $%
) are then square integrable functions of the variables $x\in {\cal E}_{%
\overline{0}{\Bbb R}}^{-}$ (or $x\in {\cal E}_{\overline{1}{\Bbb R}}^{-}$).
\bibliographystyle{plain}

\end{document}